\documentclass[fleqn,usenatbib]{mnras}

\usepackage{newtxtext,newtxmath}

\usepackage{siunitx}
\usepackage[T1]{fontenc}
\usepackage{ae,aecompl}
\usepackage{color,colortbl}
\definecolor{red}{rgb}{1,0.5,0.5}
\usepackage{graphicx}	% Including figure files
\usepackage{amsmath}	% Advanced maths commands
\usepackage{amssymb}	% Extra maths symbols

\newcommand{\cfour}{C~{\small IV}}
\newcommand{\cfourmath}{\text{C{\tiny IV}}}

\title[Metal-tracing the faint sources of reionisation]{The role of galaxies and AGN in reionising the IGM - II: metal-tracing the faint sources of reionisation at $5\lesssim z\lesssim6$}

\author[R. A. Meyer et al.]{
Romain A. Meyer,$^{1}$\thanks{E-mail: r.meyer.17@ucl.ac.uk}
Sarah E. I. Bosman,$^{1}$
Koki Kakiichi$^{1}$
and Richard S. Ellis$^{1}$
\\
$^{1}$Department of Physics and Astronomy, University College London, Gower Street, London WC1E 6BT, UK\\}

\date{Accepted XXX. Received YYY; in original form ZZZ}

\pubyear{2018}

\begin{document}
\label{firstpage}
\pagerange{\pageref{firstpage}--\pageref{lastpage}}
\maketitle
% Abstract of the paper
\begin{abstract}
We present a new method to study the contribution of faint sources to the UV background using the 1D correlation of metal absorbers with the intergalactic medium (IGM) transmission in a Quasi Stellar Object (QSO) sightline. We take advantage of a sample of $25$ high signal-to-noise ratio QSO spectra to retrieve $150$ triply-ionised carbon (\cfour) absorbers at $4.5\lesssim z\lesssim 6.2$, of which $37$ systems whose expected H{~\small I} absorption lie in the Lyman-$\alpha$ forest. We derive improved constraints on the cosmic density of \cfour \,at $4.3< z < 6.2$ and infer from abundance-matching that \cfour \,absorbers trace $M_{\text{UV}}\lesssim -16$ galaxies. 
Correlation with the Lyman-$\alpha$ forest of the QSOs indicates that these objects are surrounded by a highly opaque region at $r\lesssim 5 $ cMpc/h followed by an excess of transmission at $r\gtrsim 10$ cMpc/h detected at $2.7\sigma$. This is in contrast to equivalent measurements at lower redshifts where only the opaque trough is detected.
We interpret this excess as a statistical enhancement of the local photoionisation rate due to clustered faint galaxies around the \cfour \,absorbers. Using the analytical framework described in Paper I of this series, we derive a constraint on the average product of the escape fraction and the Lyman continuum photon production efficiency of the galaxy population clustered around the \cfour \,absorbers, $\log \langle f_{\text{esc}}\xi_{\text{ion}}\rangle /[{\rm erg^{-1}~Hz}] =  25.01^{+0.30}_{-0.19}$. This implies that faint galaxies beyond the reach of current facilities may have harder radiation fields and/or larger escape fractions than currently detected objects at the end of the reionisation epoch.
\end{abstract}


%%%%%%%%%%%%%%%%%%%%%%%%%%%%%%%%%%%%%%%%%%%%%%%%%%

%%%%%%%%%%%%%%%%% BODY OF PAPER %%%%%%%%%%%%%%%%%%

\section{Introduction} \label{sec:intro}

Cosmic reionisation has been the object of much scrutiny in the last decades, resulting in remarkable progress on the determination of its timing. The period during which most of the cosmic neutral hydrogen gas was reionised is now constrained to lie in the redshift interval $6 \lesssim z \lesssim 15$  \citep{Planck16}. Evidence suggests that during this time it is actively star-forming galaxies that produced most of the required ionising photons \citep[e.g.][]{Robertson13,Puchwein18}, although some debate currently exists around a possible  contribution of active galactic nuclei (AGN) at $z\sim 6$ \citep{Giallongo15,DAloisio17,Chardin17, Parsa18}. 

The major challenge in investigating the physical processes governing cosmic reionisation is the difficulty of probing the nature of the ionising radiation from sources at $z>7$.
Deep surveys with the \textit{Hubble Space Telescope} combined with the magnification of gravitational lensing by large clusters have recently pushed the census of galaxies to $z\sim 10$ \citep[e.g.][]{Bouwens07,Ellis13,Bouwens15,Atek15,Ishigaki15,McLeod15,Livermore17, Ishigaki18,Atek18}. However, at redshifts beyond $z\gtrsim 7$, only a handful of galaxies and active galactic nuclei have been spectroscopically confirmed and studied in any detail \citep[e.g.][]{Mortlock11,Laporte15,RobertsBorsani16,Laporte17a,Laporte17b,Banados18}. 
A major impasse in understanding the role of star-forming galaxies is the uncertain fraction of ionising photons that can escape into the intergalactic medium (IGM). Using the demographics of galaxies observed out to redshifts $z\simeq$10 and the likely spectral energy distributions of young, metal-poor stellar populations, \cite{Robertson15} estimate an average escape fraction of $f_{\text{esc}}\simeq$10-20\% is required. Direct measures of the escape fraction are only possible at redshifts below $z\simeq$3 because Lyman continuum (LyC) photons emitted from star-forming galaxies are increasingly absorbed by the IGM at higher redshift.

In the meantime, the number and quality of quasi-stellar object (QSO) sightlines probing the IGM at the end of the reionisation era has multiplied quickly following systematic searches in the Sloan Digital Sky Survey \citep[SDSS,][]{Jiang16}, Panoramic Survey Telescope and Rapid Response System data\citep[Pan-STARSS,][]{Kaiser10,Banados16}, Dark Energy Survey - Visible and Infrared Survey Telescope for Astronomy (VISTA) Hemisphere Survey \citep[DES-VHS,][]{Reed15}, Subaru High-$z$ Exploration of Low-Luminosity Quasars survey \citep[SHELLQS,][]{Matsuoka16} VISTA Kilo-Degree Infrared Galaxy Survey \citep[VIKING,][]{Venemans13,Carnall15} and UKIRT Infrared Deep Sky Survey \citep[UKIDSS,][]{Venemans07,Mortlock09,Mortlock11}. QSO sightlines offer a powerful tool to study the IGM at the end of the reionisation era \citep[e.g.][]{Fan06,Becker15rev} and the cosmic metal enrichment history \citep[e.g.][]{RyanWeber2009,Becker09,DOdorico10,DOdorico13}. As a result of the increased number of sightlines now available \citep{Bosman18}, a large scatter in the IGM opacity at $z\sim 6$ has been revealed with individual cases showing unexpected large opaque regions \citep{Becker15,Becker18}. The physical origin of these features remains unclear, although a late reionization model powered by galaxies claims to address these \citep{Kulkarni18}.

In \citet{Kakiichi18} (thereafter Paper I), we proposed a new method to directly measure the escape fraction of faint galaxies based on their clustering around luminous spectroscopically-confirmed galaxies which provide a local enhancement of transmission in the Lyman-$\alpha$ forest. By charting the distribution of galaxies close to the sightline of a bright QSO we constructed the cross-correlation with the IGM transmission, revealing a tentative indication of a statistical `proximity effect'. In the present study, we aim at measuring a similar correlation of galaxies to the IGM transmission but using a different tracer. Triply-ionised carbon (\cfour) is the most common metal absorbing species in QSO spectra \citep[][, thereafter DO13]{Becker09, DOdorico13} and, at low redshifts, associated with the metal-enriched halos of galaxies \citep[e.g.][]{Adelberger03,Adelberger05,Steidel10,Ford13,Turner2014}. At higher-redshift, little is known about the appropriate host galaxies, although some estimate these objects to have quite faint ultraviolet (UV) luminosities \citep{Becker15rev}. It is however expected that \cfour \, absorbers lie $\lesssim 100$ pkpc from their host \citep[e.g.][]{Oppenheimer09, Bird16, Keating16}. More recently, \citep{DOdorico18} reported the recent detection of a galaxy $40$ pkpc away from a DLA at $z\simeq 5.94$. Although there is no evidence for a consistent link between DLAs and \cfour\, absorbers at that redshift, the authors also report the potential detection of a weak associated \cfour\, absorption. This would support the idea that potential host galaxies can be found indeed very close to \cfour\, absorbers. Metal-tracing these faint sources should, in principle, allow us to probe the ionising capability of intrinsically faint galaxies well beyond reach of current spectroscopic facilities. Moreover, because metal absorbers lie directly on the QSO sightline, we can probe the IGM transmission around their hosts on the scales of $0.1\lesssim r \lesssim 1$ pMpc unattainable in the approach introduced in Paper I which uses nearby LBGs.

In this second paper in the series, we take advantage of a large sample of $z>5.4$ QSO spectra to study the abundance and distribution of \cfour \,absorbers. We then study the 1D correlation of these absorbers with the IGM transmission measured in the Lyman-$\alpha$ forest of the QSO to assess their impact on the IGM. Our study focuses on \cfour \,absorbers at $4.3<z<6.2$, but future possibilities include other metals and potentially studies of the redshift evolution of such correlations.

The plan for this paper is as follows. Section \ref{sec:data} introduces our observational sample of QSO spectra and the initial data reduction. Section \ref{sec:method} details our semi-automated search for \cfour \,absorbers and the result of our search. We present in section \ref{sec:results} the new constraints on the $4.3<z<6.2$ \cfour \,cosmic density derived from our large sample. We also present our measurement of the \cfour-IGM transmission 1D correlation. Section \ref{sec:models} presents two models of the said correlation. In Section \ref{sec:physics} we then discuss the nature of the \cfour\, absorbers host galaxies, and our evidence for an enhanced transmission in the IGM surrounding \cfour \,absorbers. We put a constraint on the product of the escape fraction and the LyC photon production efficiency. We conclude in Section \ref{sec:conclusion} with a brief summary of our findings and future prospects for this new method measuring the escape fraction at the end of the reionisation era. 

Throughout this paper we adopt the Planck 2015 cosmology $(\Omega_m,\Omega_\Lambda, \Omega_b, h,\sigma_8, n_s)=(0.3089, 0.6911, 0.04860, 0.6774,0.8159, 0.9667)$ \citep{Planck16}. We use pkpc and pMpc (ckpc and cMpc) to indicate distances in proper (comoving) units. 

\section{Methods}
\subsection{Observations} \label{sec:data}

\begin{table}
\caption{QSO sightlines used in this work. References: (1) \citet{McGreer15}; (2) \citet{Bosman18}; (3) \citet{Becker15}; (4) \citet{Becker06}; (5) \citet{Eilers17}; (6)  reduced from online archives in this work (see Table \ref{table:QSO_Obs}).}
\label{QSO_all}
\centering
\begin{tabular}{lrrrr}
\hline\hline
QSO name&$z$&Instrument & SNR & ref.\\
\hline
J1148+0702 & 6.419&HIRES&29.7&(1)\\
J0100+2802&6.30&XShooter&85.2&(2)\\
J1030+0524&6.28&XShooter&28.0&(1)\\
J0050+3445&6.25&ESI&24.4&(3)\\
J1048+4637&6.198&HIRES&29.2&(4)\\
J1319+0950&6.132&XShooter&96.8&(3)\\
J1509--1749&6.12&XShooter&88.9&(1)\\
J2315--0023&6.117&ESI&29.8&(3)\\
J1602+4228&6.09&ESI&33.3&(2)\\
J0353+0104&6.072&ESI&80.7&(3)\\
J0842+1218&6.07 &ESI&18.0&(6)\\
J2054--0005&6.062&ESI&39.5&(3)\\
J1306+0356&6.016&XShooter&55.8&(1)\\
J1137+3549&6.01&ESI&31.7&(2)\\
J0818+1722&6.00&XShooter&114.0&(6)\\
J1411+1217&5.927&ESI&15.9&(1)\\
J0148+0600&5.923&XShooter&128.0&(3)\\
J0005--0006&5.85&ESI&28.8&(5)\\
J0840+5624&5.844&ESI&17.6&(1)\\
J0836+0054&5.81&XShooter&93.4&(1)\\
J0002+2550&5.80&ESI&121.0&(6)\\
J1044--0125&5.782&ESI&49.2&(3)\\
J0927+2001&5.772&XShooter&73.7&(3)\\
J1022+2252&5.47&ESI& 19.0&(6)\\
J0231--0728&5.42&XShooter&115.0&(6)\\
\hline
\end{tabular}
\end{table}

\begin{table*}
\caption{QSO sightlines in addition to those listed in \citet{Bosman18}. These spectra were reduced either from Keck Observatory Archive (KOA) data for the ESI spectrograph, or from the XShooter archive. \label{table:QSO_Obs}}
\begin{tabular}{lrrrrrrr}
\hline\hline
QSO name&$z$&Instrument&$t_\text{total} (s)$ & SNR &ref.&PID&P.I. \\
\hline
J0842+1218&6.07 &ESI&2400& 18 &\citet{DeRosa11}&U085E&R. Becker\\
J0002+2550&5.80&ESI&22200&121 & \citet{Fan04}&H31aE&A. Cowie\\
&&&&& &H46aE&Kakazu\\
J1022+2252&5.47&ESI&6000& 19&--& U130Ei&G. Becker\\
J0818+1722&6.00&XShooter&20750&114 & \citet{Fan06},&084A-0550 & D'Odorico\\
 & & & & & \citet{Dodorico11} & 086A-0574& De Rosa \\
 & & & & & & 088A-0897& De Rosa \\ 
J0231--0728&5.42&XShooter&21600&115 &\citet{Becker12}&084A-0574& G. Becker\\
\hline
\end{tabular}
\end{table*}

Our sample consists of $25$ optical spectra of quasars with $z_\text{source}>5.4$ originating from the Echellette Spectrograph and Imager (ESI) on the Keck II telescope \citep{Sheinis02}, the XShooter instrument on Cassegrain UT2 \citep{Vernet11}, and the High Resolution Echelle Spectrometer (HIRES, \citealt{Vogt94}) as shown in Table \ref{QSO_all}. Out of these, 20 spectra are re-used from the quasar sample of \citet{Bosman18}. These spectra were selected for their high signal-to-noise ratios (SNR $>17$) measured over the $1265<\lambda<1275$ \AA \ range via 
\begin{equation}
\text{SNR} = \left< \frac{F}{\epsilon} \right>\cdot \sqrt{N_{60}} \, \text{,}
\end{equation}
where $F$ is the flux, $\epsilon$ is the error, and $N_{60}$ is the number of spectral bins covering $60$ km s$^{-1}$ \footnote{The scale of $60$ km/s is a convention chosen in \citet{Bosman18} as a suitable intermediate scale for their large range of QSO spectra.}. An exception is the quasar J1411+1217 which is included despite its relatively poor SNR ($=15.9$) due to the presence of a particularly broad \cfour \,absorber. Out of these objects, 7 originate in a study from \citet{McGreer15} (3 of which were independently re-reduced), 6 from \citet{Becker15}, one from \citet{Becker06}, one from \citet{Eilers17} (re-reduced), and 5 from \citet{Bosman18} (of which 2 are archival). 

Together with these $20$ sightlines, we reduced 5 additional spectra from the Keck Observatory Archive\footnote{\url{https://koa.ipac.caltech.edu/cgi-bin/KOA/nph-KOAlogin}} and the XShooter search tool for the ESO Science Archive Facility\footnote{\url{http://archive.eso.org/wdb/wdb/eso/xshooter/form}} as summarised in Table \ref{table:QSO_Obs}. The spectra were extracted optimally making use of the calibration files (flat fields and standard star exposures) available in the archives for each set of observations. After performing sky subtraction, different observations of the same sightline are combined when necessary. The implementations of optimal extraction, sky subtraction, and telluric correction used herein are outlined in more detail in \citet{Horne86}, \citet{Kelson03} and \citet{Becker12} respectively. Our final sample contains ESI spectra with either $55.9$ km/s or $74.6$ km/s resolution depending on the slit, while XShooter spectra have either a $28.0$ km/s or $34.1$ km/s resolution and HIRES spectra have $6$ km/s resolution. We have re-binned the HIRES spectra by a factor $5$ to match the XShooter resolution to facilitate the search for \cfour \, absorbers redwards of Lyman-$\alpha$.

To measure transmitted Lyman-$\alpha$ fluxes bluewards of the Lyman-$\alpha$ emission line, we fit each spectrum with a power-law continuum. This power-law is fitted over the relatively featureless wavelength interval $1270$ \AA--$1450$ \AA \ in the rest frame. We exclude pixels affected by sky lines and use two rounds of sigma-clipping with thresholds of $\vert F_\text{PL} - F_\text{obs}\vert<2\epsilon$ and $1.5\epsilon$, where $\epsilon$ is the observational error and $F_\text{PL}, F_\text{obs}$ are the values of the power-law fit and the quasar flux, respectively.

The Lyman-$\alpha$ forest of $z\gtrsim 5$ is characterised by high absorption at all wavelengths, making the continuum notoriously difficult to determine or model. Due to the sparsity of transmitted flux, we make no attempt at modelling the continuum bluewards of the  Lyman-$\alpha$ emission line beyond a power-law. At lower redshift, more advanced techniques have been used and include subtracting profiles of weak intrinsic absorption and emission lines (e.g. \citealt{Crighton11}). When comparing to such studies, a potential worry is a bias in the Lyman-$\alpha$ forest self-correlation on the scales of such features that are not removed by a simple power-law. We however verified that the self-correlation was not deviated from unity by more than 1$\sigma$ on any scale \citep[see][]{Bosman18}.

\begin{figure*}
 \centering
\includegraphics[width=0.7\textwidth]{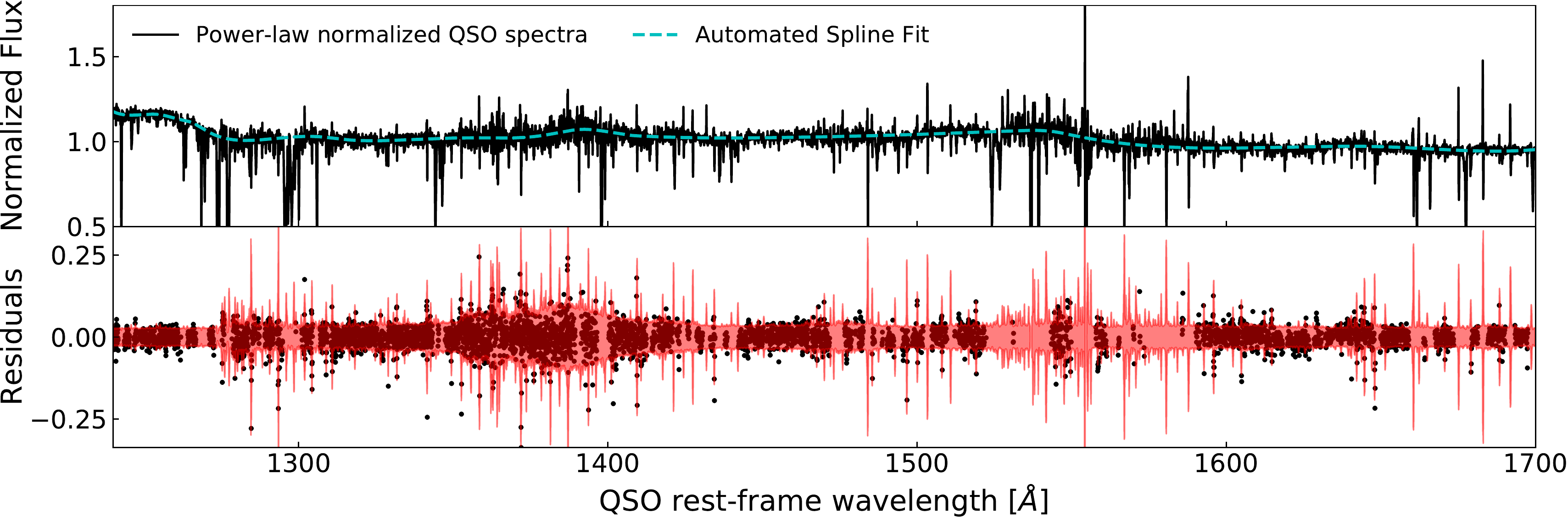} 
\includegraphics[width=0.26\textwidth]{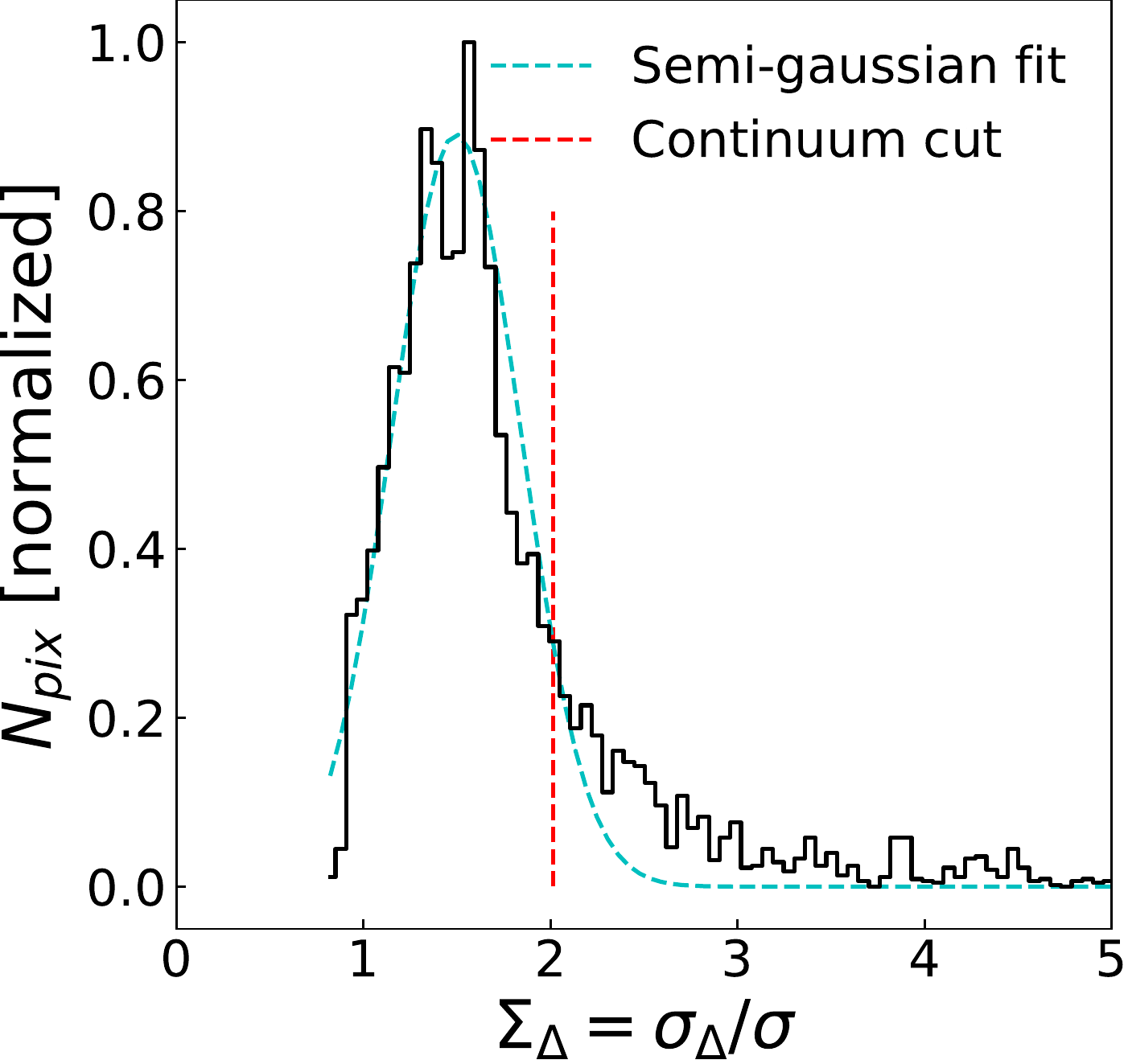} \\ \vspace{0.5cm}
\includegraphics[width=\textwidth]{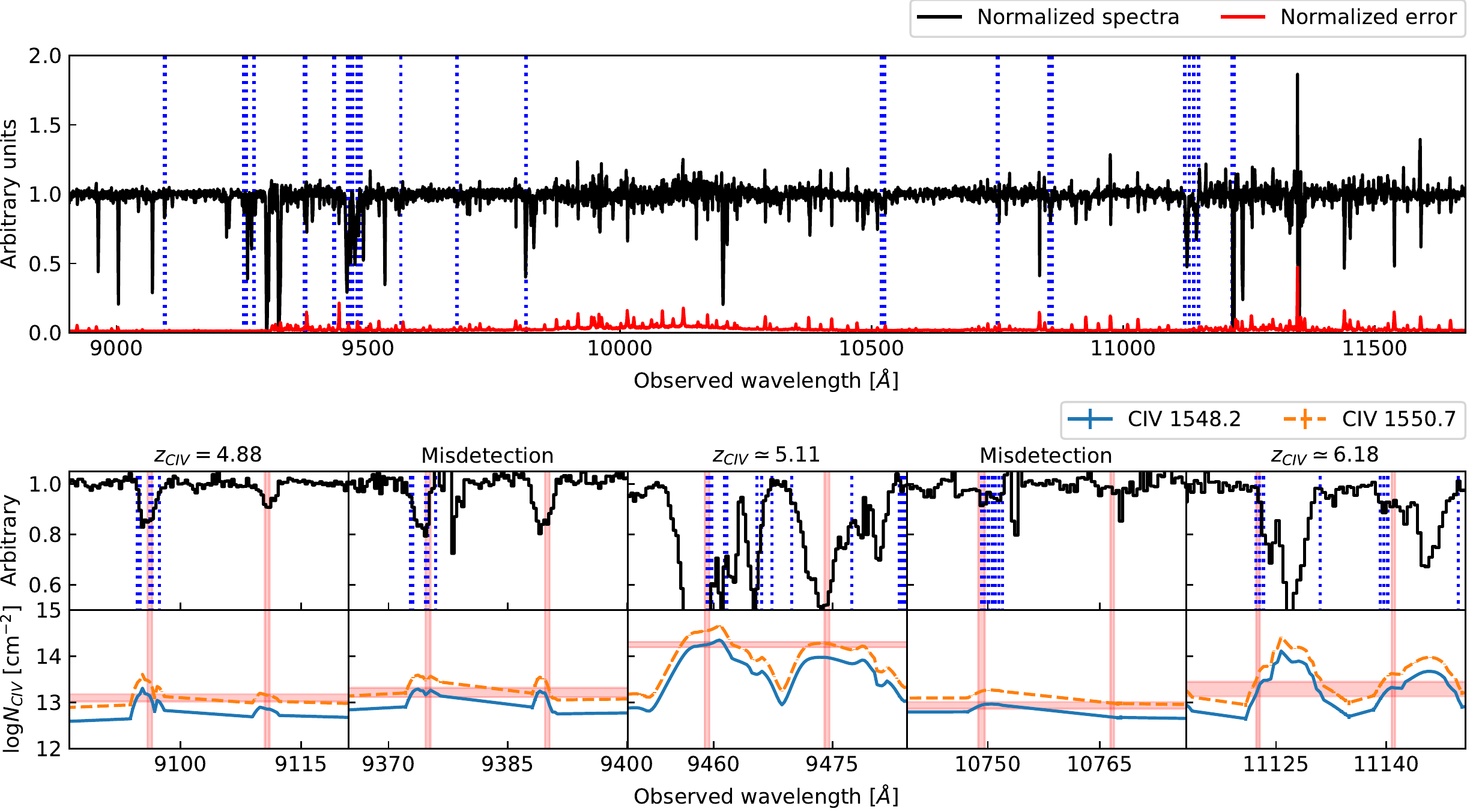}
\caption{Illustration of our automated search procedure for \cfour \, absorbers as demonstrated for the QSO J0100+2802. \textbf{Upper left panel:} Measured (power-law normalized) flux in black, fitted continuum in cyan. The residuals present very little to no slow-varying residual trend and are consistent within the $2\sigma$ errors (red). \textbf{Upper right panel:} Histogram (black) of the $\Delta F$ variance $\sigma_\Delta$ divided by the ``error array'' variance $\sigma$, and the associated threshold (red line) for non-continuum pixels, taken to be a $1.5\sigma$ cut away from the mean of a Gaussian (cyan) fitted to the values lower than the expected mean of $\sqrt{2}$. \textbf{Middle panel:} The intrinsic continuum normalized flux (black) and noise (red) are overlaid with the algorithm-identified \cfour\, absorbers (indicated by blue dotted lines) 
(see Section \ref{sec:method} for the search strategy). \textbf{Lower panel:} Zoom-in on some flagged \cfour \, candidates from the middle panel. The algorithm computes the column density for both \cfour \, 1548 \AA, 1550 \AA \, at every pixel, and flags every pair with matching column density at the correct separation, with the tolerance indicated in red. False detections and misalignments due to sigma-clipping can be easily removed by eye or when fitting with \textsc{vpfit}. \label{fig:QUICFit} }
\end{figure*}

\subsection{QSO Broad Emission Lines and continuum fitting}\label{sec:quicfit}

Previous systematic searches for \cfour \,or metal absorptions in the continuum of $z\simeq 6$ QSO \citep[e.g.][]{RyanWeber2009,Becker2011, Simcoe2011,DOdorico13,Bosman2017, Codoreanu18} proceed by removing first the power-law continuum of the QSO and secondly fitting the Broad Emission Lines (BEL). The BEL fit is usually performed with splines in an iterative ``by-eye'' process where the user supervises the selection of knot points. Since our sample contains $25$ QSO spectra, we developed an automated QSO continuum spline fitting (redwards of Lyman-$\alpha$ only) further described below. 
The first step is to determine which part of the spectra are devoid of narrow emission lines or Broad Absorption Lines.  A ``fit-by-eye" procedure would select such regions as representative of the QSO continuum. In fact, human users select regions where \textit{the pixel-to-pixel flux variation is consistent with the error array}. Mathematically, we expect that for a slowly varying spectrum with high enough resolution, the pixel-to-pixel flux difference is distributed as
\begin{equation}
\Delta F_{i} = F_{i} - F_{i+1} \sim \mathcal{N}(0,2\sigma_i^2) \, \text{ , }
\end{equation}
where $F_i$ is the flux recorded at pixel $i$, $\sigma_i$ the corresponding error, and $\mathcal{N}(\mu,\sigma^2)$ the normal distribution with mean $\mu$ and variance $\sigma^2$. We thus estimate the variance of the flux difference $\sigma_\Delta$ by simply computing a running variance on the flux variation $\Delta F_i$. The running variance is taken as the square of the standard deviation of $\Delta F$ in a 40 pixel wide window centered on pixel $i$. We then take the ratio between the $\sigma_\Delta$ and the error array $\sigma$ at each pixel $i$. 
The distribution of the resulting variable,  $\Sigma_\Delta = \sigma_\Delta/\sigma$ is a Gaussian distribution of mean $\sqrt{2}$ with a tail of larger values, as expected (see upper right corner of Fig. \ref{fig:QUICFit}).
We fit a Gaussian to the low-value wing of the distribution of $\Sigma_\Delta$ and exclude all pixels at $>N\sigma$ from the mean of the fitted Gaussian from the ``continuum pixels''. $N$ is a parameter chosen by the user, and $1\leq N \leq 3$ is used for all our spectra.
We note however that in large, completely absorbed features, the pixel-to-pixel variation differs from the noise distribution array \emph{only in the wings of the absorption}. To remove pixels at the bottom of these absorption troughs, we run a matched-filter with a Gaussian kernel with a width of $5$ \AA . Any pixel with $S/N_{\text{matched-filter}}$ greater than a chosen threshold (here $>3$) should be rejected from the continuum.

The second step of the algorithm is based on the idea that the BEL that provide the complexity of the QSO continuum fitting are not a nuisance but rather a powerful indicator of where the splines knot points should be located. Based on a BEL rest-frame wavelength list given by the user, we assign a knot point on top of each BEL with some tolerance (provided by the user as well, but $\sim 10$ \AA\ is a reasonable choice), and another intermediate knot point in between each BEL knots. We minimize the $\chi^2$ of the spline fit on the previously selected continuum pixels by moving the knot points in the assigned tolerance regions to yield our final fit. An example of a resulting fit is shown in Fig. \ref{fig:QUICFit}. The automated continuum method successfully fits the BEL features as well as avoiding the regions contaminated by skylines.  The \textsc{quicfit} (QUasar Intrinsic Continuum FITter) code is publicly available at \url{https://github.com/rameyer/QUICFit}.

\subsection{\cfour \, identification} \label{sec:method}
It is possible to search for many metal absorber species redwards of the QSO Lyman-$\alpha$ emission at high-redshift, including amongst other O~{\small I}, C~{\small II}, Si~{\small II}, N~{\small V}, Si~{\small IV}, Al~{\small II}, Al~{\small III}. However \cfour\, is the most useful due to its ubiquity and the fact that it is reliably identifiable as a doublet. Once the BEL and power-law continuum of the QSO are subtracted, the processed spectrum is searched for \cfour \, doublets. We use a semi-automated identification algorithm.

Following \cite{Bosman2017}, we fit an inverted Gaussian profile to the optical depth $\tau = -\log F$ every $\Delta v \sim 10$ km s${}^{-1}$ interval, and apply $3$ iterations of 2-$\sigma$ clipping to fit the most prominent feature only. We then estimate the column density for each \cfour\, transition following the apparent optical depth method \citep{SavageSembach91}
\begin{equation}
N_{\cfourmath} =  \frac{m_e c}{\pi e^2 f\lambda} \int \ln G(v) \text{d} v \, \text{ , }
\end{equation}
where $G(v)$ is the fitted inverted Gaussian profile, $f = 0.095$ or $0.19$ is the oscillator strength for the  $\lambda = 1548.2, 1550.8$ \AA\ transitions, respectively. 

We select all pairs of Gaussian absorption profiles with $\log N_{\cfourmath} > 12.5$ and with a discrepancy in redshift $\Delta z < R/c$, where $R$ is the resolution in km s$^{-1}$ and a discrepancy in column density $\Delta \log N_{\cfourmath}$ within $2\sigma$ of the Gaussian fitting errors. We demonstrate in Fig. \ref{fig:QUICFit} the full fitting and search procedure on the sightline towards J0100+2802. The discrete nature of the search in wavelength space and the $\sigma$-clipping operations sometimes produces false detections or unreliable estimates of the column density of our absorbers. In order to account for these issues, these flagged candidates are then fitted with \textsc{vpfit} \citep{VPFIT} to confirm their nature and derive precise column densities. 

\begin{figure}
\centering
\includegraphics[height=0.25\textheight]{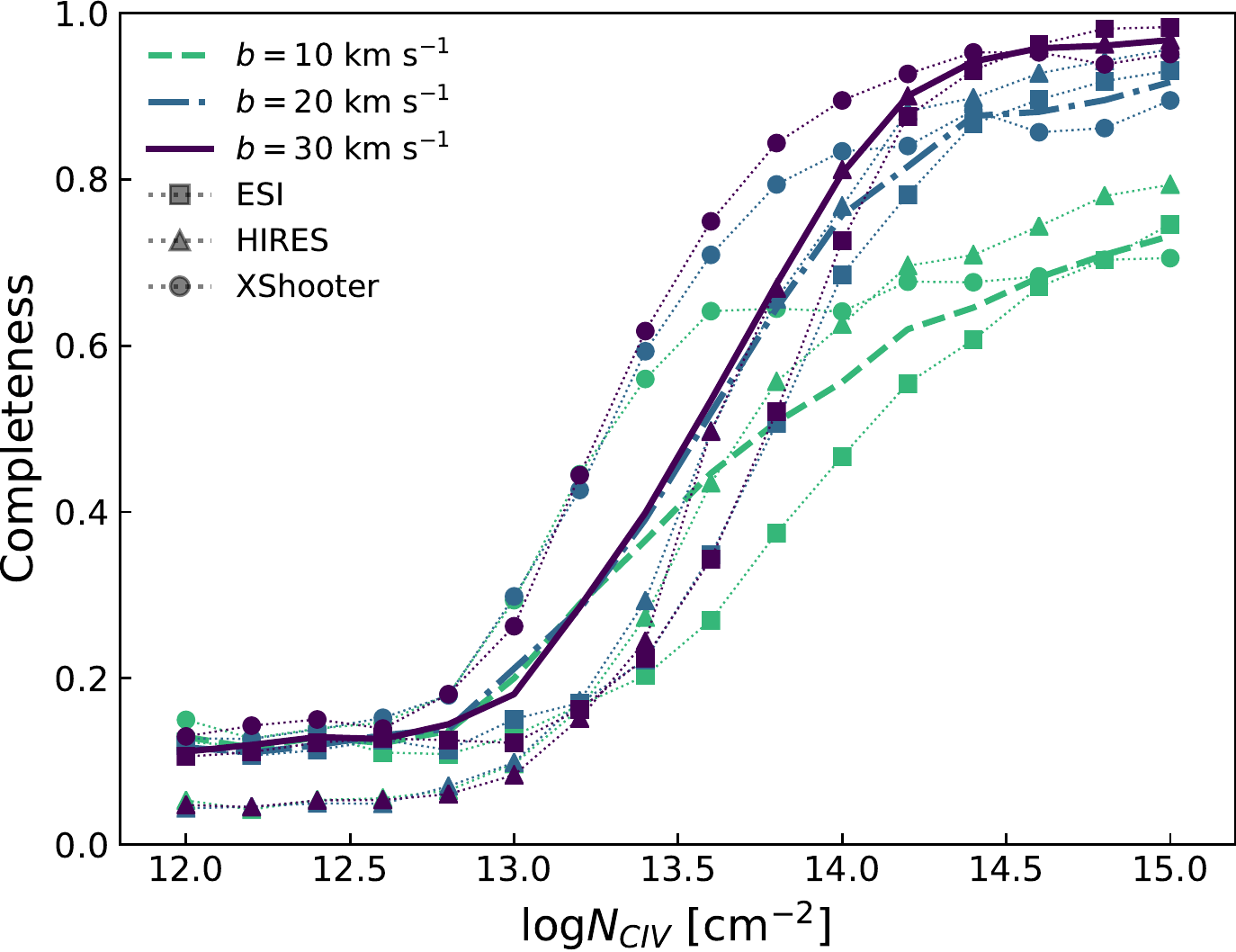}
\caption{Completeness of our \cfour \, semi-automated retrieval computed by the insertion of $1000$ \cfour \, doublets at random redshifts in the ESI (squares), HIRES (triangles) and XShooter (dots) for different Doppler parameters $b=10,20,30$ km s$^{-1}$ (green, blue, purple). The thin lines represent the average values per instrument, whereas the bold lines represent the weighted average values for our sample.  \label{fig:completeness}}
\end{figure}

To estimate the completeness of our search, we insert $1000$ mock \cfour \, absorbers for each of $b=10,20,30$ km s${}^{-1}$ Doppler parameters and 0.2 dex increments in column density from $N_{\cfourmath} = 10^{12}$ to $10^{14}$ cm$^{-2}$. We achieve a $90$\% completeness level around $\log N_{\cfourmath}\simeq 14.0$ for all $3$ instruments, assuming a Doppler parameter of $b=30$ km s$^{-1}$ (see Fig. \ref{fig:completeness}). This completeness is in good agreement with previous \cfour \, searches cited beforehand given the resolution and SNR of the QSO spectra at hand. 

\begin{figure}
\centering
\includegraphics[height=0.25\textheight]{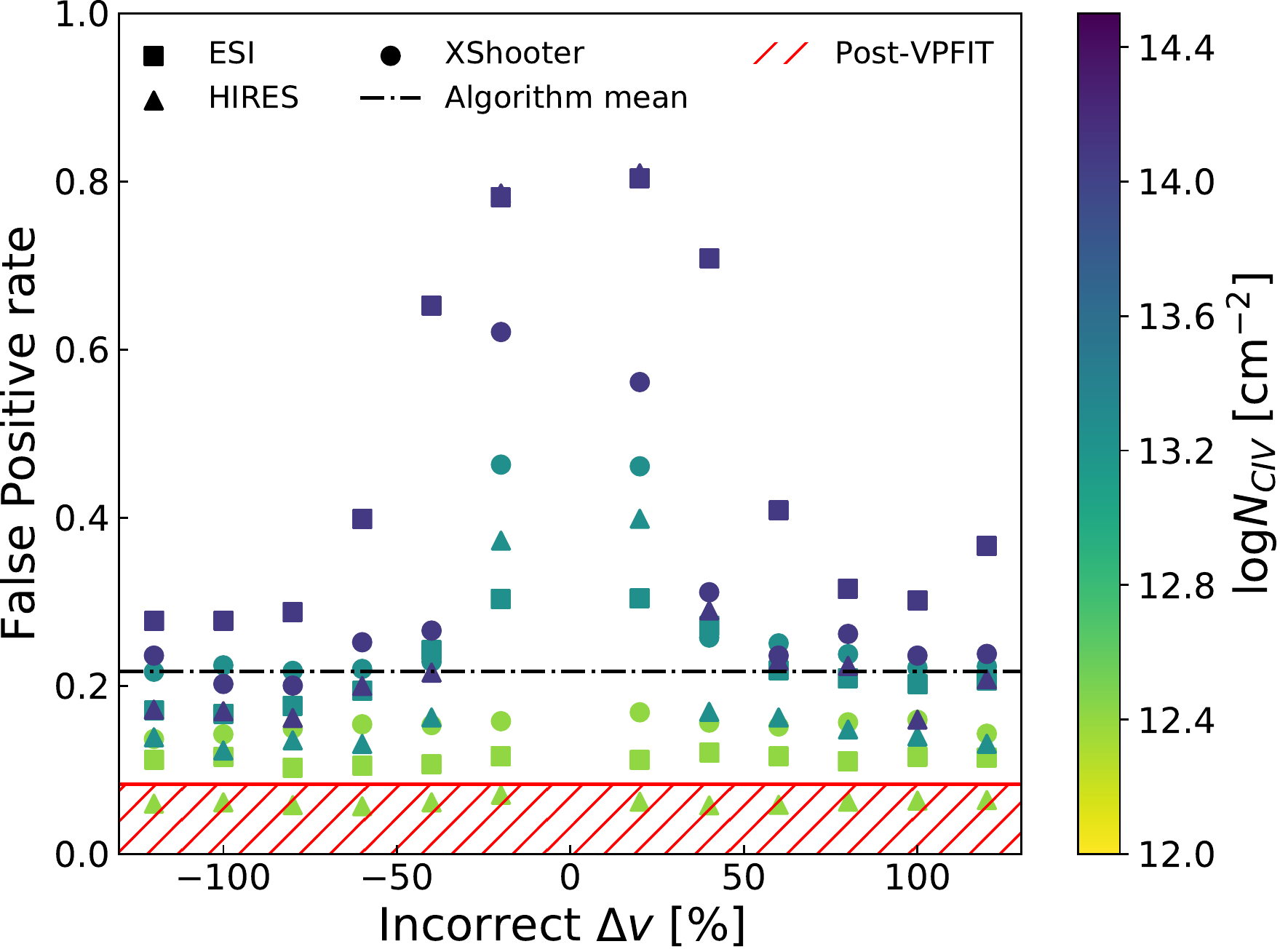}
\caption{False positive rate of the algorithmic detection rate of \cfour \, candidates for various velocity displacements, instruments and column densities (coloured dots according to the right hand scale bar). The higher densities are more contaminated due to the tolerance on the optical fitting and the broad absorption wings of theses systems. The weighted mean false positive rate for our particular sample is indicated in dotted black, whereas the allowed values of the final false positive rate in our sample after visual inspection is shown in hatched red.\label{fig:FPrate} }
\end{figure}

\begin{table}
\caption{Average completeness for the \cfour\, search. The completeness is a weighted average of the completeness values of Fig. \ref{fig:completeness} with $b=20$km/s for our specific set of QSO  sightlines (HIRES:2,  ESI:12, XShooter:13).} \label{table:completeness}
\centering
\begin{tabular}{rcccccc}
\hline\hline
$\log N_{\cfourmath} / [\text{cm}^{-2}] $ \vline & $ 13.0$ & $13.2$ & $13.4$ & $13.6$ & $13.8$ & $14$  \\
Completeness  \vline & $ 0.18$ & $0.27$ & $0.36$ & $0.49$ & $0.60$  &$0.70$ \\
$\log N_{\cfourmath} / [\text{cm}^{-2}] $ \vline & $14.2$ & $14.4$ & $14.6$ & $14.8$ & $15.0$ &  \\ 
Completeness \vline & $0.78$ & $0.82$ & $0.84$ & $0.85$ & $0.87$  &\\
\hline
\end{tabular}
\end{table}

\begin{figure*}
\centering
\includegraphics[height=0.4\textheight]{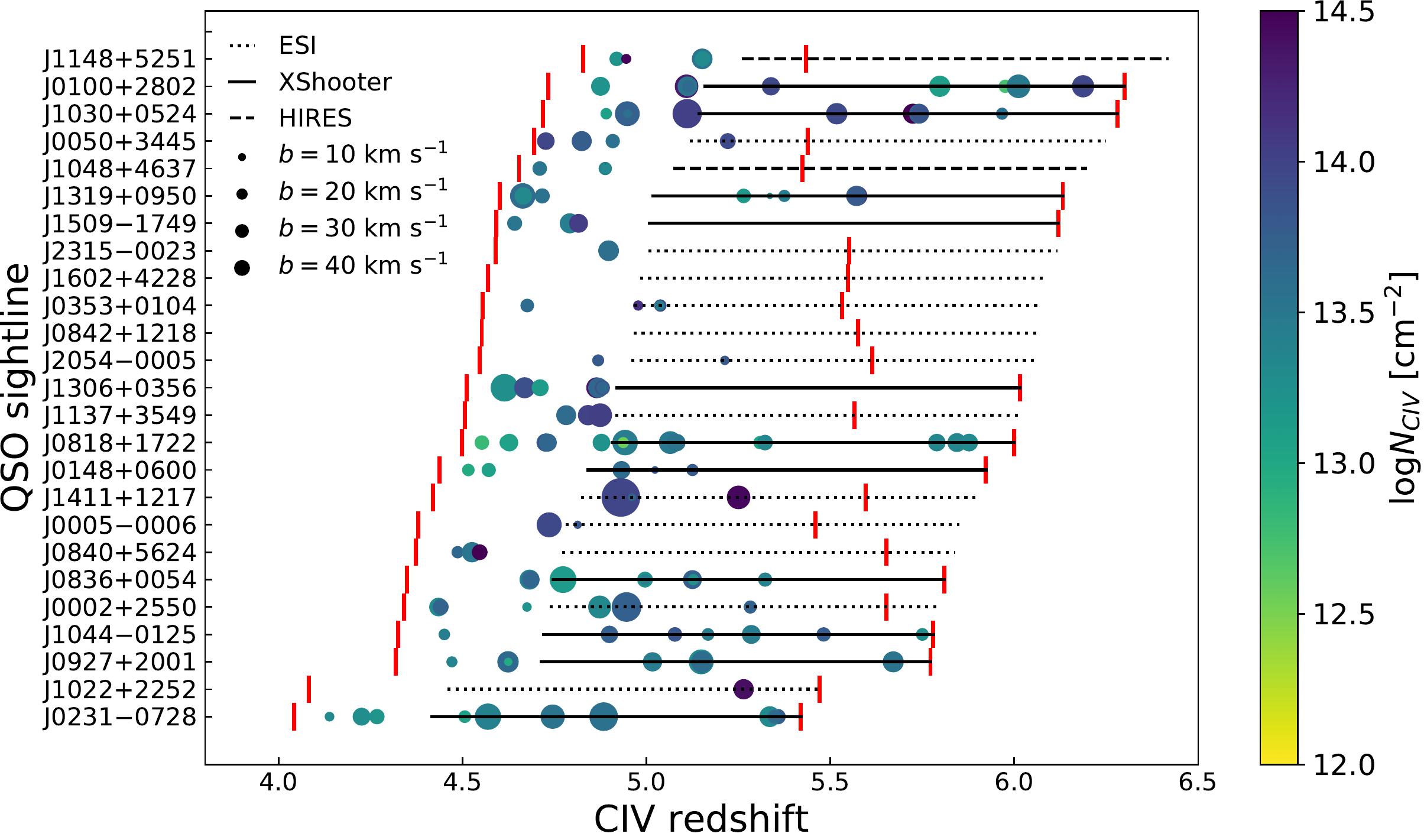}
\caption{Redshift distribution of \cfour \, absorbers for $25$ QSO sightlines. The black lines show the full range of the Lyman-$\alpha$ forest for each QSO and the style of the line indicates the instrument. The red bars indicate the redshift range of our CIV search. The absorbers are shown as coloured dots whose size and colour coding reflect their Doppler $b$ parameter and column density, respectively. \label{fig:search_CIV}}
\end{figure*}

An important issue is the possibility of false positives which would weaken the sought-after correlation. To assess the fraction of false positives in the candidates flagged by our algorithm, we first run the algorithm to search for doublet emission lines instead of absorption lines. This procedure should record no detections, but we sometimes recorded one or two detections per sightline due to glitches in the spectra or large residuals from continuum correction. Both kind of false positives were rejected when fitting with a Voigt profile with \textsc{vpfit} as they do not show the characteristic shape of genuine \cfour \, absorbers.  We also insert mock absorbers with incorrect velocity spacing to assess the sensitivity of our algorithm to spuriously aligned absorption lines. We show in Fig. \ref{fig:FPrate} the false positive rate for different velocity offsets of the two \cfour \, transitions from $25\%$ to $150\%$ relative error on the correct velocity spacing $\Delta v$ of the doublet. We find that the mean false positive rate is $\simeq 22 \%$ for the algorithm search. We compute the mean false positive rate by weighting the false positive rates of each instrument by the number of QSO observed.
Given that the candidates are inspected by eye while being fitted with \textsc{vpfit}, we expect the final false positive rate to be lower because some false positives are discarded. However, assessing the efficiency of a visual inspection is difficult. We compare our search with DO13 on $6$ matching sightlines where we found $56$ \cfour \, absorbers detections (see Section \ref{sec:comparatives_searches}). Although we were more cautious than DO13 and rejected some of their absorbers, the fact that we found only two additional detections also argues against our technique generating false positives. Thus if the remainder of the $150$ \cfour \, sample is pure at the algorithm level ($< 22 \%$), then the false positive rate of our final sample is expected to be $ \lesssim 8\%$. 

Finally, our search results in $150$ \cfour \, absorber detections at $4.5< z< 6.3$. Fig. \ref{fig:search_CIV} shows a graphical representation of the \cfour \, absorbers found. Appendix \ref{appendix:table_CIV} presents a complete list of our detected \cfour \, absorbers. 

\subsection{Comparison with previous \cfour \, searches} \label{sec:comparatives_searches}
We briefly compare our results for those QSO sightlines already searched for \cfour \, by previous authors to assess the purity of our method. All other detections on other sightlines are new detections and are listed in Table \ref{appendix:table_CIV} as well as velocity plots in Appendix \ref{appendix:vel_CIV} for the reduced sample lying in the Lyman-$\alpha$ forest. Based on this comparison, we estimate our search to be in agreement, if somewhat more conservative, than previous \cfour \, searches. We note that small differences in the column densities (up to $\sim 0.2-0.3$) are easily explained by the continuum fitting differences. We also sometimes fit fewer components than previous searches. The excellent agreement in the total cosmic density of \cfour \, between previous studies and our measurement shows these are minor issues driven by noise and different spectra.

\paragraph*{J0818+1722}: We retrieve all \cfour \, absorbers found previously by DO13 between $4.5<z<5.3$, with $\Delta z < 0.01$. We note however that we fitted one component less to the \cfour \, system at $z\simeq4.726$. This however has no impact on our analysis as we cluster systems with $\Delta z < 0.01$.

\paragraph*{J0836+0054}: We detect the same systems at $z=4.68, 4.99$,$z = 5.12$ and $z=5.32$ as DO13. We fit two components less for the $z=4.99$ to keep only the clear detections.

\paragraph*{J0840+5624}: We report $3$ new systems at $z=4.49,4.53,4.55$, which were not in the redshift range searched by \citet{RyanWeber2009}.

\paragraph*{J1030+0524 / J1319+0950}: We detect the same systems between $4.65<z<5.6$ as DO13, to which we add new detections at $z=5.11$ in the sightline of J1030+0524 and at  $z=5.34$ in the sightline of J1319+0950. We believe the detections were made possible by the quality of our XShooter spectra.

\paragraph*{J1306+0356}: We detect the same systems between $4.5<z<5$, with the exception of $z=4.723$ and $z\simeq4.885$, which were both blended with sky lines in DO13 and that were not retained here.  

\paragraph*{J1509-1749}: We recover the same systems between $4.6<z<6$, with the exception of $z=5.7690$ that we consider to be blended with a sky line in DO13's analysis. We also find that the $z=5.915$ absorber is probably due a spurious alignment of lines.

\section{Results}\label{sec:results}
\subsection{Cosmic mass density of \cfour} \label{sec:statistics_hosts}
The first physical result that can be readily derived from any sample of metal absorbers line is the comoving cosmic mass density as a function of redshift. This measurement provides valuable insight into the history of the metal enrichment of the Universe. Our large sample of \cfour \, absorbers is used to place new constraints on the cosmic density of \cfour \, at $5<z<6$. The comoving mass density of \cfour\ is computed as 
\begin{equation}
\Omega_{\cfourmath}=\frac{H_0 m_{\cfourmath}}{c \rho_{\text{crit}}} \int N_{\cfourmath}f(N_{\cfourmath}) \text{d}N_{\cfourmath} \approx \frac{H_0 m_{\cfourmath}}{c \rho_{\text{crit}}}\frac{\sum N_{\cfourmath}}{\Delta X},
\end{equation}
where $f(N_{\cfourmath}) $ is the \cfour \, column density function, $m_{\cfourmath}$ is the mass of \cfour\, ion and $\rho_{\rm crit}= 1.88 \times 10^{-29} h^2~\text{g cm}^{-3}$ is the critical density of the Universe, $\Delta X$ is the total absorption path length searched by our survey. The summation runs over all \cfour\, absorbers in the range of column densities of interest. The error is estimated as the fractional variance \citep{Storrie-Lombardi96}
\begin{equation}
\left(\frac{\delta \Omega_{\cfourmath}}{\Omega_{\cfourmath}}\right) = \frac{\sum (N_{\cfourmath})^2}{(\sum N_{\cfourmath})^2}\, \text{ . }
\end{equation}

\begin{table*}
\caption{\cfour \, number line densities and corresponding cosmic densities for the $2$ redshift-selected samples. Columns refer to: (1) redshift interval of sample (2) \cfour\, column density (3) Comoving path length (4) incidence rate per path length (5) cosmic mass density with completeness correction (see Fig. \ref{fig:completeness} and Table \ref{table:completeness}).} \label{table:CIV_stats}
\label{table:CIV_cosmic_param}
\centering
\setlength{\tabcolsep}{8pt} % Default value: 6pt
\renewcommand{\arraystretch}{1.2} % Default value: 1
\begin{tabular}{rcccc}
\hline\hline
Redshift & $\log N_{\cfourmath}$ & $\Delta X$ & $\text{d}\mathcal{N}/\text{d}X^{\rm a}$ & $\Omega_{\cfourmath}[\times 10^{-8}]^{\rm a}$ \\
\hline
$4.3-5.3$ & $13.8-15.0$ & $76.3$& $0.60~(0.39)$ & $1.06\pm 0.18~(0.75\pm 0.16)$ \\ 
$5.3-6.2$ & $13.8-15.0$ & $21.6$ & $0.34~(0.28)$ & $0.72\pm0.32~(0.62\pm0.31)$ \\ 
\hline
\multicolumn{5}{l}{$^{\rm a}$ The bracketed values are without completeness correction.}
\end{tabular}
\end{table*}

\begin{figure}
\hspace{-0.2cm}
\includegraphics[width=0.48\textwidth]{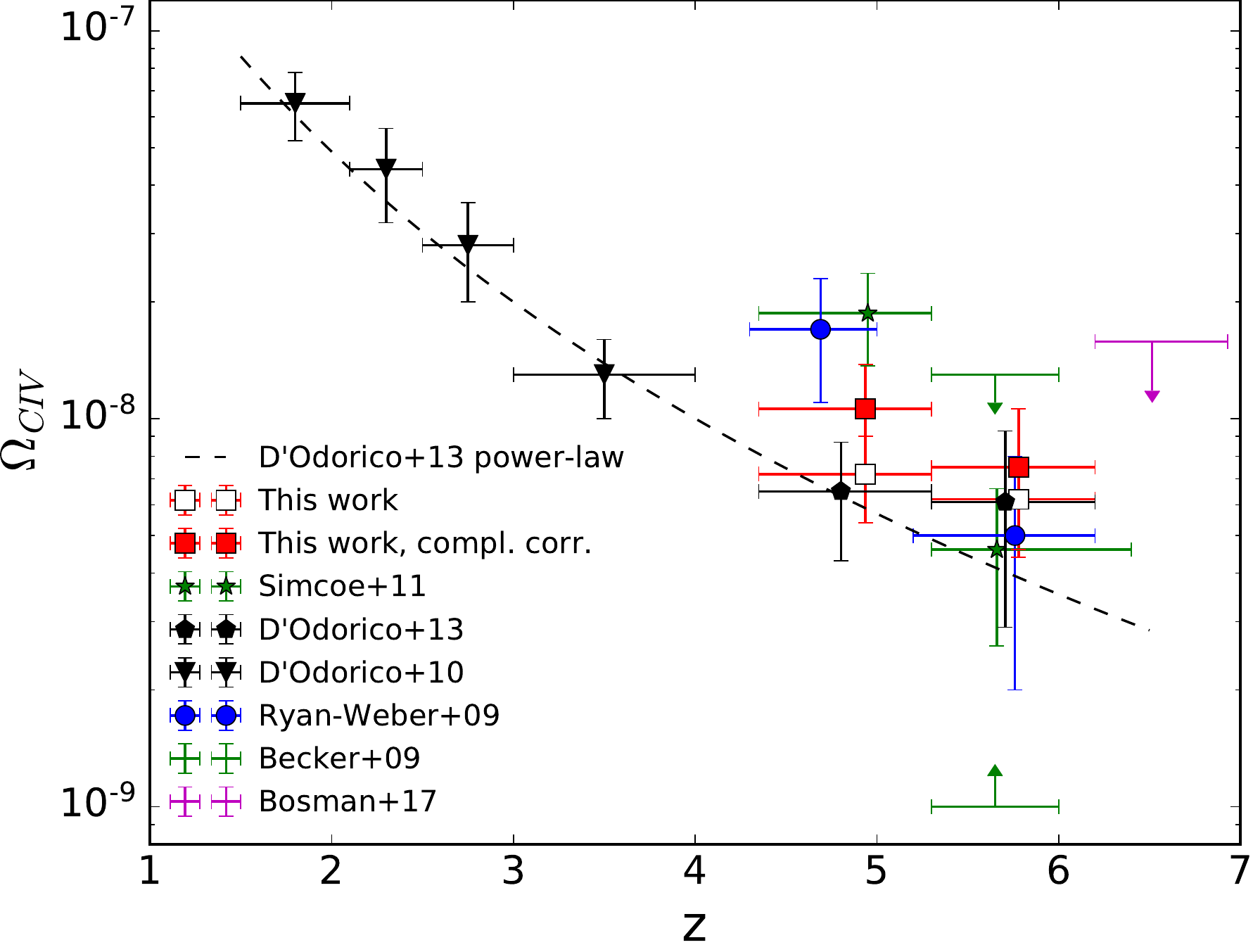}
\caption{Evolution of the cosmic density of \cfour \, at $1.5\lesssim z \lesssim 6.2$ \citep[][black triangles,black pentagons]{DOdorico10,DOdorico13} and the corresponding power-law fit $\Omega_{\cfourmath} = (2\pm 1)\times 10^{-8}[(1+z)/4]^{-3.1\pm0.1}$ (black dotted line). Our measurements (filled and empty red squares, with and without completeness correction)  are in good agreement with \citet{DOdorico13}  as well as other previous studies: \citet[][blue dots]{RyanWeber2009},  \citet[][green $95\%$ confidence interval]{Becker09}, \citet[][dark green stars]{Simcoe2011} and \citet[][magenta upper limit]{Bosman2017}. The density error bars indicate $1\sigma$ confidence intervals. \label{fig:cosmic_density}}
\end{figure}

We present our measurement on Fig. \ref{fig:cosmic_density} alongside previous results in the literature. We chose a \cfour\, absorber sample with selection criteria $13.8\leq \log N_{\cfourmath} <15.0$ at the two redshift intervals, $4.3<z<5.3$ and $5.3<z<6.2$, to facilitate a fair comparison with DO13's extensive dataset across all redshifts. We note the agreement with their evolution which indicates a general decline in cosmic density with redshift. We note that although different observations and recovery pipelines differ on the exact list of absorbers between studies, the cosmic density measurements are very similar. We list the values with and without complete correction (Table \ref{table:completeness}) for different redshift intervals in Table \ref{table:CIV_stats}. The completeness correction does not significantly change the overall decline of the cosmic density in our redshift interval.  

This decline of $\Omega_{\cfourmath}$ reflects both (\textit{i}) the build up of total carbon budget at decreasing redshifts, as more metal is ejected into the circum-/inter-galactic medium around star-forming galaxies by outflows, and (\textit{ii}) the changing ionisation state of carbon due to the evolving spectral shape of the UV background \citep[see e.g.][ for a review and references therein]{Becker15rev}. We will discuss the chemical enrichment and other properties of \cfour-host in Section \ref{sec:physics} after presenting their 1D correlation with the IGM transmission.

\subsection{The observed 1D correlation of \cfour \, with IGM transmission}

\begin{figure}
\centering
\includegraphics[width=0.47\textwidth]{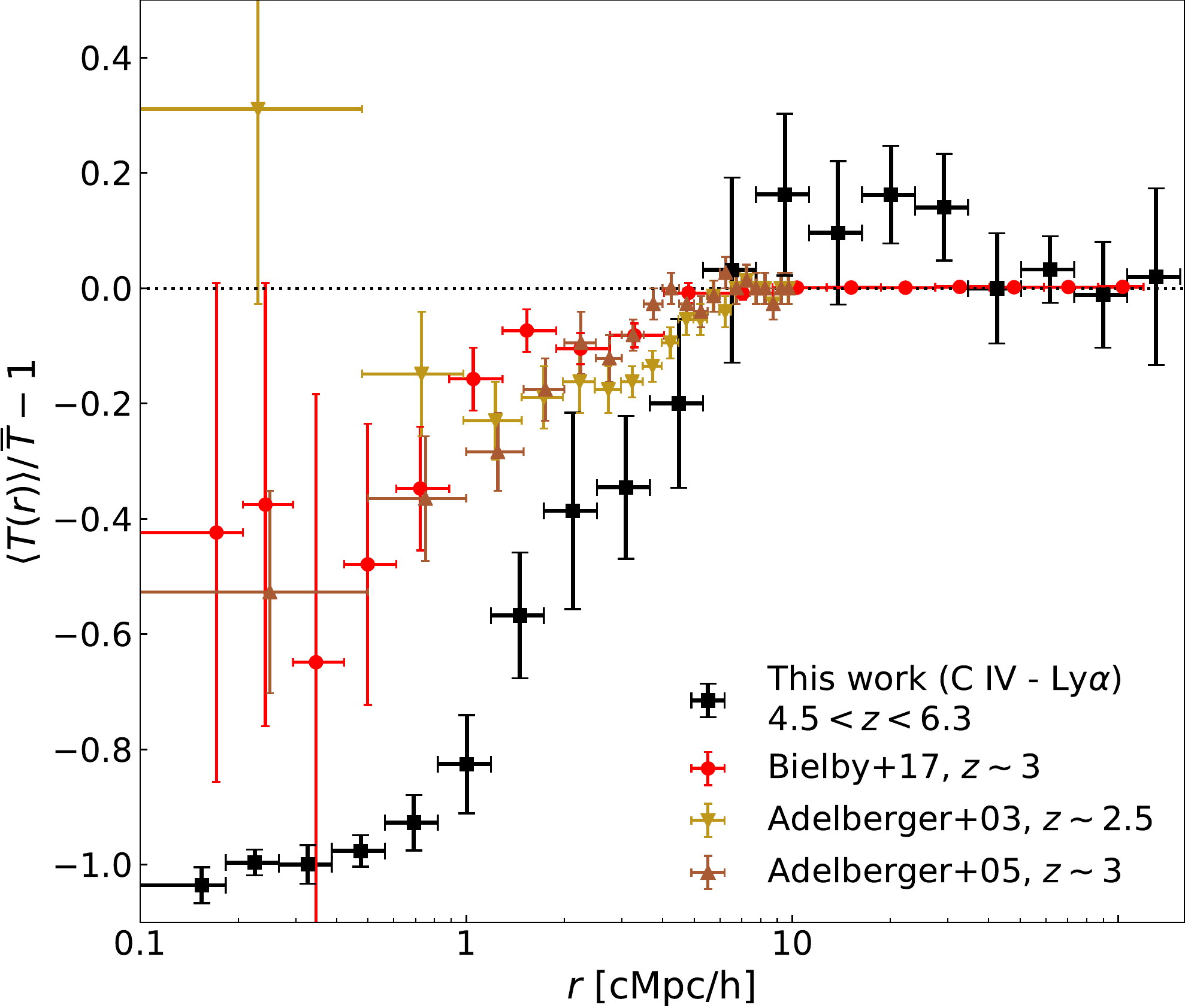}
\caption{The measured 1D correlation between \cfour \, absorbers and the IGM transmission (black) is compared to previous studies on LBG-IGM correlation: \citet[][yellow, orange crosses]{Adelberger03,Adelberger05}, studies at $z\sim 2.5$ and $z\sim 3$ and the VLT LBG survey at $z\sim 3$ \citep[][red crosses]{Bielby17}. The central trough is enhanced at high redshift and the excess on the large scales is in contrast to the flat profile seen in low-redshift data. 
\label{fig:comparison_lowz}}
\end{figure}

The main result of this paper is the 1D correlation between our \cfour \, absorbers and the IGM transmission shown in Fig. \ref{fig:comparison_lowz}. We compute the correlation using \citet{Davis1983} estimator 

\begin{equation}
\hat \xi_{\cfourmath-\rm Ly\alpha}(r) = \frac{D_{\cfourmath}D_{\text{T}}}{R_{\cfourmath}D_{\text{T}} } -1 \, = \frac{\langle T(r)\rangle_{\cfourmath}}{\overline{T}} - 1  \text{ , }
\end{equation}
where the $D_{\cfourmath}$ is the pixel corresponding to the H{~\small I} Lyman-$\alpha$ absorption at the redshift of a detected \cfour\, absorber, and $D_T = \langle T(r) \rangle_{\Delta r}$, is the transmission in the forest of the QSOs at a distance $r$ from the \cfour \, absorber. To make the comparison with previous studies easier, we provide a equivalent formulation where $\langle T(r)\rangle_{\cfourmath}$ is the average Lyman-$\alpha$ transmission $T$ at a distance $r$ of the redshift \cfour \, absorbers, and $\overline T$ is the average transmission at the redshift $z$ in the QSO line-of-sight (LOS) studied. Operationally, for each $D_{\cfourmath}$, we take the transmission in neighbouring pixels $D_T$ and compute the LOS Hubble distance. Similarly, the average transmission is estimated from a random distribution of absorbers with redshifts $R_{\cfourmath}$. The random redshifts $R_{\cfourmath}$ are generated by oversampling $50$ times the redshift interval $z_{\cfourmath}\pm 0.1$ around each observed \cfour \, absorber detected at $z_{\cfourmath}$ in each LOS, so as to reproduce the observed redshift distribution and sightline-to-sightline transmission variance. We note that the conversion from velocity space to Hubble distance is subject to a caveat due to peculiar velocities on small scales further discussed in Section \ref{sec:caveats}. We weight the pixels with the inverse variance as we perform the mean to bin the correlation function linearly or logarithmically depending on the analysis, and we bin the observed and random absorbers in a consistent manner. In order to have sensible values of the transmission and correct for any misfits of the power-law continuum (see Section \ref{sec:data}), we remove pixel artifacts by excluding those with $T<-2e$ and $T-e>1$, where $T$ is the transmission and $e$ the corresponding measurement error. We also only use pixels between $1045$ and $1176\,$ \AA\, in the QSO rest-frame to avoid the QSO Lyman-$\beta$ and -$\alpha$ intrinsic emission. To estimate the error we choose a Jackknife test given our modest sample of sightlines. We draw $500$ subsets of half of the  \cfour \, sample, generate accordingly the random samples and compute the correlation for these subsets. The variance of the $500$ draws is then used as an estimate of our errors. We note that this method is more conservative than Poisson or bootstrap errors, and converges with an increasing sample size.

Not all \cfour \, absorbers detected in the sightline of the QSOs are suitable for this measurement. We define here the sample of \cfour \, absorbers redshifts used for the correlation with the IGM transmission, named Sample $\alpha$. First and foremost, the corresponding H{~\small I} Lyman-$\alpha$ should be between $1045$ and $1176\,$ \AA\, in the QSO rest-frame to avoid the QSO Lyman-$\beta$ and -$\alpha$ intrinsic emission. Secondly, as we are using \cfour \, as a tracer of galaxies we take only the redshift of the strongest absorber for systems with multiple components when they are $\delta v < 100$ km s$^{-1}$ apart. This avoids systems with multiple components multiply sampling the same part of the QSO forest and thus biasing the measurement. Finally, as our estimator requires a proper measurement of the transmission in the Lyman-$\alpha$, it cannot produce a sensible measurement where the average flux in the Lyman-$\alpha$  forest falls below the sensitivity of the spectrograph , producing lower limits on the correlation that are not easily interpreted. We hence remove \cfour \, absorbers lying in saturated end regions of the Lyman$-\alpha$ forest where the average transmission over $\Delta z = 0.1$ is less than the average $1\sigma$ error level of the flux measurement. We emphasize that this last step only removes the end part of two QSO forests in which \cfour\, absorbers sit, and exclude only $\sim 10\%$ of Sample $\alpha$.

The selection described above leaves $37$ \cfour \, systems absorbers suitable for the correlation measurement to which we add the $z=5.738$ absorber on the LOS of J1148+3549 from \citet{RyanWeber2009}, which was detected in a NIRSPEC spectra in wavelengths unaccessible to our ESI spectra. The average redshift of Sample $\alpha$ is $\langle z_{\cfourmath}\rangle = 5.18 $ and the average column density $ \log \langle N_{\cfourmath}\rangle \simeq 13.8 $. We show the 6 lower-redshift detections of Sample $\alpha$ on Fig. \ref{fig:examples_cfour}. The whole sample is presented in Appendix \ref{appendix:vel_CIV}.

\begin{figure*}
\centering
\includegraphics[height=0.24 \textheight]{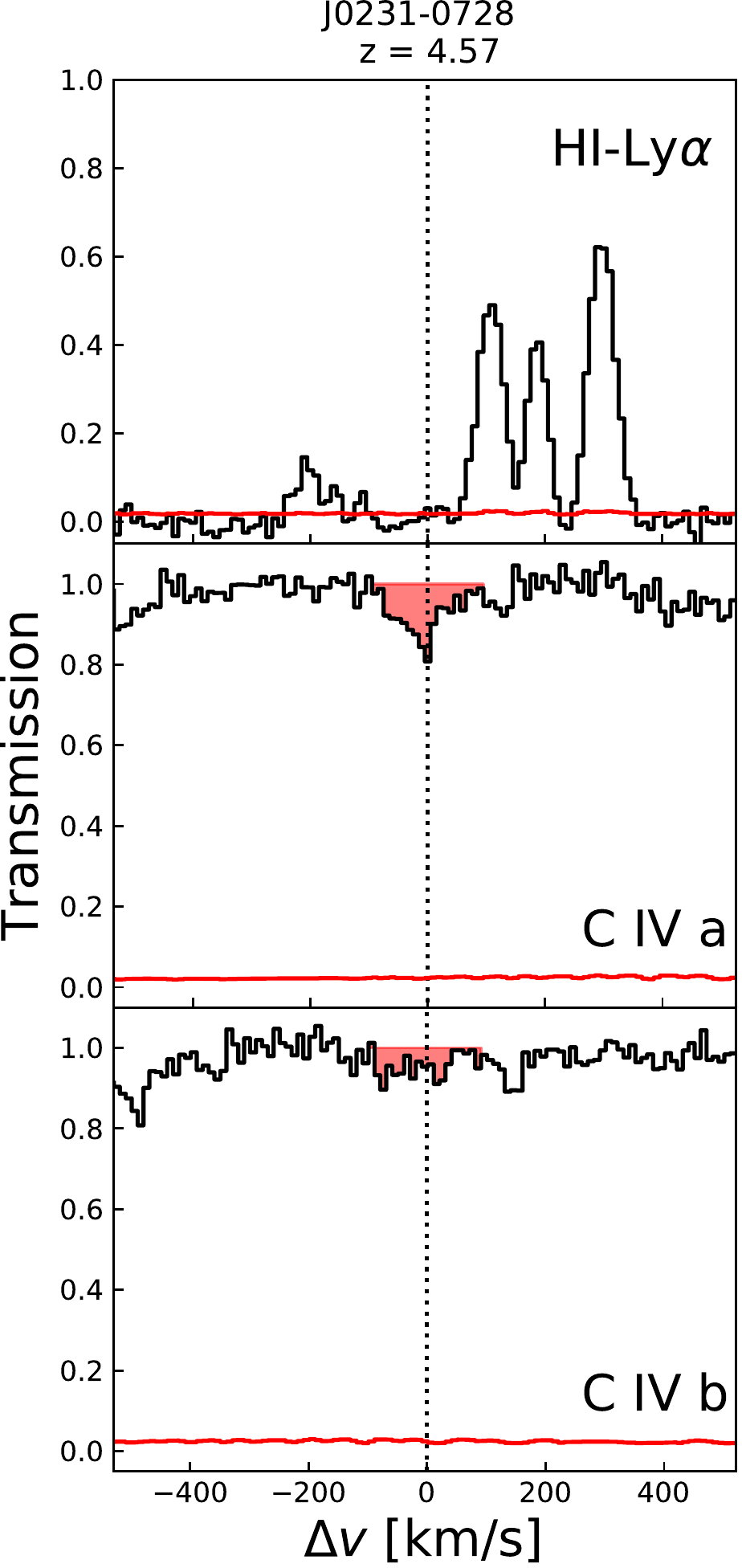}
\includegraphics[height=0.24 \textheight]{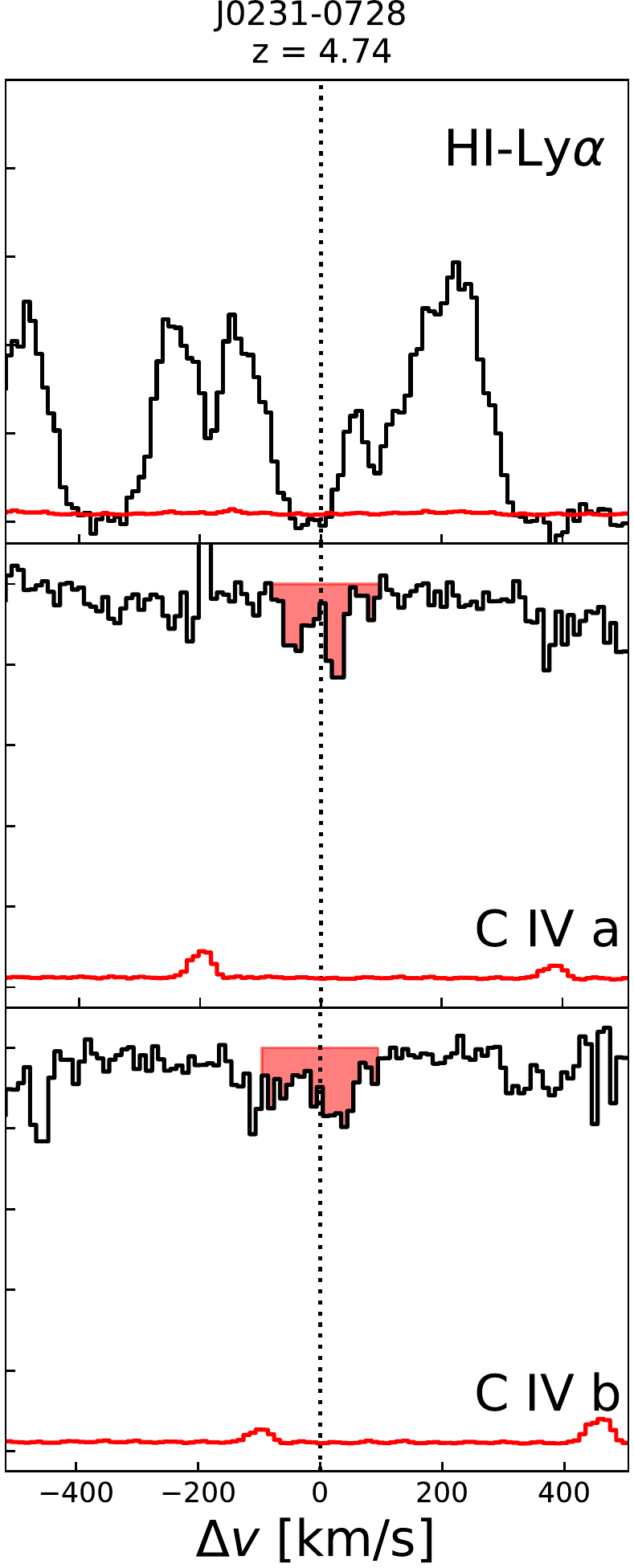}
\includegraphics[height=0.24 \textheight]{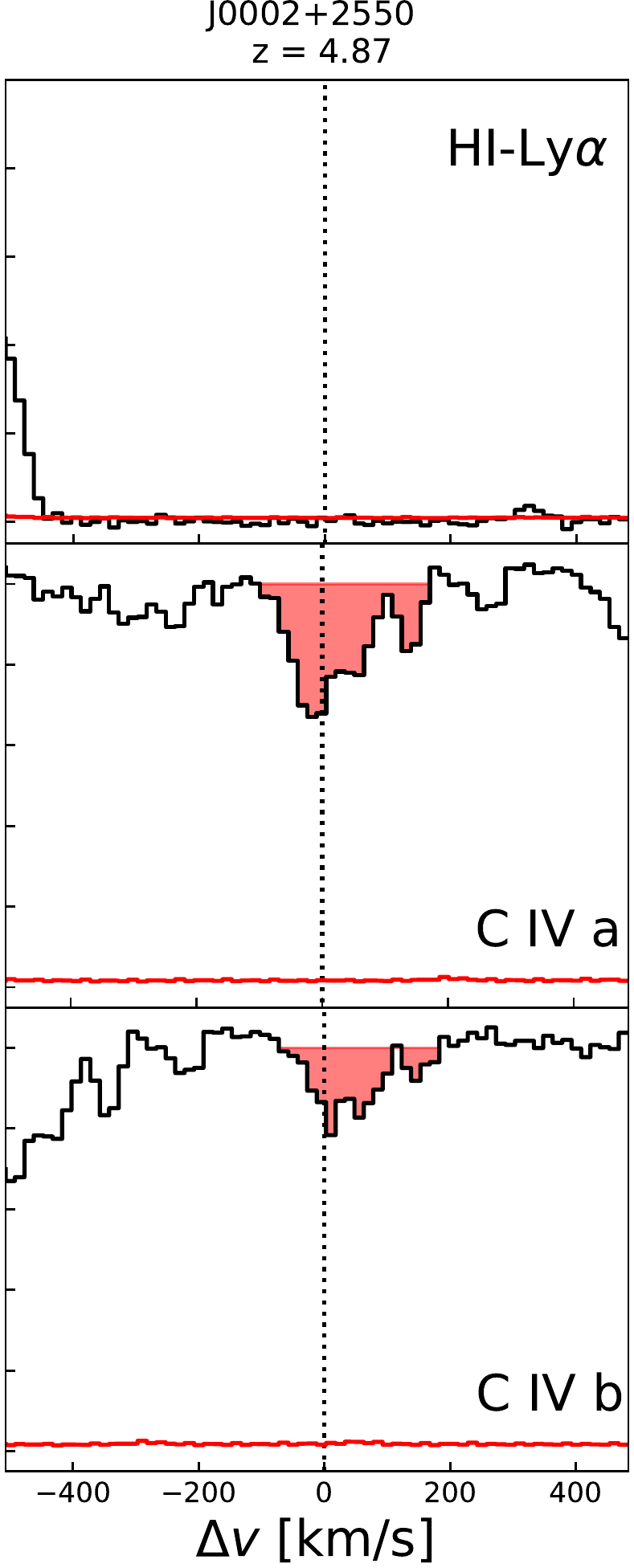}
\includegraphics[height=0.24 \textheight]{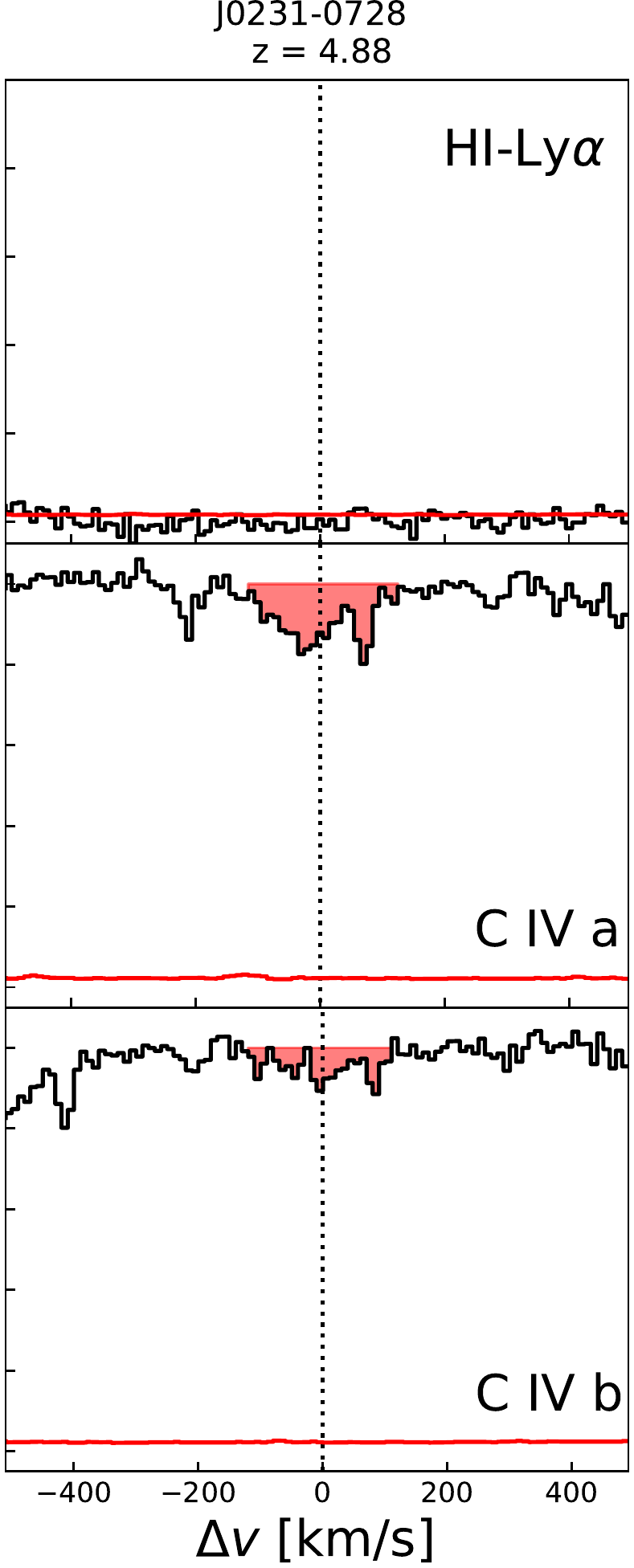}
\includegraphics[height=0.24 \textheight]{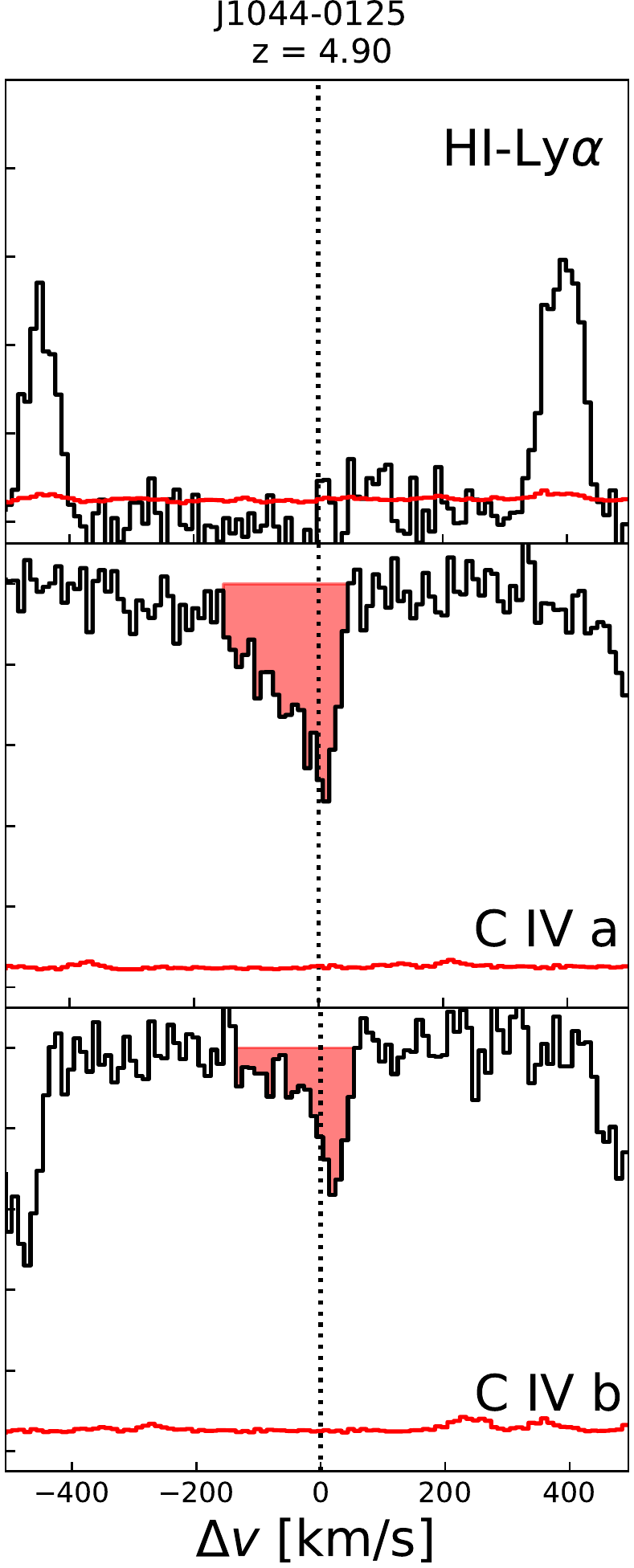}
\includegraphics[height=0.24 \textheight]{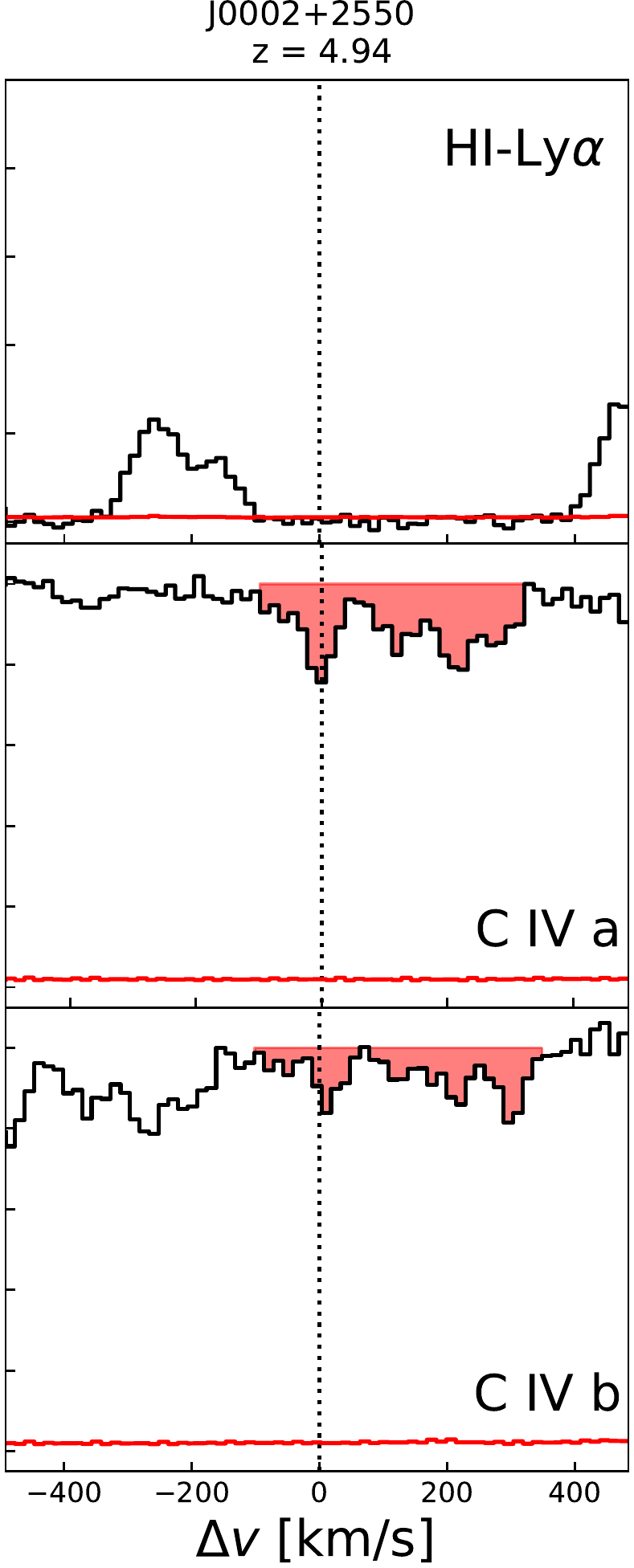} 
\caption{First lower-redshift detections of our \cfour \, sample detected with a corresponding H{~\small{I}} Lyman-$\alpha$ absorption in the QSO Lyman-$\alpha$ forest. The entire sample of $37$ such absorbers is presented in Fig. \ref{fig:mosaic_transmission}.\label{fig:examples_cfour} } 
\end{figure*}

The 1D correlation signal reveals an excess absorption at $r_c\lesssim 5$ cMpc/h around \cfour \, absorbers at $4.5<z<6$. This excess absorption is also found around $z\sim 3$ Lyman Break Galaxies (LBGs) \citep{Adelberger03,Adelberger05,Bielby17} (see Figure \ref{fig:comparison_lowz} for a comparison). The excess absorption is detected on the same scales ($\sim 5$ cMpc/h) for both $z\sim 3$ LBGs and $z\sim 5.2$ \cfour s, but the excess absorption seems much stronger in the latter objects. The absorption excess is perhaps emerging more clearly due the overall opaqueness Lyman-$\alpha$ forest at $z\sim 5.2$. The trough also shows the expected co-spatiality of \cfour \, and Lyman-$\alpha$ absorption. There is also an excess of transmission at $10 \lesssim r \lesssim 30$ cMpc/h around \cfour \,. This excess on large scales was detected around spectroscopically confirmed LBGs in Paper I. The significance of the excess on $10-30$ cMpc/h scales is $2.7\sigma$ for the average transmission $T\simeq 0.14$ versus a null mean flux. We note that the signal goes to zero at large distances, indicating that the excess is unlikely to be caused by the wrong normalisation of the 1D \cfour-IGM correlation. We discuss the physical implications in Section \ref{sec:physics}. 

\section{Modelling the \cfour -IGM correlation}\label{sec:models}

In order to interpret the data, we use the linearised version of the model introduced and discussed in detail in Paper I. The precepts for the LBG-IGM cross-correlation discussed in Paper I can be easily modified for \textit{any} tracer of galaxies, and the dense, ionised \cfour \, gas is an ideal candidate.

Supposing that galaxies hosted by dark matter haloes eject their material by galactic winds, chemically enriching the surrounding IGM environment. \cfour\, absorbers act as a tracer of galaxies.  At $z\gtrsim5$, patchy reionisation can produce the fluctuations in the UV background affecting the Ly$\alpha$ forest transmission around galaxies. Therefore, the \cfour-IGM correlation will reflect both  (\textit{i}) the correlation between matter and galaxies and (\textit{ii}) the enhanced UV background around the \cfour-host galaxies. Following Appendix B of Paper I, we showed the expected flux transmission at a comoving distance $r$ from the position of a \cfour\,  absorber is 
\begin{equation}
T(r) = \exp (-\tau_{\alpha}(r)) = \overline T (1+\xi_{\cfourmath-\rm Ly\alpha}(r) ) \text{  ,  }
\end{equation}
where $\overline T$ is the average transmitted flux in the Lyman-$\alpha$ forest at the redshift of interest, and the \cfour-IGM correlation along the LOS is then given by 
\begin{equation}
 \xi_{\cfourmath-\rm Ly\alpha}(r) \approx  b_{\cfourmath}b_\alpha \xi_{\rm m}^{\text{lin}}(r,\mu=1) + b_{\cfourmath}b_\Gamma \xi_\Gamma^{\rm lin}(r) \, \text{ . } \label{eq:linear_model}
\end{equation}
The contribution of the matter correlation around galaxies is quantified with the two bias factors of \cfour-host galaxies $b_{\cfourmath}$ and the Lyman-$\alpha$ forest $b_\alpha$. The redshift-space sightline linear matter correlation function $\xi^{\rm lin}_m(r,\mu=1)$ follows from the real-space matter correlation function \citep{Hamilton92},
\begin{equation}
\xi^{\rm lin}_m(r,\mu=1)=\xi_0(r)P_0(\mu)+\xi_2(r)P_2(\mu)+\xi_4(r)P_4(\mu)
\end{equation}
with $\xi_0(r)=[1+(\beta_{\cfourmath}+\beta_\alpha)/3+\beta_{\cfourmath}\beta_\alpha/5]\xi(r)$, 
$\xi_2(r)=[(2/3)(\beta_{\cfourmath}+\beta_\alpha)+(4/7)\beta_{\cfourmath}\beta_\alpha][\xi(r)-\bar{\xi}(r)]$, and $\xi_4(r)=(8/35)\beta_{\cfourmath}\beta_\alpha[\xi(r)+(5/2)\bar{\xi}(r)-(7/2)\bar{\bar{\xi}}(r)]$ where $\xi(r)$ is the linear matter correlation function in real space, and $\beta_{\cfourmath}$ and $\beta_\alpha$ are the redshift space distortion (RSD) parameters \citep{Kaiser87}. The RSD parameter of \cfour \, is set to $\beta_{\cfourmath} \approx \Omega_m^{0.6}(z)/b_{\cfourmath} \approx 1/b_{\cfourmath}$. The Lyman-$\alpha$ forest RSD parameter $\beta_\alpha$ is found to be relatively constant at lower redshift from observations in the range $\beta \simeq 1.2-1.7$ \citep{Slosar11,Blomqvist15,Bautista17}. We set $\beta_{\alpha} = 1.5$ as a fiducial value. The Ly$\alpha$ forest bias is chosen such that the model 1D Ly$\alpha$ forest power spectrum is consistent with observations from \citet{Viel2013}, which leads to the fiducial value of $b_\alpha=-1.3$.  

On large scales, the contribution of the enhanced UV background becomes increasingly important. The mean photoionisation rate $\bar{\Gamma}$ of the IGM from star-forming galaxies depends on the the population average of the product of LyC escape fraction and and LyC photon production efficiency $\langle f_{\rm esc}\xi_{\rm ion}\rangle$, 
\begin{equation}
\bar{\Gamma}\propto\langle f_{\rm esc}\xi_{\rm ion}\rangle\lambda_{\rm mfp}\int^{M_{\rm UV}^{\rm lim}}_{-\infty} L_{\rm UV}(M_{\rm UV})\frac{dn}{dM_{\rm UV}}dM_{\rm UV}\text{ , } 
\end{equation}
where $dn/dM_{\rm UV}$ is the UV luminosity function \citep{Bouwens15} and $M_{\rm UV}^{\rm lim}$ is the limiting UV magnitude of galaxies that contribute to the UV background. We adopt the value of mean free path $\lambda_{\rm mfp}$ of \citet{Worseck2014}. The effect of the UV background is then modelled through the bias factor  $b_\Gamma$ defined by
\begin{equation}
b_\Gamma=\frac{\int\Delta_b P_V(\Delta_b)\tau_\alpha(\bar{\Gamma},\Delta_b)e^{-\tau_\alpha(\bar{\Gamma},\Delta_b)}}{\int\Delta_b P_V(\Delta_b)e^{-\tau_\alpha(\bar{\Gamma},\Delta_b)}} \, \text{ , } 
\end{equation}
where $\tau_\alpha(\bar{\Gamma},\Delta_b)\simeq5.5\Delta_b^2(\bar{\Gamma}/10^{-12}{\rm s^{-1}})^{-1}(T/10^4{\rm K})^{-0.72}[(1+z)/6]^{9/2}$ is the Lyman-$\alpha$ optical depth at the mean photoionisation rate (we assume a uniform temperature of $T=10^4$ K), $P_V(\Delta_b)$ is the volume-weighted  probability distribution function of baryon overdensity $\Delta_b$ \citep{Pawlik2009}. The correlation function of the UV background with galaxies is 
\begin{equation}
\xi_{\Gamma}^{\rm lin}(r)= \langle b_g(<M^{\rm lim}_{\rm UV}) \rangle_L \int \frac{k^2 \text{d}k}{2\pi^2}R(k \lambda_{\rm mfp}) P^{\rm lin}_m(k) \frac{\sin kr}{kr} \, \text{ , } \label{eq:linear_UVB_fluct}
\end{equation}
where $R(k\lambda_{\rm mfp})=\arctan(k\lambda_{\rm mfp})/(k\lambda_{\rm mfp})$, $P_m^{\rm lin}(k)$ is the linear matter power spectrum, and $\langle b_g(<M_{\rm UV}) \rangle_L$ is the luminosity-weighted bias of ionising sources above $M^{\rm lim}_{\rm UV}$, which is evaluated with the same procedure as in Paper I. We use the halo occupation number framework with the conditional luminosity function which parameters are fixed by simultaneously fitting the \citet{Bouwens15} $z\sim5$ luminosity function and the \citet{Harikane16} LBG correlation functions, resulting in the best-fit parameters $(M_{\rm UV,0},\log M_h^\ast, \gamma_1,\gamma_2,\sigma_c,\log\phi_0,\alpha_s,\beta_s)=(-22.29, 11.90, 2.39, -0.06, 0.2\mbox{(fixed)}, -1.49, -1.26, 0.85)$ (see Paper I for more details).

Overall, the linear model used to describe the observed \cfour-IGM correlation contains five parameters; two to describe the UV background  $(\langle f_{\rm esc}\xi_{\rm ion}\rangle, M_{\rm UV}^{\rm lim})$, one to describe the halo bias of \cfour-host galaxies $b_{\rm CIV}$, and two to describe the matter fluctuations in the Lyman-$\alpha$ forest $(b_\alpha, \beta_\alpha)$. Our fiducial linear model leaves the first three parameters free $(\langle f_{\rm esc}\xi_{\rm ion}\rangle, M_{\rm UV}^{\rm lim},b_{\cfourmath})$, but the full five parameter model including $(b_\alpha, \beta_\alpha)$ is examined in Appendix \ref{appendix:params}. 

\begin{figure*}
\centering
\includegraphics[width=0.65\textwidth]{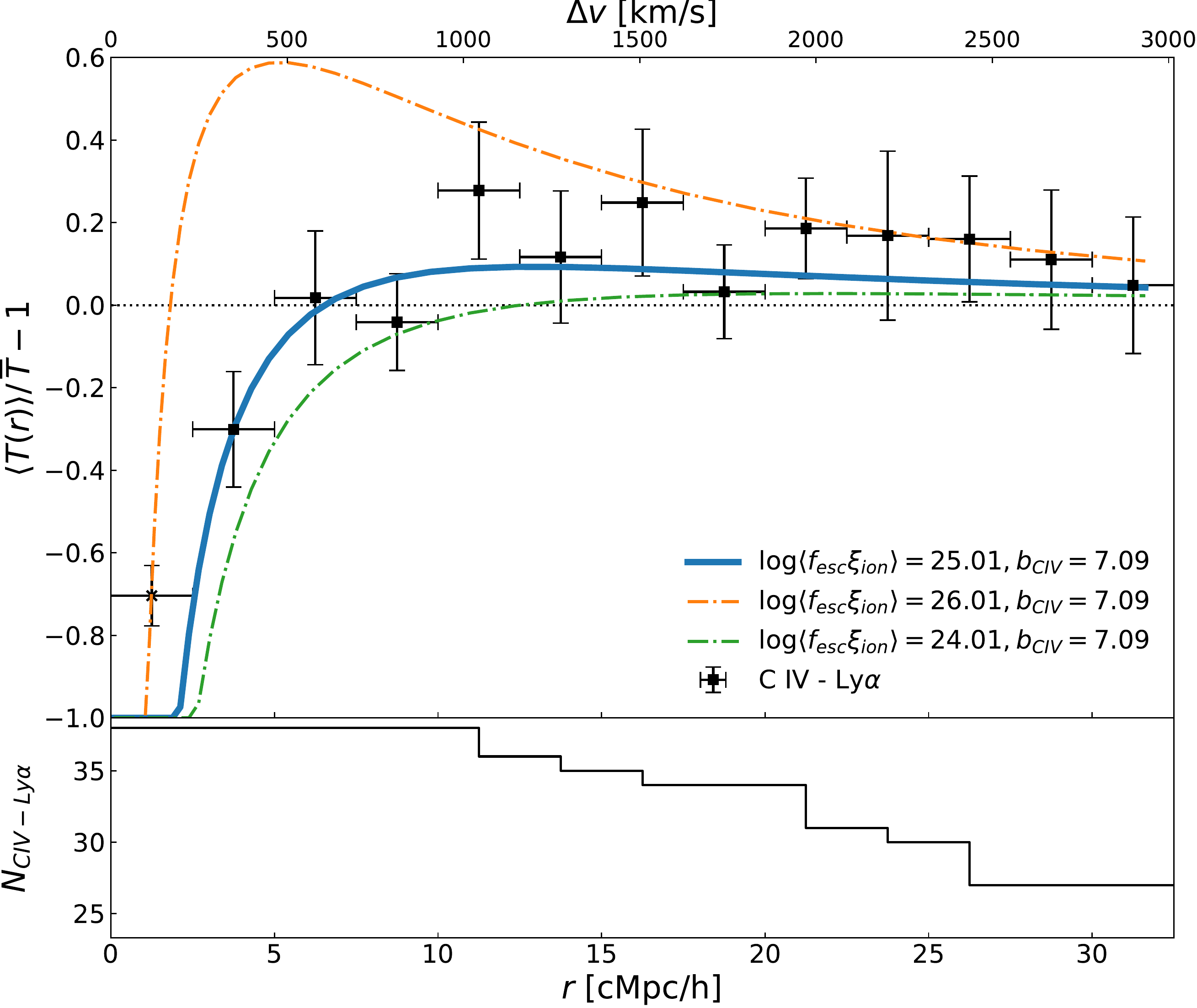}\caption{\textit{Upper panel: } The measured correlation between \cfour \,and the IGM transmission, linearly binned, compared to our models. We show various linear models of the correlation for illustrative purposes. Blue line (maximum likelihood model, see Fig. \ref{fig:posterior_linear}) : $(\log \langle f_{\text{esc}}\xi_{\text{ion}} \rangle /[{\rm erg^{-1}~Hz}]=25.01,M_{\text{UV}}^{\text{lim}} = -10.82, b_{\cfourmath}, b_\alpha = -1.3, \beta_\alpha=1.5 )$. We then change the parameter of interest, the photon production product to show the impact on the correlation. Orange semi-dotted line: $\log \langle f_{\text{esc}}\xi_{\text{ion}} \rangle /[{\rm erg^{-1}~Hz}] =26.01$. Green semi-dotted line: $\log \langle f_{\text{esc}}\xi_{\text{ion}} \rangle /[{\rm erg^{-1}~Hz}] =24.01$. The limitation of the linear model on the scale where it produces unphysical values of $\xi_{\cfourmath-{\rm Ly}\alpha}(\lesssim 3 \text{cMpc/h}) < -1$. In order to limit the impact on the fit, we capped the value of the linear model at $-1$ and excluded the first point from the fit. \textit{Lower panel: } Number of \cfour \, absorbers in Sample $\alpha$ contributing to the measure at different scales.
\label{fig:cross_correlation} }
\end{figure*}

We fit the linear model to the linearly binned correlation using the  Markov chain Monte Carlo affine sampler from the \textsc{emcee} package  \citep{emcee}. In doing so, we assume a flat prior in all three parameters in the following ranges: $23<\log \langle f_{\text{esc}}\xi_{\text{ion}} \rangle /[{\rm erg^{-1}~Hz}]<27$, $-20< M_{\text{UV}}^{\text{lim}} <-8$, $0< b_{\cfourmath} <10^3$. The priors are quite broad and encompass all plausible physical values. We run the sampler for $5000$ steps with $24$ walkers, discarding $500$ steps for burn-in and fixing the scale parameter to ensure the acceptance rate stays within $0.3<r<0.5$. The walkers are initialized in a Gaussian sphere with variance $\sigma=0.1$ at different locations in the allowed parameter space, without any noticeable change to our results. We exclude the first datapoint at $r=2.5 \pm 2.5$ cMpc/h from the fit and we cap the linear model at $-1$ because it does not hold on very small scales, predicting unphysical correlation values $\xi(r\lesssim 2.5\text{cMpc/h})<-1$. The model is evaluated at $\langle z_{\cfourmath}\rangle=5.18$. The resulting fit and the possible inference on the three parameters is discussed in the next section.

\section{Physical implications}\label{sec:physics}

We discuss the physical implications of our measurements of the \cfour-IGM 1D correlation at $z\simeq5$.  The two main features of the correlation are (\textit{i}) an excess Lyman-$\alpha$ forest absorption on small scales $r<5$ cMpc/h suggestive of the gas overdensity around \cfour\, absorbers and indicative evidence of the outskirt of the CGM around the $z>5$ galaxies, and (\textit{ii}) an excess IGM transmission on large-scale ($r\gtrsim10$ cMpc/h) which is consistent with an enhanced UV background around \cfour\, powered by galaxy clustering with a large ionising photon budget as predicted in Paper I. In Fig. \ref{fig:cross_correlation} we show the observed \cfour-IGM correlation overlaid with our linear model presented before. The large scale excess transmission of the correlation is reasonably well captured by the model despite its simplicity, confirming the foregoing interpretation. Clearly any interpretation, and subsequent inference on the exact escape fraction and spectral hardness,  will be subjected to the uncertainties due to our modest sample size and theoretical model. These are addressed in Section \ref{sec:caveats}. 

\begin{figure}
\hspace{-0.5cm}
\includegraphics[width=0.5\textwidth]{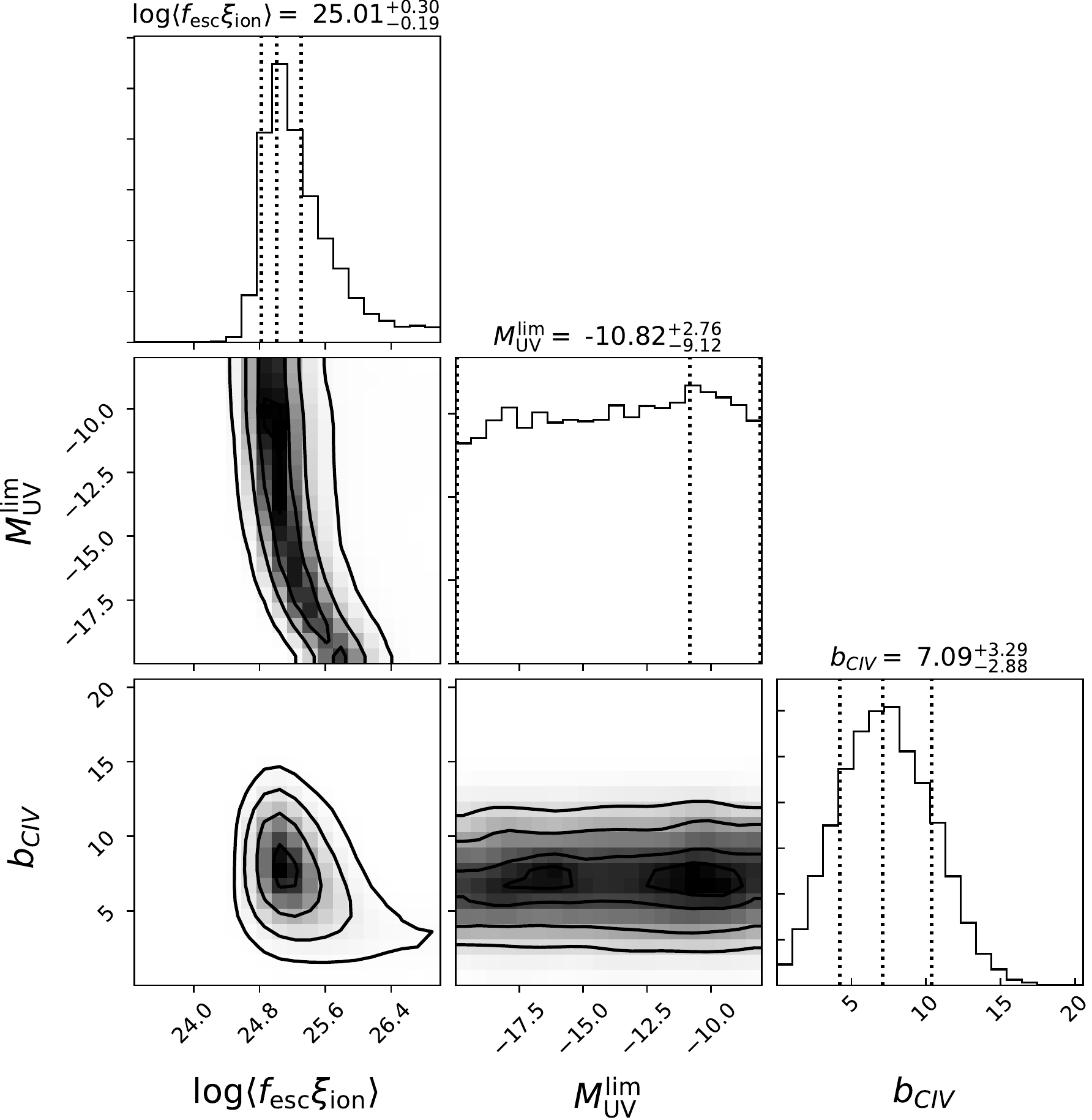}
\caption{Posterior distribution of the parameters in our linear model of the correlation. The quoted numbers of the 1-D histogram give the $1\sigma$ credibility interval. Fitting the data with our linear model put  constraints on the product $\log \langle f_{\text{esc}} \xi_{\text{ion}} \rangle/[{\rm erg^{-1}~Hz}]= 25.01^{+0.30}_{-0.19}$ and the bias of the \cfour \, haloes $b_{\cfourmath} =7.09^{+3.29}_{-2.86}$. The limiting UV magnitude of UV contributors is left unconstrained. It is indeed the parameter that affects the least the correlation. \label{fig:posterior_linear}}
\end{figure}

We show the posterior probability distribution of the parameters for our likelihood in Fig. \ref{fig:posterior_linear}. We find that the observed large-scale IGM transmission excess requires a large population-averaged product of LyC escape fraction and spectral hardness parameter, $\log \langle f_{\text{esc}} \xi_{\text{ion}} \rangle/[{\rm erg^{-1}~Hz}] = 25.01_{-0.19}^{+0.30}$ (for the fiducial model fit) where the quoted error is the $1\sigma$ credibility interval. Even using a conservative modelling approach using the full five parameters with flat priors, the observed level of the large-scale excess seems to indicate a large value $\log \langle f_{\text{esc}} \xi_{\text{ion}} \rangle/[{\rm erg^{-1}~Hz}] \gtrsim 24.7$ ($1\sigma$ limit) (see Appendix \ref{appendix:params} for full details). The limiting UV magnitude of the ionising sources is unconstrained in the prior range. At face value, the best-fit value of \cfour\, bias $b_\cfourmath=7.09^{+3.29}_{-2.89}$ appears somewhat large corresponding to a host halo mass of $ 11.3 \lesssim \log M_{\rm h}/{\rm M_\odot}\lesssim 12.6$. However, this value is degenerate with other Lyman-$\alpha$ forest parameters unknown at $z>5.0$, permitting values as small as $b_\cfourmath=3.5^{+2.0}_{-1.0}$.
The corresponding host halo mass of $ 10.4 \lesssim \log M_{\rm h}/{\rm M_\odot}\lesssim 11.6$ is then consistent with the data. Therefore, the host halo mass of \cfour\, absorbers is loosely constrained to lie between $ 10.4 \lesssim \log M_{\rm h}/{\rm M_\odot}\lesssim 12.6$.

\subsection{The properties of \cfour \, hosts, faint galaxies and feedback}

We can use abundance matching to compute the halo mass of the \cfour \, hosts. We find that the sightline number density of our sample of \cfour \, absorbers with $\log N_{\cfourmath} > 13.0$ is $\text{d}\mathcal{N}/\text{d}X = 4.38 \pm 0.42 ~ (4.42\pm0.85)$ for absorbers at $4.3<z<5.3\, (5.3<z<6.2)$, where the quoted errors are Poisson and we have applied a completeness correction.

We then compute the comoving density of the absorbers. As galactic outflows enrich the gas around galaxies out to a distance $R_{\cfourmath}$ and with a \cfour\, covering fraction $f_{\rm c}$, assuming a one-to-one relation between \cfour\, absorbers and dark matter haloes, the incidence rate is
\begin{eqnarray}
\frac{d\mathcal{N}}{dX}=\frac{c}{H_0}\int_{M_{\rm h}}^\infty \langle  f_{\rm c}\pi R^2_{\cfourmath}\rangle \frac{dn}{dM'_{\rm h}}dM'_{\rm h} \text{ , } 
\end{eqnarray}
where $\langle  f_{\rm c}\pi R^2_{\cfourmath}\rangle$ is the population-averaged physical cross section of metal enriched gas, $dn/dM_{\rm h}$ is a halo mass function in comoving units, and the comoving density is given by $\phi \equiv \int_{M_{\rm h}}^\infty\frac{dn}{dM'_{\rm h}}dM'_{\rm h}$. A conservative maximal enrichment radius for \cfour \, is $r \lesssim 100 $ pkpc, which has been derived from simulations of high-redshift galaxies by \citet{Oppenheimer09}. \citet{Keating16}  have found that the maximal enrichment radius could be twice as small, with most metal-rich outflows traveling less than 50 pkpc from their host galaxy. Assuming a single physical cross-section for the whole sample, we derive likely UV luminosities and masses for \cfour \, hosts with a conservative enrichment radius $25 < R_{\cfourmath} < 100$ pkpc with $f_{\rm c}=1$. For our fiducial choice of $5.3<z<6.2$ absorbers, $d\mathcal{N}/dX = 4.42$, and a physical cross enrichment radius of $R_{\cfourmath} \simeq 100$ pkpc, we find $\phi \simeq 3.2 \times 10^{-2}$ cMpc$^{-3}$.

This translates to likely halo masses of $M_{\text{h}} \gtrsim 10^{10} M_\odot$ \citep{Murray13}. We can also translate the abundance to luminosities of $ M_{\text{UV}} \lesssim -16$ \citep[e.g.][]{Bouwens15}. Clearly a smaller enrichment radius by weaker outflow and/or clumpy distribution of metal enriched gas ($f_c<1$) requires lower mass, more abundant haloes as the hosts of \cfour \,. With improved measurements and analysis, in the future we hope to invert this argument such that, given an independent measure of \cfour-host halo mass from the correlation, it will be possible to infer the cross-section of metal enriched gas, i.e. the properties of galactic outflows around near reionisation-era galaxies.

Our study offers an interesting new insight on \cfour \, absorbers as we view them in absorption with respect to the Lyman-$\alpha$ forest at $z\sim 5$. We note that all \cfour \, absorbers in our Sample $\alpha$ always fall into highly opaque troughs with no Lyman-$\alpha$ transmission, which is surprising given that velocity shifts can occur between the \cfour \, and the expected associated Lyman-$\alpha$ line, and thus in principle flux could be detected in some cases (see Fig. \ref{fig:examples_cfour} and \ref{fig:mosaic_transmission}). When multiple \cfour \, absorbers are detected in the same trough, the separation is at least $0.002 < \Delta z < 0.005 $. Pairs of \cfour \, absorbers sharing the same trough could be two distinct \cfour -enriched clouds, corresponding to separations of $200-400$ ckpc/h. Alternatively this might imply outflows with speeds around $\sim 25 - 100$ km s$^{-1}$. 
The absence of Lyman-$\alpha$ transmission at the redshift of every \cfour\, absorber and the average separation between absorbers within the same opaque region could potentially serve as a test for models which aim at reproducing the distribution and velocity of metals in the early Universe. 

\subsection{Escape fraction, spectral hardness and UV background }

Our inference on the mean ionising photon production rate and escape fraction product of the galaxies clustered around \cfour \, and likely contributing to reionisation is $\log \langle f_{\text{esc}} \xi_{\text{ion}} \rangle /[{\rm erg^{-1}~Hz}] = 25.01^{+0.30}_{-0.19}$. This would imply a $\gtrsim 50\%$ escape fraction if adopting the ``canonical'' values of the LyC production efficiency of $\log \xi_{\text{ion}}/ [{\rm erg^{-1}~Hz}] = 25.2 - 25.4   $ found for $z = 4 - 5$ LBGs \citep{Bouwens16}. However, recent studies of intermediate redshift Lyman-$\alpha$ Emitters (LAEs) \citep{Nakajima16,Nakajima18} and CIII] emitters \citep{Stark15,Stark17} have found higher LyC production rates in the range at $\log \xi_{\text{ion}} /[{\rm erg^{-1}~Hz}]\simeq 25.5-25.8$. Our estimated value for the product would still then imply a high mean escape fraction e.g. $\langle f_{\text{esc}}\rangle \simeq 0.32~(0.16)$ for a fiducial $\log \langle\xi_{\text{ion}}\rangle/[{\rm erg^{-1}~Hz}]=25.5~(25.8)$. Conceivably, our sub-luminous sources clustered around \cfour \, absorbers may have an even harder ionising spectrum if these sources are expected to have an average escape fraction of $\lesssim 10 \%$ as is found at lower redshift.

In Paper I we found an escape fraction $\langle f_{\text{esc}}\rangle\gtrsim 10 \%$ with a fiducial $\log\langle\xi_{\text{ion}}\rangle/[{\rm erg^{-1}~Hz}]=25.2$ for faint galaxies clustered around LBGs, with a $\log \langle f_{\text{esc}}\xi_{\text{ion}}\rangle$ product $1$ dex lower than our inference from the \cfour-IGM 1D correlation. At face value both the escape fraction and the spectral hardness of the galaxies probed by \cfour \, absorbers seems to be increased with respect to the population probed by LBGs in Paper I. Meanwhile, we found several O~{\small{I}} absorbers \citep{Becker06} in the vicinity of a confirmed LBGs, while none of our detected \cfour \, absorbers had a confirmed bright counterpart in the LOS towards J1148+5251. These two clues likely suggest that the population traced by LBGs and \cfour \, absorbers is different. 

If \cfour \, systems correspond to $M_{\text{UV}} \lesssim -16$ galaxies, then by using them as tracers we are likely selecting overdensities less massive and fainter than those traced by LBGs and LAEs.
A harder spectrum might then be attributed to a faint clustered population, consistent with the trend of a harder ionising spectrum $\xi_{\text{ion}}$ in fainter galaxies recently reported by \citet{Nakajima16,Nakajima18}. We can however not exclude that a high average escape fraction is solely driving our high value of $\log \langle f_{\text{esc}}\xi_{\text{ion}}\rangle$.

Finally, our results have interesting implications for the observability of galaxies associated with \cfour-hosting halos. \cfour \, is a highly ionised ion, indicating the presence of radiation at the $\sim$4 Ryd level in the immediate vicinity of the hosts. While at low redshift this radiation is provided by the mean UVB, at $z=5$ the large-scale excess transmission seems to indicates that the collective radiation by large-scale galaxy overdensities around \cfour \, absorbers becomes important.
Together with our relatively large halo masses for \cfour \, hosts ($ 10.4 \lesssim \log M_{\rm h}/{\rm M_\odot} < 12.6$), this seems to indicate that \cfour \, absorption should be tracing galaxy overdensities. However, the independent evidence from searches for emission counterparts to metal absorbers at high redshift is sparse and conflicting \citep{Diaz11,Cai17}. 
To date, no direct emission counterparts of \cfour \, absorbers have been found at $z>5.0$.
In spite of this, \citet{Diaz14,Diaz15} found an overdensity of LAEs within 10 cMpc/h of two quasar sightlines containing \cfour \, absorbers.
This is broadly consistent with the picture in which the strongest ionisation takes place in small galaxies, implying the likely hosts of \cfour \, would be fainter LAEs within overdensities of brighter objects.

\subsection{Alternative interpretations and caveats}
\label{sec:caveats}

We have assumed for our analysis that \cfour \, absorbers are good tracers of galaxies. Due to the \cfour \, wind velocity and the spatial distance between the gas and the host galaxy, this assumption is however only true on somewhat large scales. The typical outflow speed \citep[e.g.][]{Steidel10} is $\sim 200$ km/s, meaning that at $z\sim 5.5$, the maximum distance a metal can travel over the age of the Universe is about $\sim 0.2$ pMpc. In simulations where a more careful modelling of the distribution of metals is done, \cfour \, is expected to travel on average $\lesssim 100$ pkpc  away from the progenitor galaxy \citep{Oppenheimer09,Bird16, Keating16}. The recent detection of a galaxy at $\sim 40$ pkpc from a \cfour \, absorber in the sightline of J2310+1855 further strengthen this point \citep{DOdorico18} . Given the expected spatial offsets ($\lesssim 0.2 pMpc$), and the wind speeds involved ($\sim 200$ km/s), it is fair to argue that the redshift of \cfour \, is a proxy for the redshift of the host galaxy with an error $<400$ km/s. We note that this is comparable with the typical difference between the systemic redshift and the one derived from Lyman-$\alpha$ emission for galaxies in Paper I. Thus \cfour \, reasonably traces galaxies on scales $\gtrsim 0.6 $ pMpc at redshift $z\sim 5.5$. This impacts only the two innermost bins of the correlation in Fig. \ref{fig:cross_correlation}, but the innermost bin is excluded from the fit for reasons exposed above. Hence we conclude \cfour \, is therefore a suitable tracer of galaxies for the purpose of the 1-D correlation with the IGM transmission where a transmission excess is expected to show a positive signal on scales much greater than the redshift-space offset between \cfour \, and its host ($10-30$ cMpc).

An alternative interpretation of the transmission excess seen at $r\gtrsim 10$ cMpc/h in Fig. \ref{fig:cross_correlation} is shifted Lyman-$\alpha$ flux from a associated galaxies. This, however, would imply a mean velocity shift of $\Delta v_{\mbox{\tiny Ly}\alpha} \gtrsim 1000$ km s$^{-1}$, which is a very high value considering the results of previous studies \citep[e.g.][]{Adelberger03,Steidel10,Erb14,Stark17}. The addition of the physical offset between \cfour \, gas and the progenitor galaxy could potentially add to this shift, but we have no reason to believe that spatial offsets of \cfour \, and velocity offsets of Lyman-$\alpha$ should conspire to influence significantly the observed flux in the Lyman-$\alpha$ forests of QSOs.

Measuring the correlation of any population with the IGM transmission is subject to uncertainties. First of all, the sample is subject to cosmic variance even with a size of $38$ objects. Indeed, some sightlines present up to $\sim 5$ \cfour \,absorbers with $\log N_{\cfourmath} > 13$  where some are devoid of them in the redshift range searched (see Fig. \ref{fig:search_CIV}). This, in conjunction with the fact that the Lyman-$\alpha$ forest at $5<z<6$ can show large deviations from the mean opacity \citep{Bosman18}, yields a noisy correlation even with our sample.

Two other sources of errors are the possible contamination of the Lyman-$\alpha$ forest of the QSO by weak emission and/or metal absorption lines from \cfour \, host galaxies or nearby galaxies. The first should only contribute at most in a few bins of Fig. \ref{fig:cross_correlation}, given the $200$-$250$ km s$^{-1}$ winds of \cfour \,clouds.  
As all our observations were carried out with $0.5''-1'' \times 11''$ slits (XShooter) and $0.5''-1'' \times 20''$ slits (ESI), the \cfour \, hosts most of the time do not fall in the slit if they are believed to be within $< 100$ pkpc of the \cfour \, cloud. We hence expect no significant contamination from an associated Lyman-$\alpha$ emitting galaxy in the QSO Lyman-$\alpha$ forest. 
Metal absorption lines (e.g. Si{~\small III} $1206$) can only reduce the signal observed and thus would not affect our claimed excess transmission on large scales.
The large redshift interval sampling is likely to smear the signal as there may be a rapid evolution in the population
of \cfour at $z\sim 5$. 
If \cfour \, traces many distinct populations at once, the signal could be indeed mixed across species and redshifts, but the detection of an excess transmission still holds.

Although surprisingly effective for a first interpretation of the \cfour -IGM correlation, our model has a number of shortcomings. It is firstly a linear model and thus the small scales may contain in large modelling uncertainties due to nonlinear effects.  This shortcoming on the small scales is probably best illustrated by the unphysical values derived in the $r\lesssim 2.5$ cMpc/h region. This model also requires a measurement of the bias and RSD parameter of the Lyman-$\alpha$ forest at $z\gtrsim 5$. To illustrate this, we have left $b_\alpha, \beta_\alpha$ as a free parameter with flat prior in $-3<b_\alpha<0, 3<\beta_\alpha<0$ to see the effect on the inferred parameters (see Appendix \ref{appendix:params}). We notice that $b_\alpha$ is in near perfect degeneracy with $b_{\cfourmath}$ and $\log \langle f_{\text{esc}}\xi_{\text{ion}}\rangle$.
Although the inferred $\log \langle f_{\text{esc}}\xi_{\text{ion}}\rangle$ using a flat prior is consistent within $1\sigma$ of our result presented above, a substantial uncertainty still remains. This linear model would hence benefit from a reliable measurement of the Lyman-$\alpha$ bias parameters at $z>5$. Clearly one possible way to circumvent this issue is to directly compare the 1D correlation measurement with hydrodynamical (radiative transfer) simulations calibrated against Lyman-$\alpha$ forest observables at the same redshift. We are planning to investigate such approach in future work, but we here limit ourselves to the linear model for the sake of brevity. 

\section{Conclusion and future work} \label{sec:conclusion}
The 1D correlation of metals with the IGM transmission offers a promising tool to test different models of reionisation and requires high-resolution spectroscopy of a fair number of bright sources at the redshift of interest. This measurement enables the indirect study of objects aligned with and hence outshined by $z\sim 6$ QSO. We have therefore conducted a semi-automated search for \cfour \,absorbers in order to study how these absorbers can trace potential sources of ionising photons and gathered the largest sample of \cfour \, absorbers at $4.3<z<6.2$. We have updated the measurements of \cfour \,cosmic density, confirming its rapid decline with redshift. Through abundance-matching arguments, we have identified \cfour \, as being associated with $M_{\rm UV} \lesssim -16$ faint galaxies in $\log M_h/{\rm M_\odot} \gtrsim 10$ haloes. 

We have detected excess {H~\small{I}} absorption in the Lyman-$\alpha$ forest at the  redshift of $z\sim 5-6$ \cfour \,absorbers, at a similar scale to that of the IGM absorption around lower redshift LBGs. We have also detected an excess transmission at $2.7 \sigma$ on larger scales in the correlation of \cfour \, with the IGM transmission. We interpret this excess as a signal of the reionisation process driven by galaxies clustered around \cfour \, absorbers. Using the model developed in Paper I, we have put constraints on the product of the escape fraction and the LyC photon production efficiency $\log \langle f_{\text{esc}}\xi_{\text{ion}}\rangle =  25.01^{+0.30}_{-0.19} $. Although caveats about the observation and the modelling remain, we have shown that \cfour \, absorbers trace different galaxies than the ones clustered around LAEs (Paper I), with either higher spectral hardness or possibly larger escape fractions. 

More QSO sightlines are needed to fully sample cosmic variance and provide a better measurement of the correlation. Larger numbers of sightlines and absorbers would not only improve the statistics but also allow a study of the redshift evolution of the escape fraction and LyC production efficiency of the probed galaxies. We point out that a decrease of the cosmic density of \cfour \,makes it more difficult to trace the same objects at all redshifts. However at higher redshift different metal absorbers such as Mg~{\small II} or Si~{\small IV} could be used to trace galaxies. In doing so, we would probe as well different ionising environments and possibly different galaxy populations. Eventually, the degeneracy with the spectral hardness of our measurement can be broken by harvesting our large sample of aligned metal absorbers to probe their ionisation state using forward modeling. 
Radiative transfer simulations with non-uniform UVB including the tracking and modelling of the different metal gas phases could reproduce the correlation, provided large enough boxes and sightlines can be produced in a reasonable amount of time. This opens new avenues into the question driving this series: the nature of the sources of reionisation.

\section*{Acknowledgments}
We thank the anonymous referee for comments that have significantly improved the manuscript. RAM, SEIB, KK, RSE acknowledge funding from the European Research Council (ERC) under the European Union's Horizon 2020 research and innovation programme (grant agreement No 669253). We thank R. Bielby for kindly sharing the measurements of the $z\sim 3$ LBG-IGM correlation. We thank G. Kulkarni, L. Weinberger, A. Font-Ribera for useful discussions. 
Based on observations collected at the European Organisation for Astronomical Research in the Southern Hemisphere under ESO programme(s) 084A-0550, 084A-0574, 086A-0574, 087A-0890 and 088A-0897.
This research has made use of the Keck Observatory Archive (KOA), which is operated by the W. M. Keck Observatory and the NASA Exoplanet Science Institute (NExScI), under contract with the National Aeronautics and Space Administration.
Some of the data used in this work was taken with the W.M. Keck Observatory on Maunakea, Hawaii, which is operated as a scientific partnership among the California Institute of Technology, the University of California and the National Aeronautics and Space Administration. This Observatory  was  made  possible  by  the generous financial support of the W. M. Keck Foundation. The authors wish to recognize and acknowledge the very significant cultural role and reverence that the summit of Maunakea has always had within the indigenous Hawaiian community. We are most fortunate to have the opportunity to conduct observations from this mountain.
The authors acknowledge the use of the UCL Legion High Performance Computing Facility (Legion@UCL), and associated support services, in the completion of this work.

%%%%%%%%%%%%%%%%%%%%%%%%%%%%%%%%%%%%%%%%%%%%%%%%%%
%%%%%%%%%%%%%%%%%%%% REFERENCES %%%%%%%%%%%%%%%%%%

%%%%%%%%%%%%%%%%%%%%%%%%%%%%%%%%%%%%%%%%%%%%%%%%%%
%%%%%%%%%%%%%%%%% APPENDICES %%%%%%%%%%%%%%%%%%%%%

\appendix

\section{Properties of all retrieved \cfour \,systems}
\label{appendix:table_CIV}
We hereby give the tabulated fitted redshift, column density, Doppler parameter and the corresponding errors of all our detected \cfour \, absorbers in each QSO sightline studied. 
\begin{table}
\scriptsize
 \caption{Complete list of \cfour \,absorbers with redshift, Doppler parameter $b \, [\text{km s}^{-1}]$, column density $\log N \, [\text{cm}^{-2}]$ and associated errors.Fixed parameters in \textsc{VPFIT} are indicated with a $^\dagger$. \label{table:CIV}}
 \begin{tabular}{lrrrrrr}
 \hline\hline
 QSO & $z$ & $\Delta z$ & $b$  & $\Delta b$ & $\log N $ & $\Delta \log N$ \\ \hline 
 J0002+2550 &  &  &  &  &  \\ 
 \hline
& 4.434586 & 0.000101 & 43.31 & 10.31 & 13.317 & 0.054 \\
& 4.440465 & 0.000091 & 31.22 & 8.24 & 13.696 & 0.049 \\
& 4.441937 & 0.000077 & 21.93 & 7.62 & 13.708 & 0.042 \\
 &  4.675162  &  0.000042  &   8.22   &   7.25 & 13.184 &    0.099 \\
%& 4.675161 & 0.000115 & 8.81 & 20.82 & 13.202 & 0.253 \\
 &4.870726 &   0.000066  &  21.65   &   7.59 & 13.579  &  0.031\\
 &4.872056  &  0.000097   &  5.74    &  3.19 & 13.498  &  0.208 \\
 &4.873713   & 0.000102   &  6.00$^\dagger$   &  -- &  13.109  &  0.080 \\ 
%& 4.870899 & 0.000339 & 40.57 & 19.70 & 13.627 & 0.310 \\
%& 4.872737 & 0.001700 & 69.21 & 84.88 & 13.335 & 0.620 \\
 &4.941568$^\dagger$  & - &   10.87 &     9.71&  13.387  &  0.117 \\
 &4.943826  &  0.000090  &   5.00    &  4.16  &13.356 & 0.267 \\
 &4.945804 $^\dagger$  & -  &  43.39 &  10.10 &13.417  &  0.051 \\
%& 4.941508 & 0.000143 & 20.33 & 19.78 & 13.358 & 0.095 \\
%& 4.945804 & 0.000307 & 120.00 & 24.48 & 13.747 & 0.068 \\
%& 5.282356 & 0.000137 & 19.61 & 16.89 & 13.710 & 0.136 \\
 & 5.282356  & 0.000070 &  19.67   &  8.62  & 13.709 &   0.069 \\
J0005-0006 &        &  &  &  &  &    \\ \hline
& 4.735463 & 0.000114 & 83.32 & 8.75 & 13.944 & 0.035 \\
 & 4.812924 & 0.000088 & 6.64 & 3.67 & 13.832 & 0.400 \\
J0050+3445 &          &  &  &  &  &  \\ \hline
 & 4.724040 & 0.000057 & 9.78 & 5.43 & 13.664 & 0.178 \\
& 4.726409 & 0.000043 & 38.82 & 4.01 & 13.979 & 0.024 \\
  & 4.824044 & 0.000090 & 51.15 & 7.91 & 13.763 & 0.040 \\
  & 4.908403 & 0.000140 & 24.06 & 16.53 & 13.559 & 0.087 \\
 & 5.221126 & 0.000196 & 31.09 & 19.24 & 13.935 & 0.132 \\
J0100+2802 &          &  &  &  &  &  \\ \hline
  & 4.875143 & 0.000119 & 44.59 & 9.08 & 13.218 & 0.065 \\
  & 5.109066 & 0.000122 & 73.39 & 5.95 & 14.321 & 0.037 \\
  & 5.111580 & 0.000412 & 49.73 & 30.36 & 13.530 & 0.255 \\
  & 5.113629 & 0.000050 & 18.18 & 4.71 & 13.639 & 0.046 \\
  & 5.338458 & 0.000059 & 42.39 & 3.94 & 13.955 & 0.033 \\
  & 5.797501 & 0.000184 & 58.14 & 13.37 & 13.118 & 0.072 \\
  &5.975669 & 0.000158 & 20.63 & 13.10 & 12.742 & 0.120 \\
 & 6.011654 & 0.000268 & 73.94 & 16.94 & 13.488 & 0.078 \\
  & 6.184454 & 0.000362 & 31.35 & 22.12 & 13.061 & 0.256 \\
   & 6.187095 & 0.000121 & 64.42 & 7.57 & 13.971 & 0.039 \\
J0148+0600 &          &  &  &  &  &  \\ \hline
  & 4.515751 & 0.000067 & 17.57 & 7.02 & 12.982 & 0.078 \\
   & 4.571214 & 0.000090 & 23.70 & 8.34 & 13.060 & 0.087 \\
  & 4.932220 & 0.000915 & 40.41 & 28.81 & 13.614 & 1.591 \\
   & 5.023249  & 0.000056   &  6.00$^\dagger$ & - &  13.811    &0.189 \\
  %& 5.023234 & 0.000063 & 5.57 & 4.31 & 13.864 & 0.632 \\
 & 5.124792 &  0.000042 &   27.09 & 3.02  &14.084 &   0.039 \\
%  & 5.124537 & 0.000220 & 9.76 & 9.92 & 13.813 & 0.333 \\
%& 5.125154 & 0.000338 & 16.62 & 15.61 & 13.777 & 0.374 \\
J0231-0728 &          &  &  &  &  &  \\ \hline
 & 4.138342 & 0.000017 & 9.32 & 2.88 & 13.315 & 0.037 \\
 & 4.225519 & 0.000069 & 40.31 & 6.24 & 13.283 & 0.046 \\
 & 4.267255 & 0.000063 & 28.12 & 6.14 & 13.200 & 0.053 \\
   & 4.506147 & 0.000067 & 18.27 & 7.42 & 13.093 & 0.068 \\
  & 4.569167 & 0.000224 & 93.48 & 16.61 & 13.407 & 0.066 \\
 & 4.744957 & 0.000130 & 78.70 & 9.93 & 13.552 & 0.044 \\
  & 4.883691 & 0.000205 & 111.96 & 14.74 & 13.566 & 0.048 \\
 & 5.335394 & 0.000240 & 56.00 & 16.76 & 13.299 & 0.102 \\
 & 5.357714 & 0.000044 & 28.37 & 3.55 & 13.749 & 0.031 \\
& 5.345812 & 0.000088 & 25.35 & 6.99 & 13.361 & 0.068 \\
 & 5.348184 & 0.000043 & 21.49 & 3.75 & 13.647 & 0.038 \\
J0353+0104 &          &  &  &  &  &  \\ \hline
  & 4.675911 & 0.000078 & 21.61 & 12.12 & 13.638 & 0.063 \\
 & 4.977485 & 0.000124 & 10.81 & 6.33 & 14.158 & 0.481 \\
  & 5.036105   & 0.000454 &   15.00$^\dagger$ &- & 13.465 & 0.232 \\
 & 5.037794  &  0.000222  &  15.00$^\dagger$  &- & 13.933   & 0.158 \\ 
 % & 5.037768 & 0.000343 & 17.22 & 25.78 & 13.904 & 0.301 \\
 % & 5.036180 & 0.000700 & 9.76 & 51.27 & 13.443 & 0.894 \\
J0818+1722 &          &  &  &  &  &  \\ \hline
 & 4.552517 & 0.000102 & 24.98 & 10.16 & 12.812 & 0.090 \\
  & 4.620602 & 0.000073 & 25.99 & 6.99 & 13.048 & 0.062 \\
 & 4.627338 & 0.000104 & 38.92 & 8.89 & 13.057 & 0.067 \\
  & 4.727010 & 0.000068 & 31.70 & 3.07 & 14.005 & 0.065 \\
   & 4.725780 & 0.000182 & 42.07 & 8.27 & 13.763 & 0.112 \\
  & 4.731844 & 0.000033 & 41.16 & 2.57 & 13.672 & 0.020 \\
 & 4.877661 & 0.000070 & 38.94 & 5.70 & 13.215 & 0.043 \\
  & 4.936947 & 0.000107 & 32.24  &9.12 &  12.666 & 0.075 \\ 
  &4.940176  & 0.000056 & 14.41 & 7.43 &12.773 & 0.057 \\
  &4.941255 & 0.000071 & 10.00$^\dagger$ &- & 12.783 & 0.071 \\
  &4.942578 & 0.000060 & 40.13 & 5.22 & 13.188 & 0.036 \\
 % & 4.936953 & 0.000177 & 13.27 & 23.32 & 12.562 & 0.183 \\
 %   & 4.941946 & 0.000170 & 87.01 & 12.31 & 13.446 & 0.051 \\
  & 5.064643 & 0.000207 & 68.13 & 16.13 & 13.484 & 0.081 \\
 & 5.076455 & 0.000082 & 33.61 & 6.51 & 13.563 & 0.056 \\
  & 5.082651 & 0.000078 & 37.51 & 5.90 & 13.546 & 0.047 \\
   & 5.308823 & 0.000061 & 16.45 &  6.50 & 13.031 & 0.054 \\
  %& 5.308773 & 0.000155 & 20.39 & 14.98 & 13.020 & 0.131 \\
 & 5.322429 & 0.000121 & 29.38 & 9.91 & 13.341 & 0.084 \\
 &5.789525  &  0.000111   & 38.43    &  7.49 & 13.339  &  0.062 \\ 
&5.843998   & 0.000110   & 44.91     & 7.41  &13.345  &  0.051  \\  
&5.877228   & 0.000106   & 39.41     & 7.26  &13.315  &  0.055 \\ 
\hline
\end{tabular}

\end{table}
\begin{table}
\scriptsize

\contcaption{Complete list of \cfour \,absorbers and measured properties}
\begin{tabular}{lrrrrrr}
\hline\hline
QSO & $z$ & $\Delta z$ & $b$ & $\Delta b$ & $\log N$ & $\Delta \log N$ \\ \hline 
J0836+0054 &          &  &  &  &  &  \\ \hline
 & 4.682130 & 0.000093 & 50.92 & 7.82 & 13.406 & 0.049 \\
& 4.684465 & 0.000046 & 40.42 & 4.38 & 13.608 & 0.033 \\
 & 4.686503 & 0.000029 & 33.65 & 2.53 & 13.689 & 0.022 \\
 &4.773175   & 0.000389   & 96.43   &  31.27 & 13.135  &  0.113 \\
   & 4.996702  & 0.000042& 27.12 & 3.50 & 13.569 & 0.035 \\
  % & 4.996154 & 0.000825 & 30.37 & 38.46 & 13.224 & 0.658 \\
 %& 4.996878 & 0.000132 & 13.79 & 10.97 & 13.407 & 0.410 \\
 & 5.125071 & 0.000126 & 44.86 & 7.84 & 13.729 & 0.067 \\
 & 5.127266 & 0.000066 & 13.81 & 7.10 & 13.246 & 0.072 \\
 & 5.322763 &  0.000070  &  23.54    &  5.46 & 13.400  &  0.059  \\
J0840+5624 &         &  &  &  &  &  \\ \hline
  & 4.486724 & 0.000060 & 16.99 & 9.88 & 13.674 & 0.096 \\
 & 4.525773 & 0.000144 & 52.80 & 14.54 & 13.519 & 0.058 \\
 & 4.546597 & 0.000021 & 30.20 & 4.36 & 15.395 & 0.384 \\
J0927+2001 &          &  &  &  &  &  \\ \hline
& 4.471037 & 0.000031 & 13.02 & 3.60 & 13.369 & 0.042 \\
 & 4.623510&  0.000149& 61.15& 5.56 & 13.592 & 0.076  \\
 & 4.624132 & 0.000066 & 20.94 & 8.47&  13.154 & 0.190  \\
 % & 4.623608 & 0.000119 & 59.18 & 6.33 & 13.659 & 0.056 \\
 %& 4.624174 & 0.000083 & 6.70 & 17.72 & 12.983 & 0.168 \\
   & 5.014445 & 0.000110 & 19.23 & 9.63 & 13.560 & 0.114 \\
  & 5.016426 & 0.000216 & 46.85 & 16.71 & 13.449 & 0.116 \\
 &5.149170 & 0.000799 & 75.01 & 39.93 & 13.377 & 0.271   \\
 &5.151154 & 0.000204 & 41.88 & 13.54 & 13.498 & 0.219   \\
% & 5.148801 & 0.002319 & 82.72 & 96.50 & 13.306 & 0.750 \\
% & 5.151129 & 0.000640 & 61.18 & 25.63 & 13.617 & 0.372 \\ 
 &5.671001 &   0.000187   & 57.11    & 12.31 & 13.540  &  0.073 \\
J1022+2252 &          &  &  &  &  &  \\ \hline
 & 5.264299 & 0.000152 & 52.66 & 12.47 & 14.410 & 0.078 \\
J1030+0524 &           &  &  &  &  &  \\ \hline
  & 4.890614 & 0.000164 & 14.31 & 20.24 & 13.081 & 0.162 \\
  & 4.947967 & 0.000222 & 82.22 & 15.03 & 13.738 & 0.075 \\
  & 4.948468 & 0.000068 & 7.89 & 7.27 & 13.589 & 0.245 \\
 & 5.110703 & 0.000664 & 117.36 & 34.50 & 14.023 & 0.143 \\
 & 5.517461 & 0.000102 & 53.31 &  6.76 & 13.904 & 0.045  \\
 % & 5.517445 & 0.000687 & 59.05 & 44.98 & 13.934 & 0.273 \\
 & 5.724374  &  0.000044  &  51.29   &   2.53 & 14.543  &  0.029 \\
& 5.741183 &   0.000102  &  50.90  &    6.96 & 13.748  &  0.044 \\
& 5.744256  &  0.000062  &  39.49   &   4.31 & 13.853  &  0.033 \\
& 5.966889   & 0.000076  &  16.27   &   7.02 & 13.569  &  0.074 \\
J1044-0125 &          &  &  &  &  &  \\ \hline
 &4.450487 & 0.000027 & 14.28  & 3.52 & 13.456   & 0.035\\
 &4.898284 & 0.000188 & 78.05 &  9.81 & 13.759  &  0.060\\
 &4.899596 &  0.000035 &  13.63  &  5.40 & 13.612  &  0.060\\
 &5.075408 &  0.000091 & 14.12  & 9.69  &13.433  &  0.107 \\
 &5.077315 &  0.000042 &  26.74  & 3.42 & 13.847  &  0.033\\
 &5.167313 &  0.000055 & 10.59 &  6.40  &13.493   & 0.099\\
 &5.285046 &  0.000147  & 36.47 & 11.42 & 13.477  &  0.089\\
 &5.481678 &  0.000080 & 22.20  & 6.70 & 13.775  &  0.070\\
%& 4.450501 & 0.000126 & 14.59 & 16.02 & 13.437 & 0.153 \\
% & 4.899354 & 0.000273 & 38.15 & 38.04 & 13.737 & 0.653 \\
% & 5.075415 & 0.000441 & 12.64 & 48.89 & 13.437 & 0.577 \\
%& 5.077314 & 0.000197 & 25.30 & 16.42 & 13.847 & 0.164 \\
% & 5.167303 & 0.000346 & 18.16 & 31.46 & 13.459 & 0.333 \\
%  & 5.285073 & 0.000886 & 44.99 & 66.60 & 13.450 & 0.454 \\
% & 5.481678 & 0.000504 & 24.24 & 40.86 & 13.770 & 0.405 \\
J1048+4637 &          &  &  &  &  &  \\ \hline
& 4.709835 & 0.000057 & 25.08 & 5.33 & 13.520 & 0.050 \\
 & 4.888006 & 0.000062 & 20.29 & 6.25 & 13.401 & 0.058 \\
  & 4.889219 & 0.000058 & 10.29 & 8.04 & 13.318 & 0.083 \\
J1137+3549 &          &  &  &  &  &  \\ \hline
& 4.781802 & 0.000182 & 50.43 & 17.36 & 13.622 & 0.072 \\
 & 4.841312 & 0.000108 & 53.00 & 10.08 & 14.005 & 0.042 \\
 & 4.874166 & 0.000134 & 74.27 & 10.78 & 14.033 & 0.044 \\
J1148+5251 &          &  &  &  &  &  \\ \hline
 & 4.919028 & 0.000058 & 24.72 & 4.66 & 13.192 & 0.049 \\
 & 4.944227 & 0.000058 & 8.91 & 1.09 & 17.656 & 0.123 \\
  & 4.945228 & 0.000067 & 9.85 & 4.14 & 14.480 & 0.526 \\
  &  5.152113 & 0.000121 & 54.16& 8.94 & 13.695& 0.055 \\
 % & 5.151545 & 0.002880 & 55.09 & 104.41 & 13.519 & 1.311 \\
% & 5.152666 & 0.000737 & 28.92 & 61.04 & 13.304 & 2.121 \\
J1306+0356 &          &  &  &  &  &  \\ \hline
   & 4.613924 & 0.000790 & 103.30 & 31.35 & 13.250 & 0.225 \\
 & 4.614603 & 0.000033 & 29.33 & 3.21 & 13.722 & 0.050 \\
   & 4.615828 & 0.000045 & 23.36 & 4.28 & 13.465 & 0.075 \\
   & 4.668109 & 0.000020 & 8.71 & 1.87 & 13.993 & 0.115 \\
  & 4.668741 & 0.000081 & 56.99 & 4.05 & 13.892 & 0.035 \\
  & 4.711108 & 0.000135 & 35.60 & 11.23 & 13.129 & 0.096 \\
  & 4.860508 & 0.000044 & 35.43 & 3.46 & 13.952 & 0.031 \\
  & 4.862256 & 0.000055 & 28.31 & 5.49 & 13.933 & 0.051 \\
  & 4.863239 & 0.000043 & 10.18 & 3.83 & 13.881 & 0.091 \\
   & 4.864687 & 0.000053 & 56.14 & 5.04 & 14.327 & 0.026 \\
    & 4.866813 & 0.000025 & 20.73 & 2.13 & 14.009 & 0.029 \\
   & 4.868951 & 0.000102 & 51.40 & 8.62 & 13.670 & 0.055 \\
    & 4.880126 & 0.000061 & 27.59 & 4.35 & 13.873 & 0.047 \\
   & 4.881271 & 0.000058 & 17.01 & 4.23 & 13.690 & 0.061 \\
\hline
\end{tabular}
\end{table}

\begin{table}
\scriptsize
\contcaption{Complete list of \cfour \,absorbers and measured properties}
\begin{tabular}{lrrrrrr}
\hline\hline
QSO & $z$ & $\Delta z$ & $b$ & $\Delta b$ & $\log N$ & $\Delta \log N$ \\ \hline 
J1319+0950 &          &  &  &  &  &  \\ \hline
 & 4.653406 & 0.000076 & 33.18 & 6.20 & 12.989 & 0.058 \\
   & 4.660650 & 0.000108 & 42.92 & 7.70 & 13.310 & 0.088 \\
 & 4.663610 & 0.000734 & 86.53 & 46.22 & 13.625 & 0.283 \\
   & 4.665107 & 0.000104 & 38.60 & 17.21 & 13.328 & 0.456 \\
  & 4.716591 & 0.000030 & 27.60 & 2.58 & 13.554 & 0.026 \\
  & 5.264302 & 0.000095 & 25.20 & 7.48 & 13.181 & 0.079 \\
   & 5.335365 & 0.000056 & 3.30 & 2.64 & 13.333 & 0.367 \\
    & 5.374931 & 0.000054 & 16.59 & 4.79 & 13.447 & 0.057 \\
  & 5.570341 & 0.000425 & 51.95 & 27.03 & 13.793 & 0.183 \\
   & 5.573751 & 0.000240 & 49.18 & 14.23 & 14.139 & 0.111 \\
J1411+1217 &          &  &  &  &  &  \\ \hline
  & 4.930500 & 0.000439 & 211.93 & 26.04 & 13.982 & 0.048 \\
  & 4.960547 & 0.000100 & 6.41 & 4.33 & 13.715 & 0.364 \\
 & 5.250460 & 0.000063 & 73.55 & 4.62 & 14.442 & 0.025 \\
J1509-1749 &          &  &  &  &  &  \\ \hline
 & 4.641643 & 0.000051 & 26.96 & 4.46 & 13.514 & 0.042 \\
  & 4.791660 & 0.000166 & 51.38 & 12.64 & 13.439 & 0.082 \\
 & 4.815429 & 0.000055 & 44.63 & 3.79 & 14.050 & 0.037 \\
J2054-0005 &          &  &  &  &  &  \\ \hline
& 4.868746 & 0.000079 & 15.81 & 8.89 & 13.804 & 0.143 \\
 & 5.213101 & 0.000147 & 8.67 & 5.96 & 13.828& 0.368 \\
  %& 5.213091 & 0.000235 & 8.92 & 10.16 & 13.813 & 0.568 \\
J2315-0023 &          &  &  &  &  &  \\ \hline
 & 4.897117 & 0.001163 & 55.02 & 64.70 & 13.598 & 0.478 \\
  & 4.898876 & 0.000354 & 31.61 & 22.67 & 13.854 & 0.257 \\
\hline
\end{tabular}
\end{table}

\section{Velocity plots of \cfour \,absorbers used in the correlation measurement}\label{appendix:vel_CIV}
We present in Fig. \ref{fig:mosaic_transmission} the velocity for the $37$ \cfour\, systems of Sample $\alpha$. Note the consistent presence of at least of a few completely opaque pixels at the location where the Lyman-$\alpha$ absorption at the redshift of \cfour \, is expected. Note that we have not plotted individual detections of systems with less than $\lesssim 100$ km s$^{-1}$ separation as only one redshift for the whole system was retained for the correlation measurement. However, multiple absorbers forming a system but can be easily spotted on some plots.

\begin{figure*}
\centering
\includegraphics[height=0.24 \textheight]{vplots/vplot_J0231_z_4569.pdf}
\includegraphics[height=0.24 \textheight]{vplots/vplot_J0231_z_4745.pdf}
\includegraphics[height=0.24 \textheight]{vplots/vplot_J0002_z_4871.pdf}
\includegraphics[height=0.24 \textheight]{vplots/vplot_J0231_z_4884.pdf}
\includegraphics[height=0.24 \textheight]{vplots/vplot_J1044_z_4899.pdf}
\includegraphics[height=0.24 \textheight]{vplots/vplot_J0002_z_4942.pdf} \\
\includegraphics[height=0.24 \textheight]{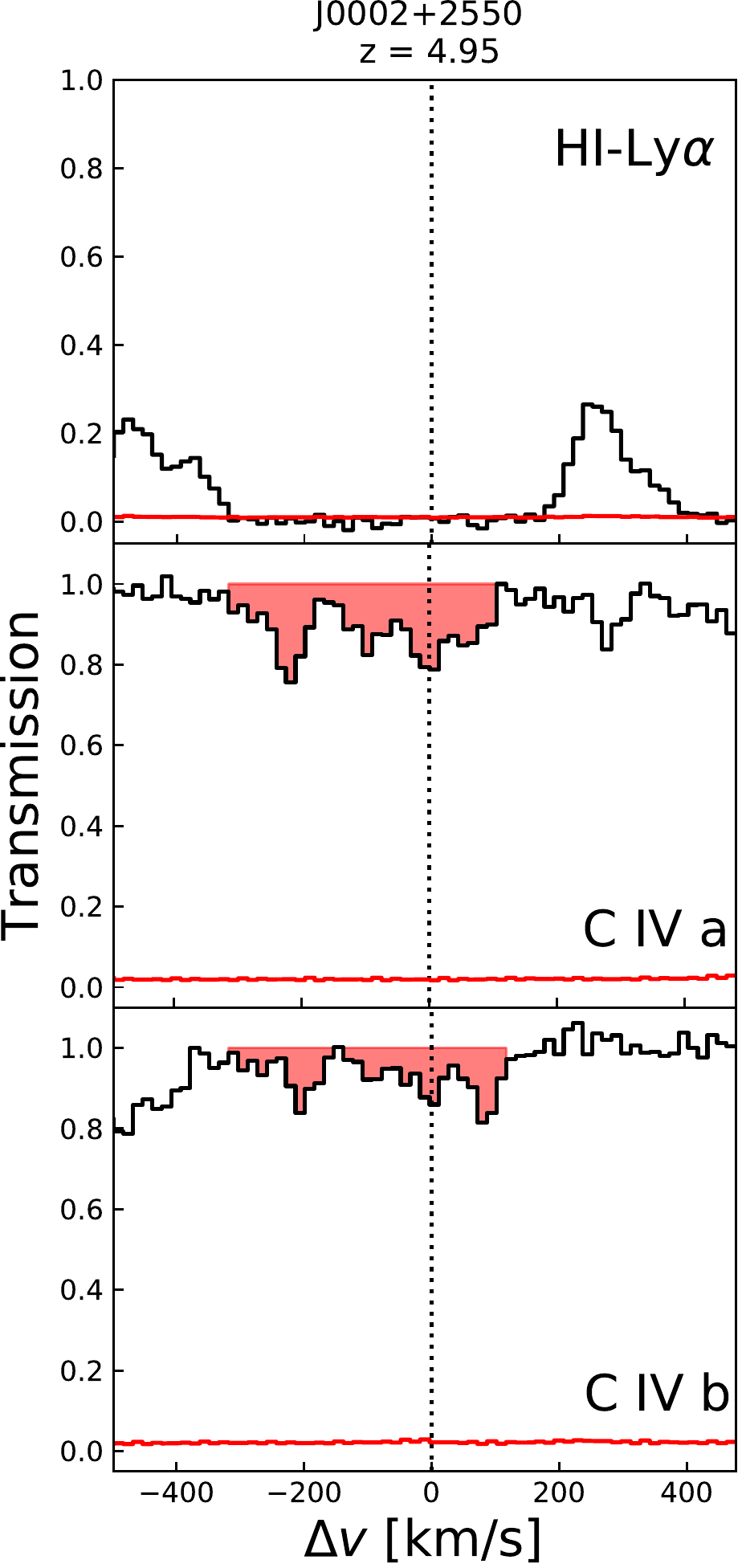} 
\includegraphics[height=0.24 \textheight]{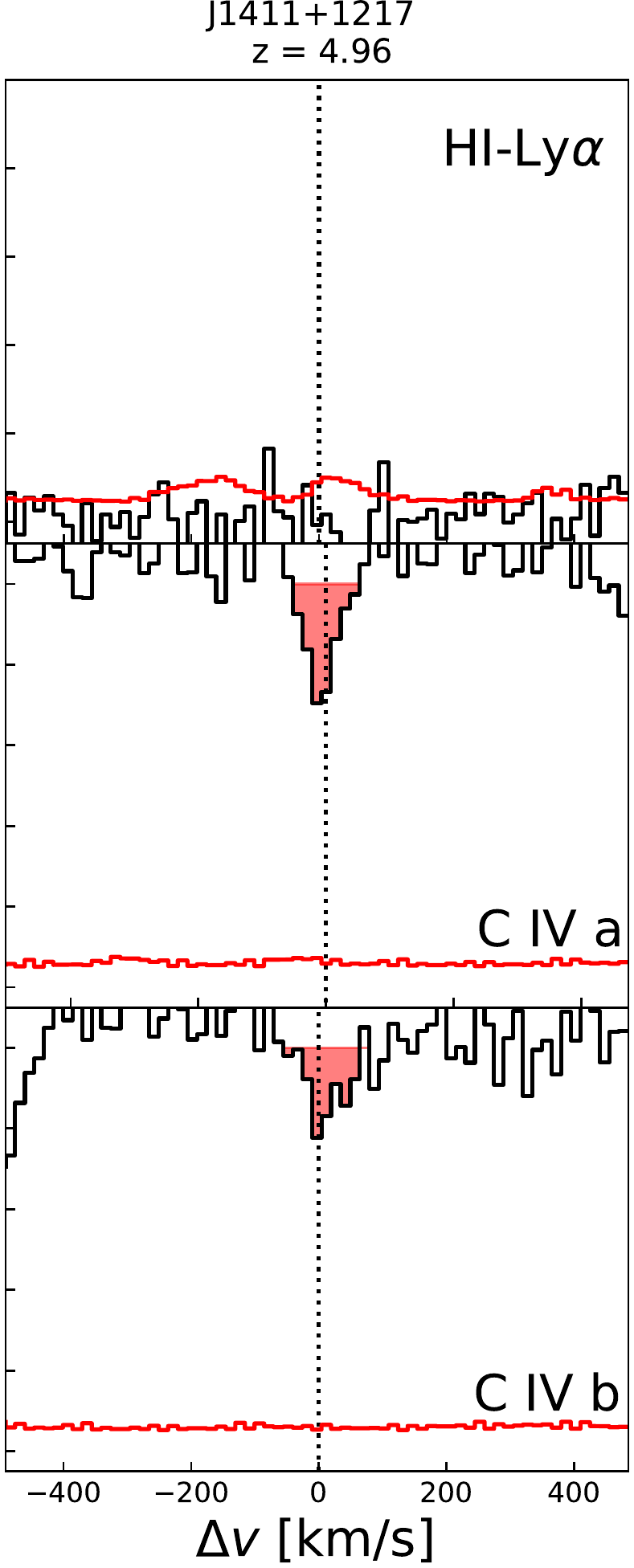}
\includegraphics[height=0.24 \textheight]{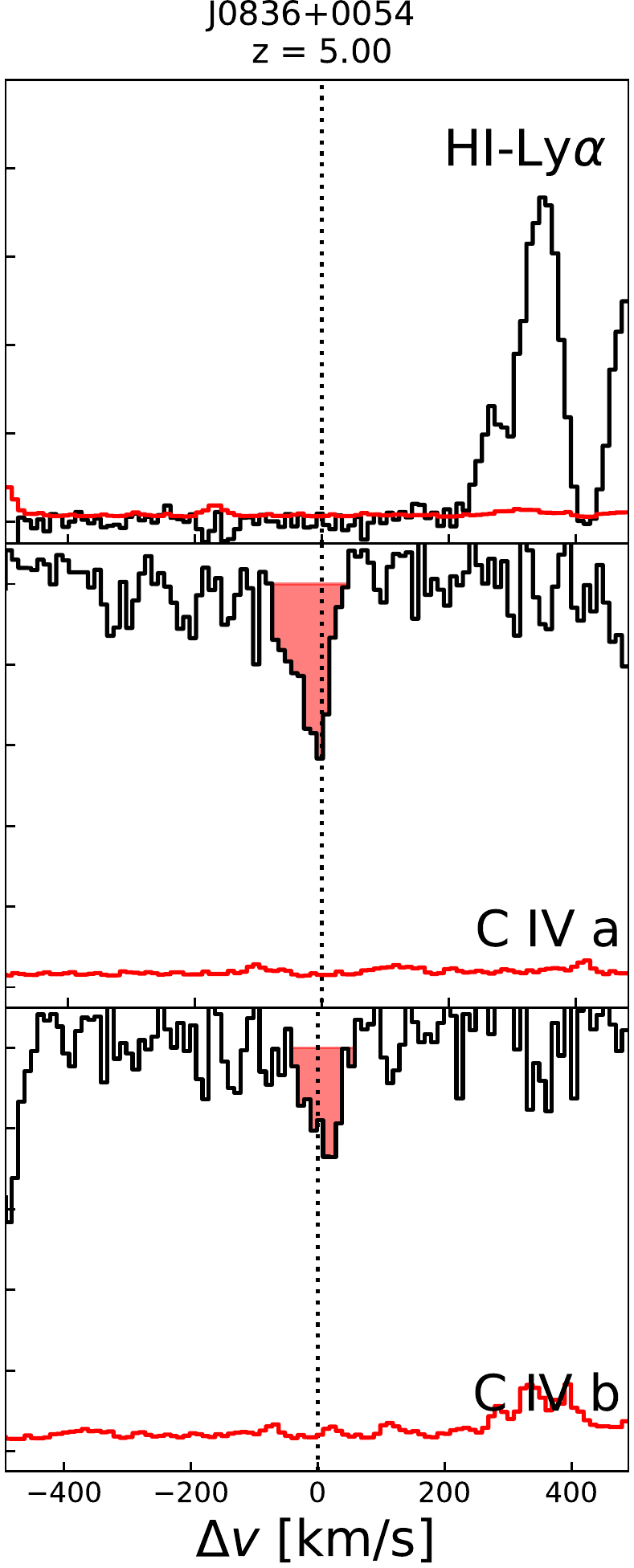}
\includegraphics[height=0.24 \textheight]{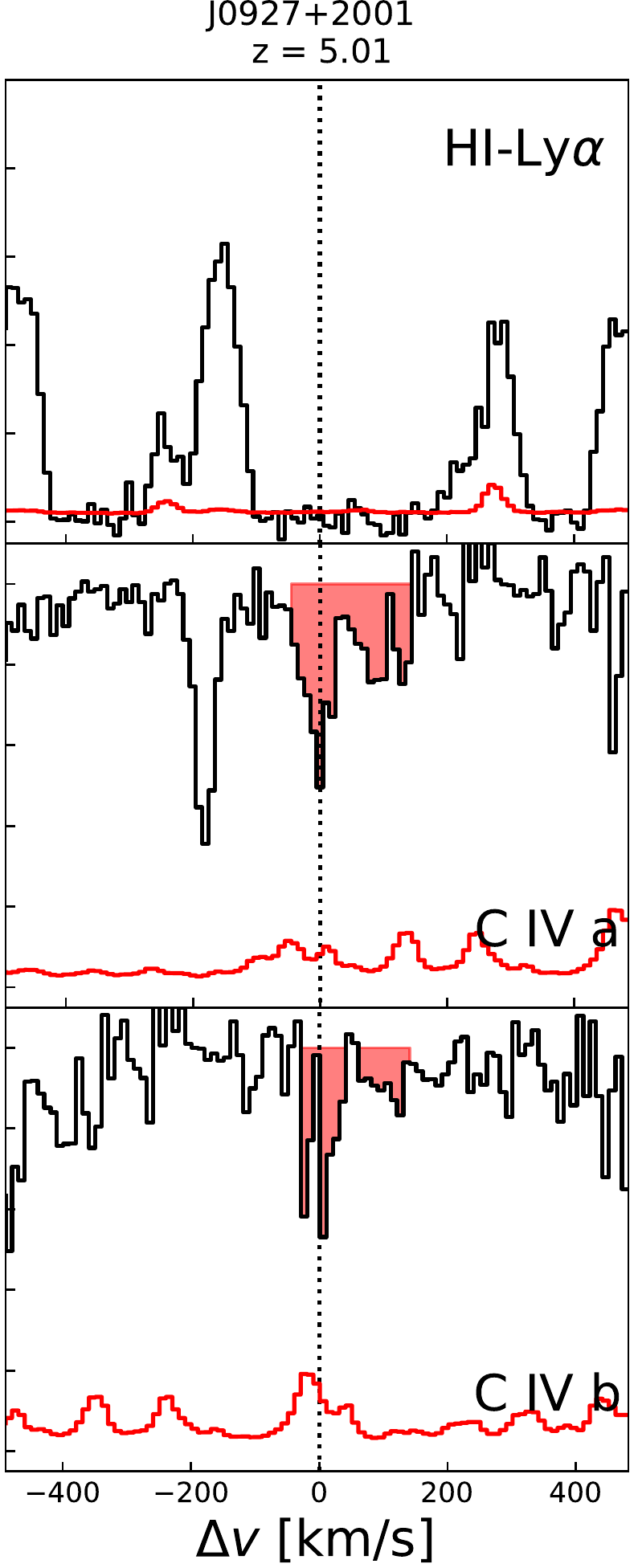}
\includegraphics[height=0.24 \textheight]{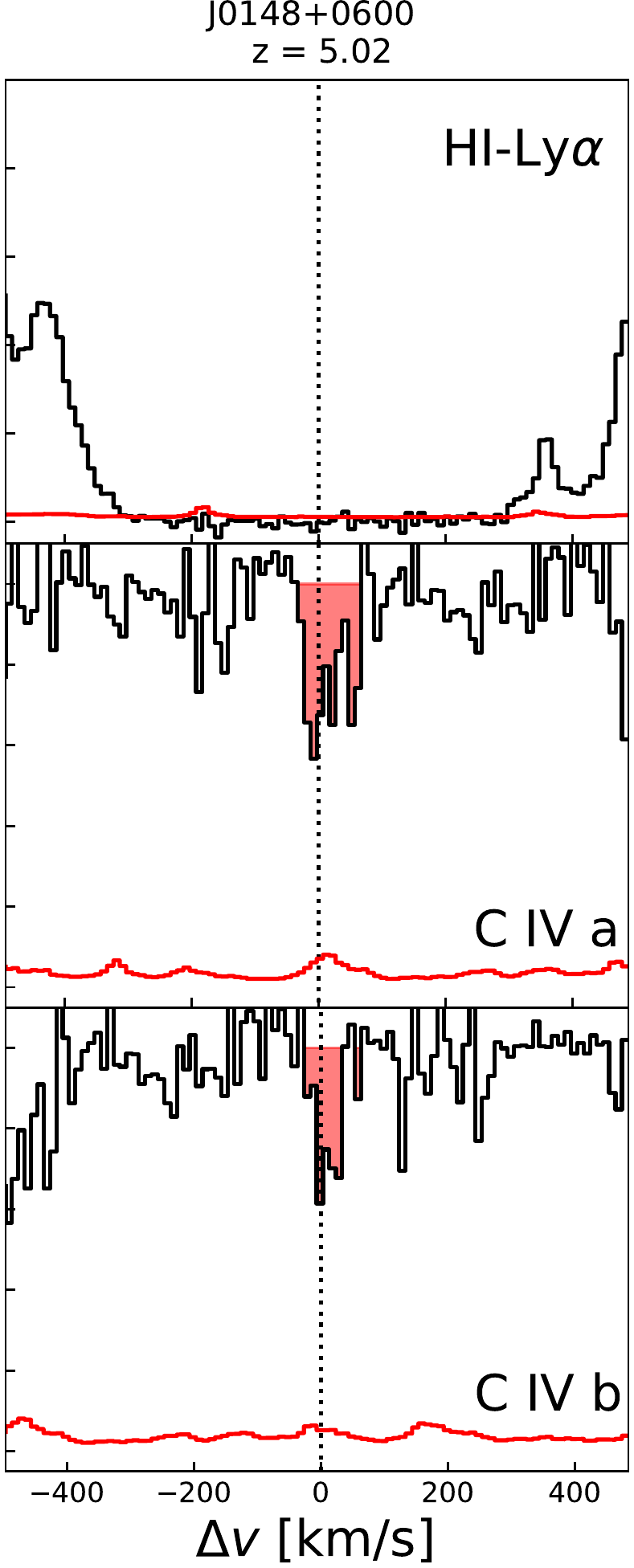} 
\includegraphics[height=0.24 \textheight]{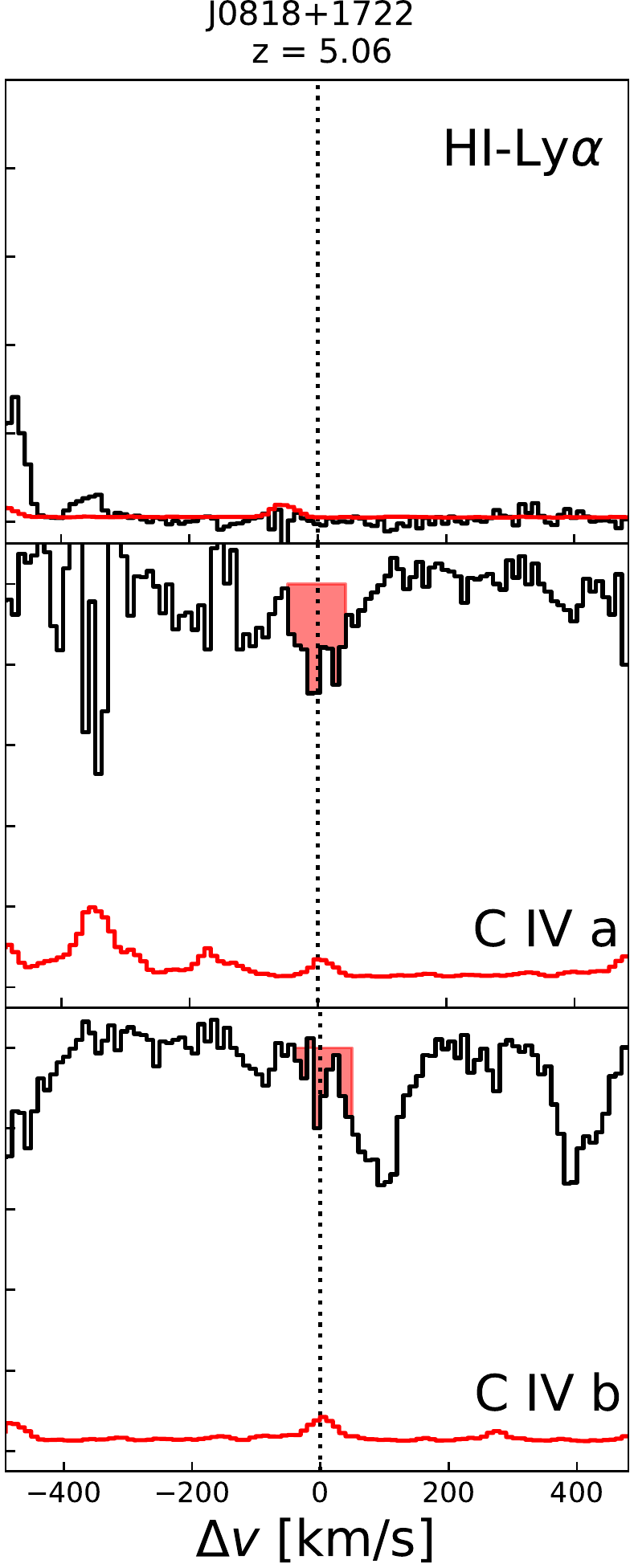} \\
\includegraphics[height=0.24 \textheight]{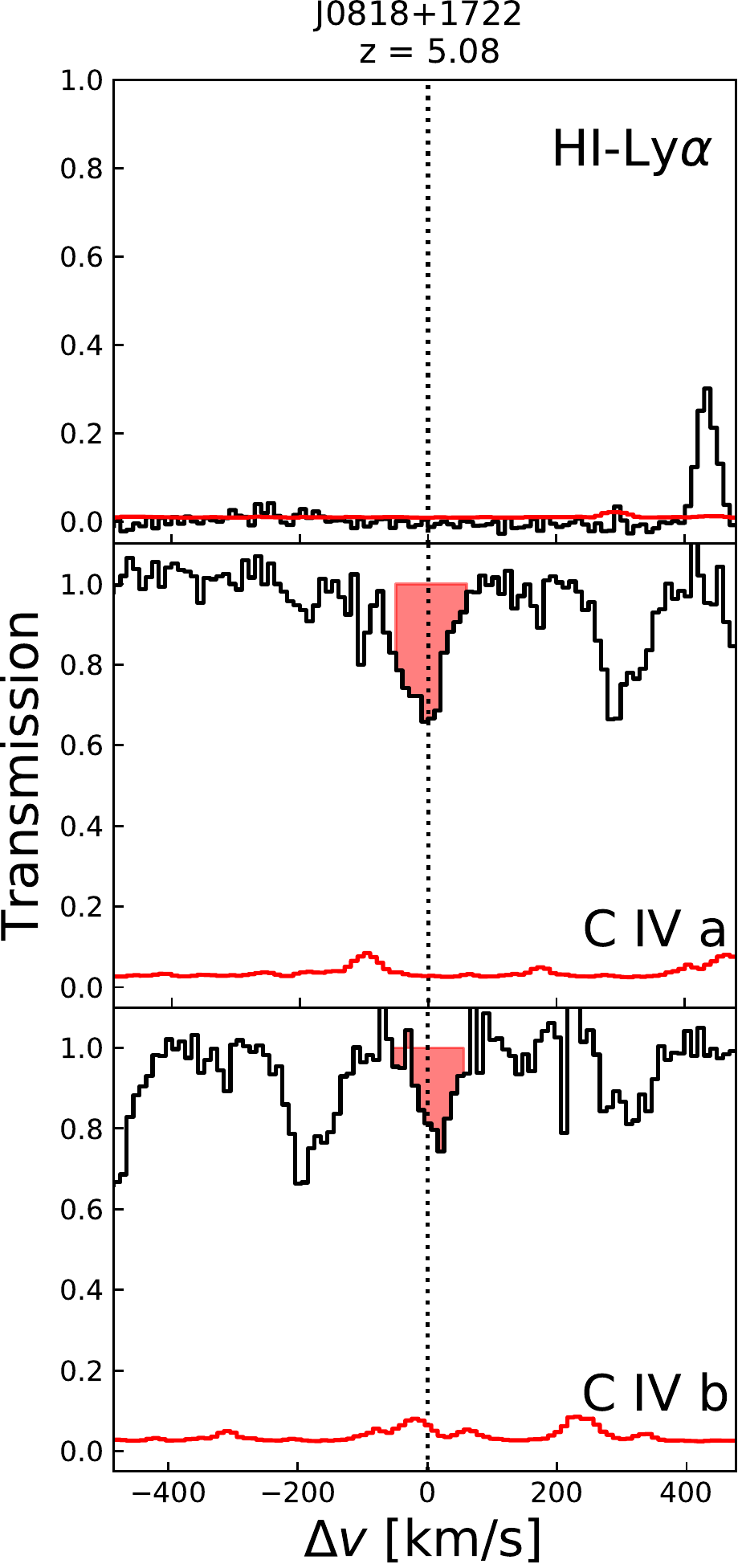}
\includegraphics[height=0.24 \textheight]{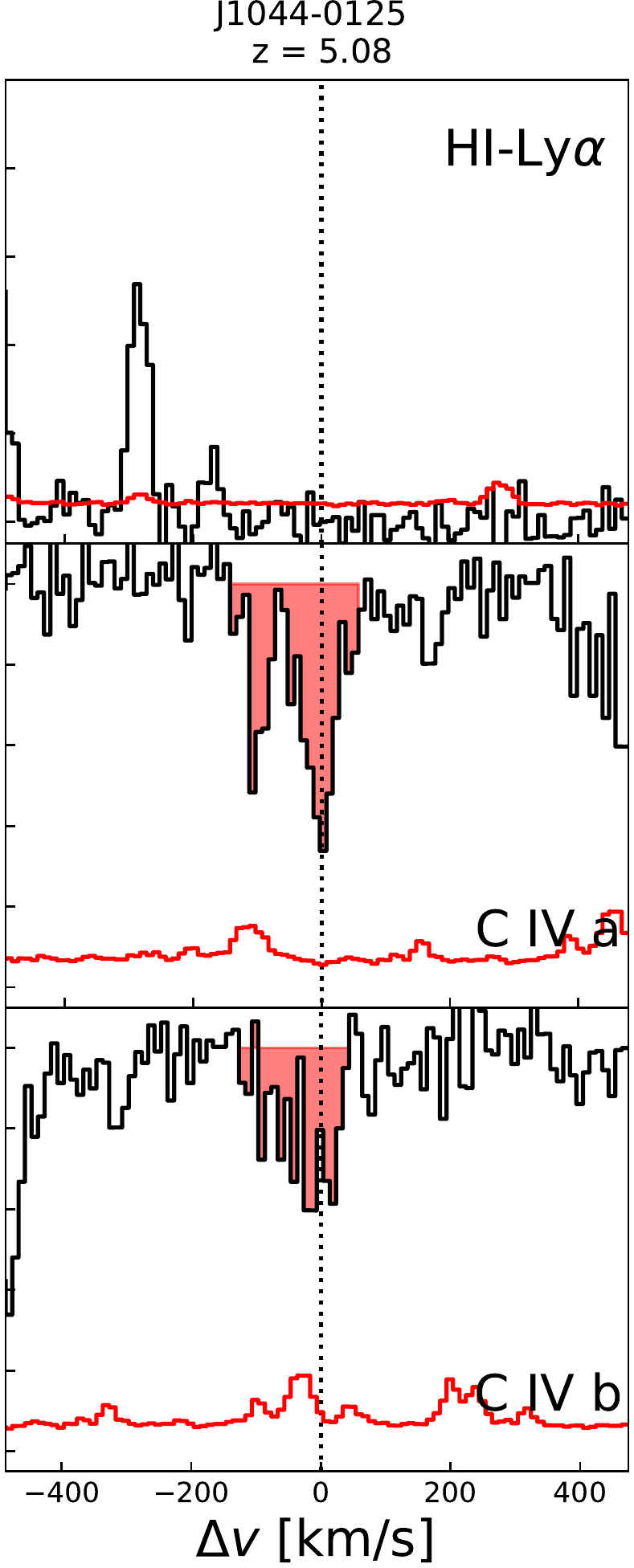}
\includegraphics[height=0.24 \textheight]{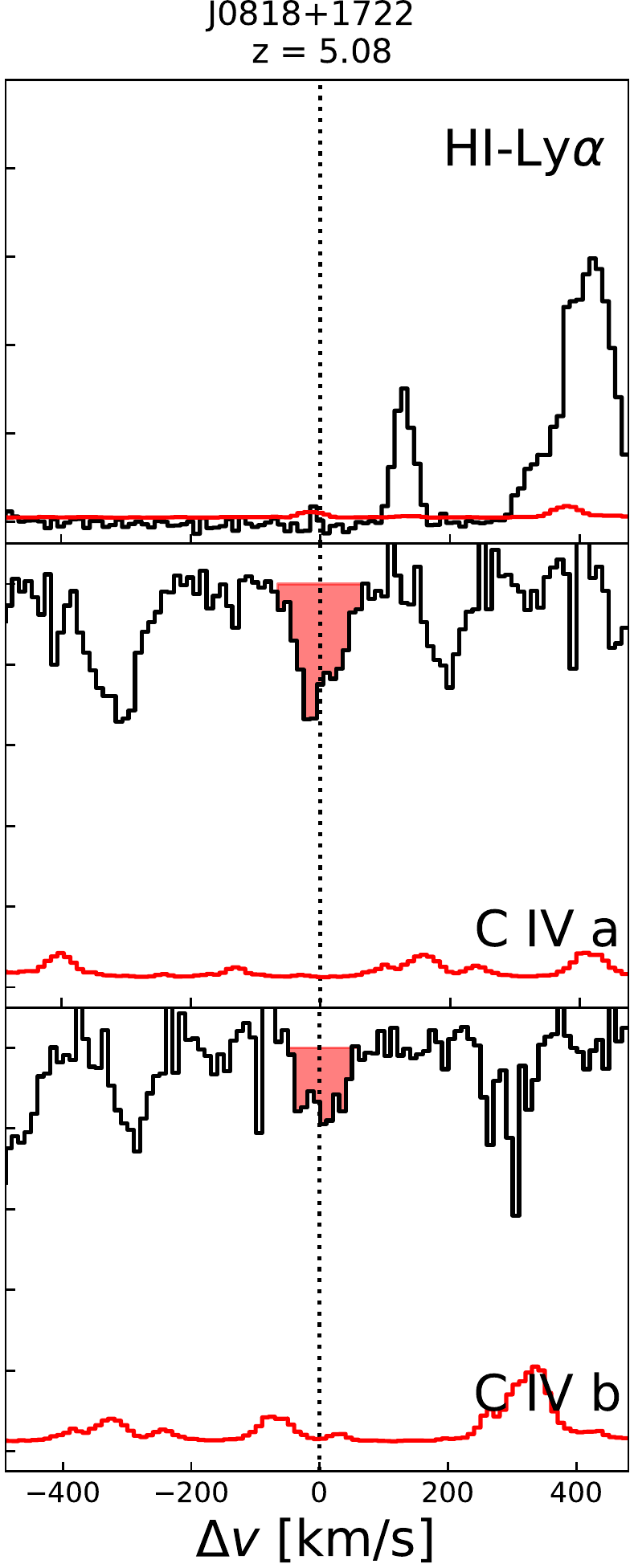}
\includegraphics[height=0.24 \textheight]{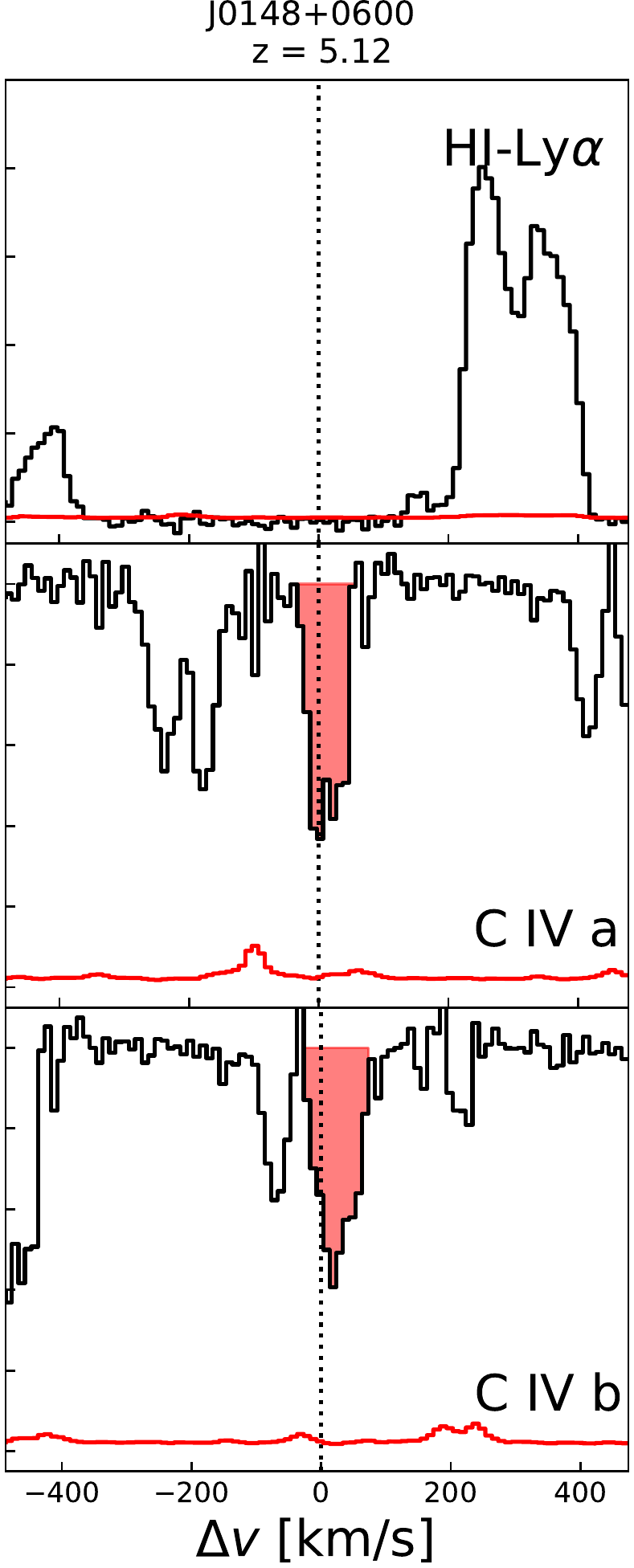}
\includegraphics[height=0.24 \textheight]{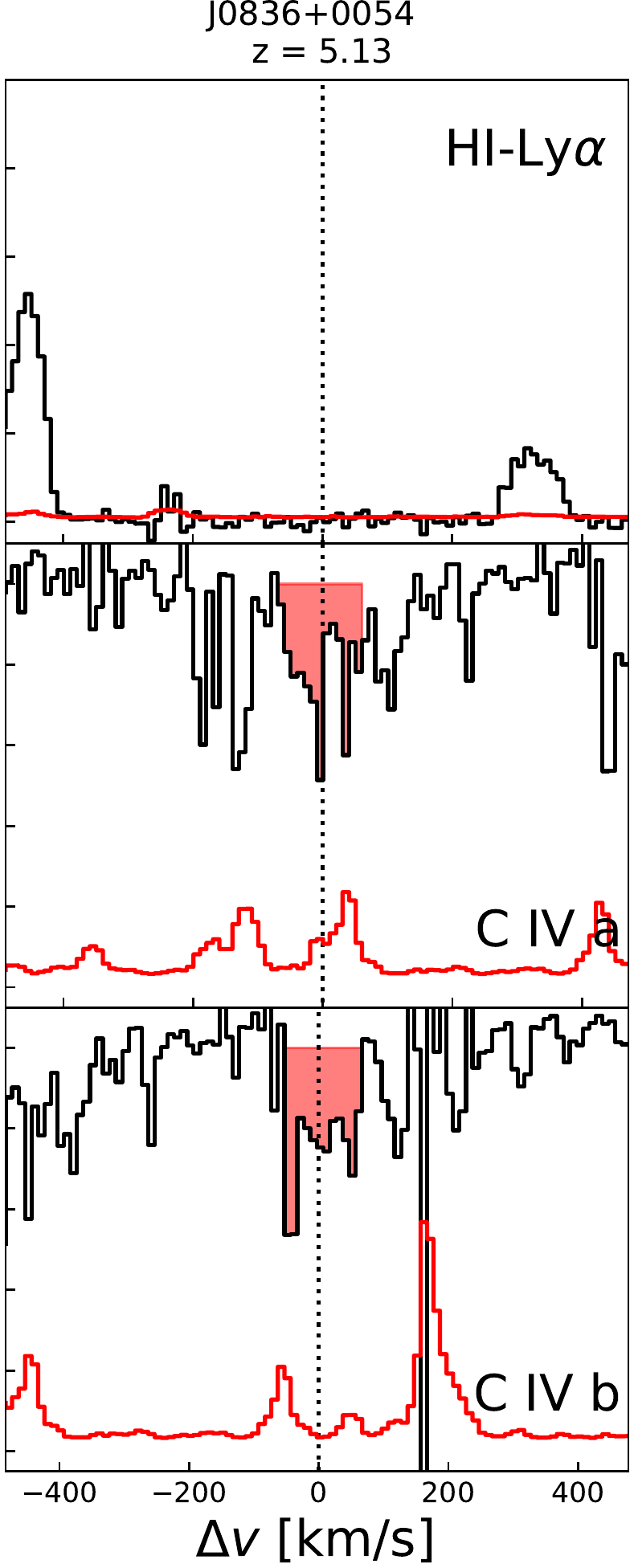} 
\includegraphics[height=0.24 \textheight]{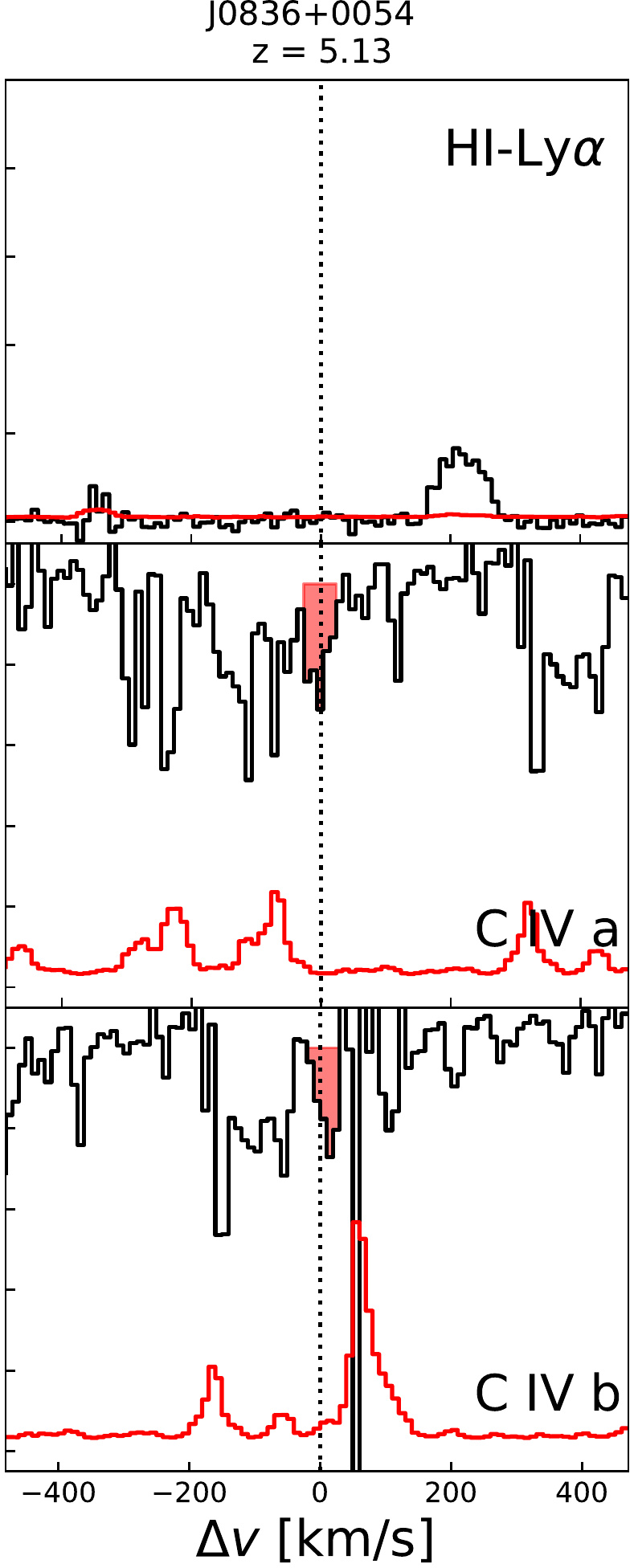} \\
\includegraphics[height=0.24 \textheight]{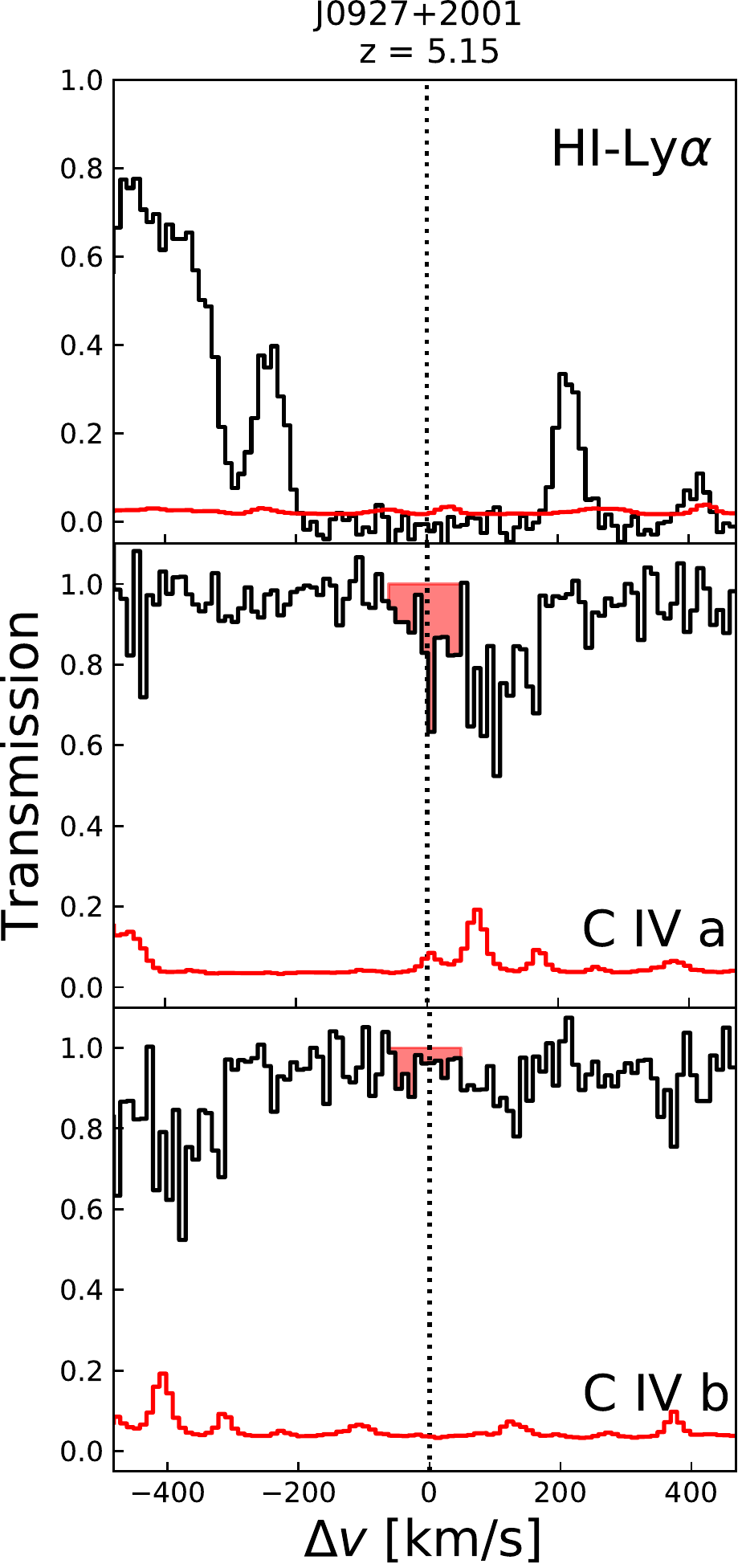}
\includegraphics[height=0.24 \textheight]{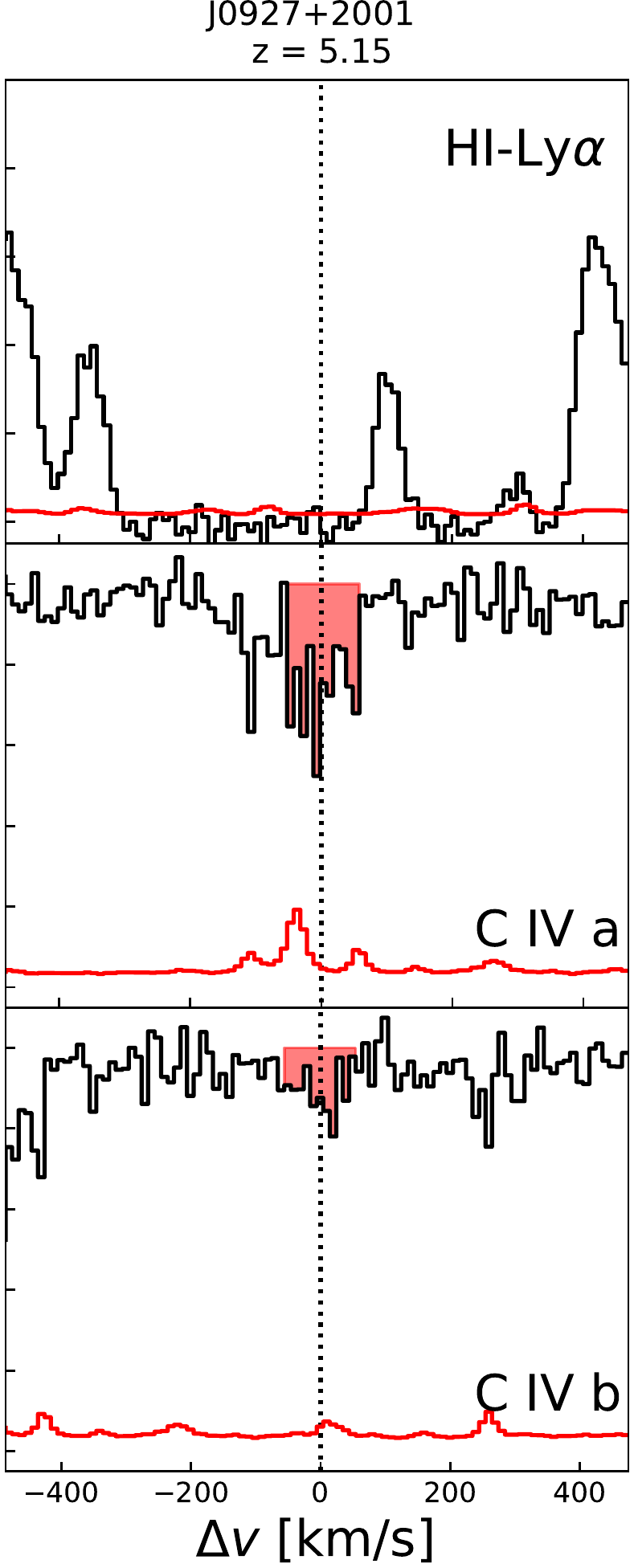}
\includegraphics[height=0.24 \textheight]{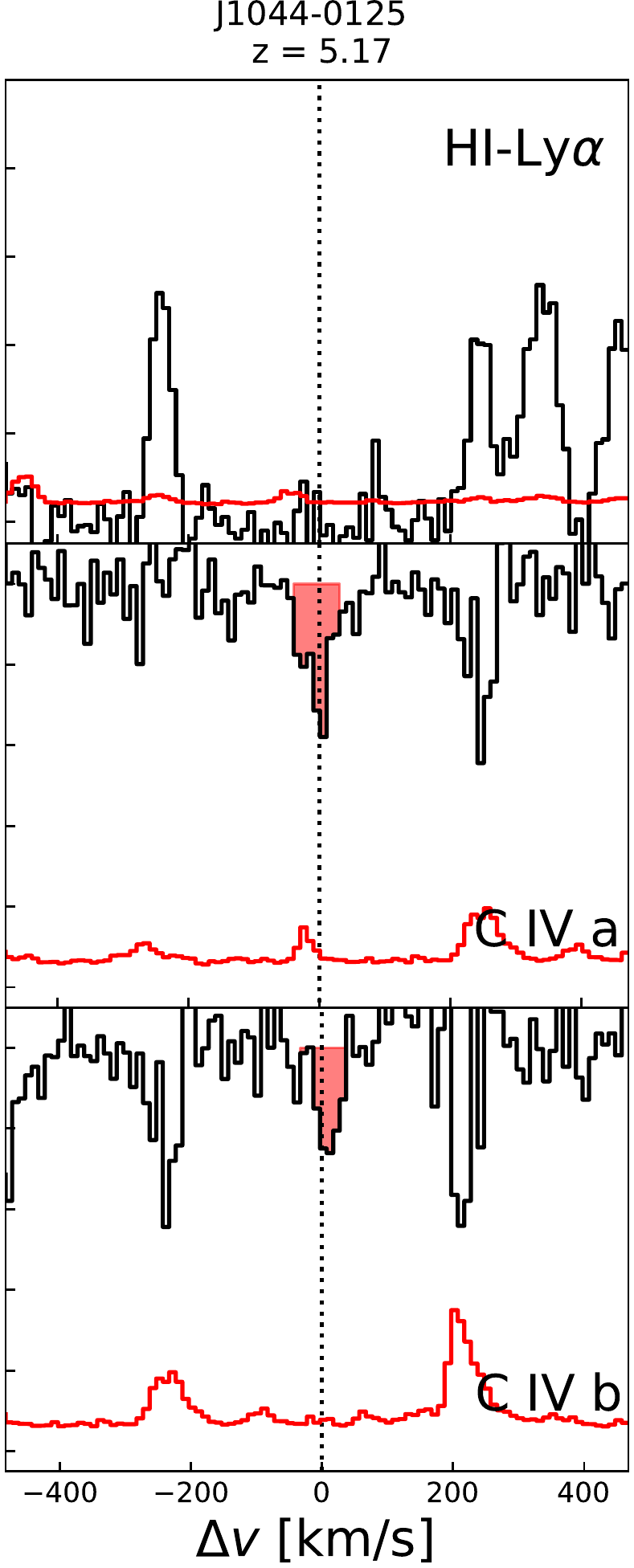}
\includegraphics[height=0.24 \textheight]{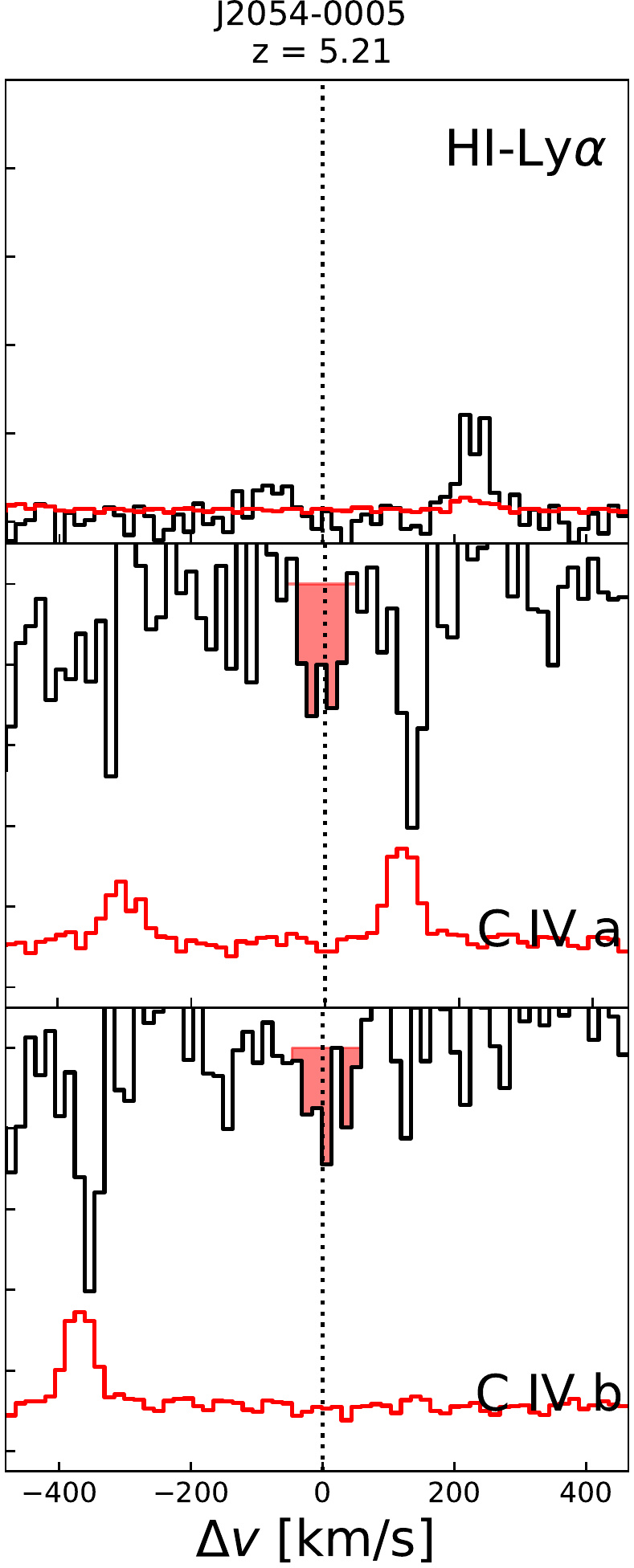}
\includegraphics[height=0.24 \textheight]{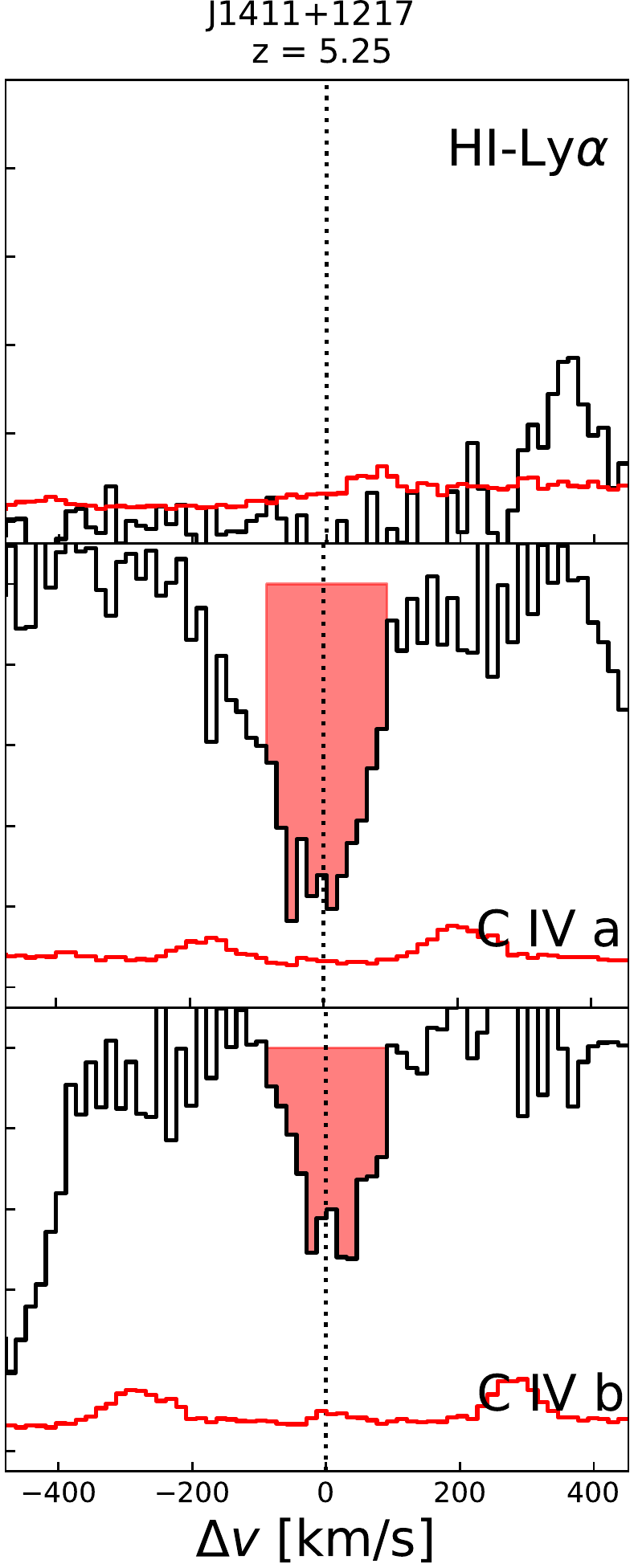}
\includegraphics[height=0.24 \textheight]{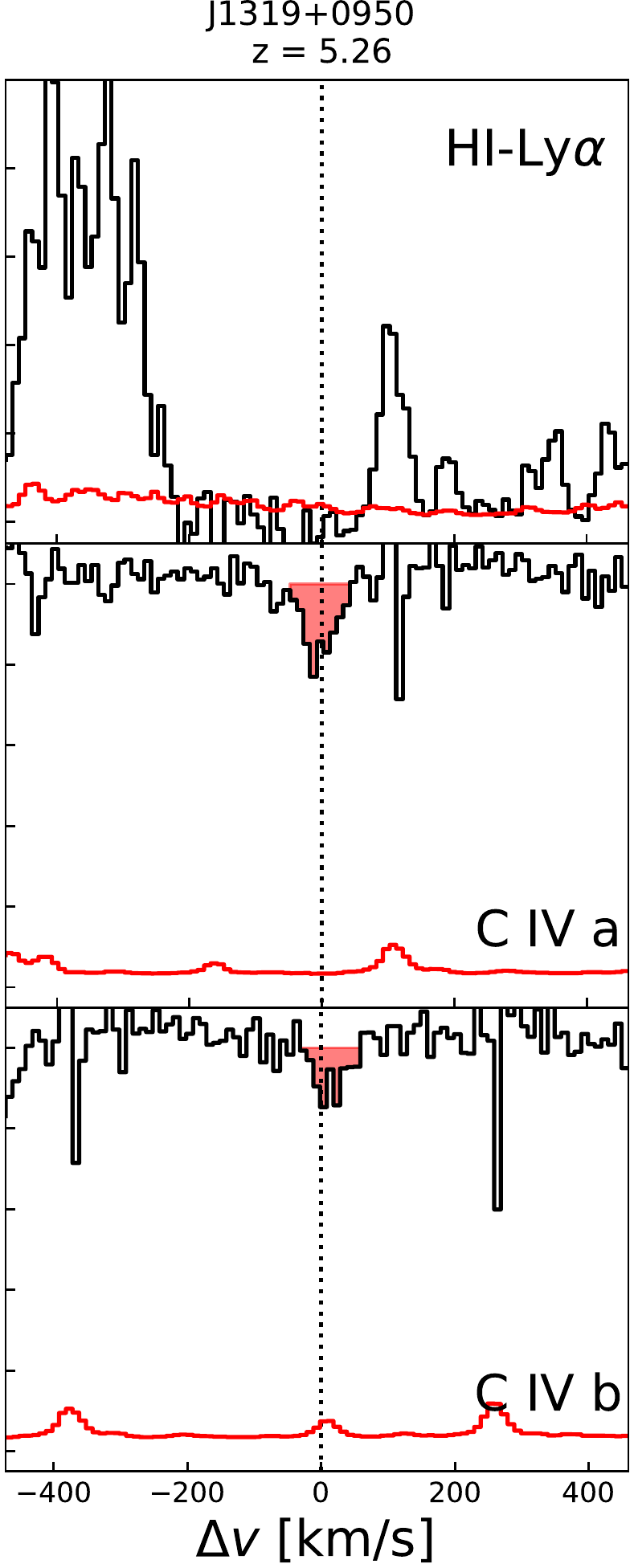}

\caption{Velocity plots of \cfour \,absorbers in Sample $\alpha$. For simplicity, absorbers with $\Delta v\lesssim 100$ km s$^{-1}$ were not plotted twice. They can be however be easily spotted in the lower \cfour \,panels. All \cfour \,absorbers land in a Lyman-$\alpha$ opaque region, often enclosed by high transmission spikes.  \label{fig:mosaic_transmission}}
\end{figure*}

\begin{figure*}
\centering
\includegraphics[height=0.24 \textheight]{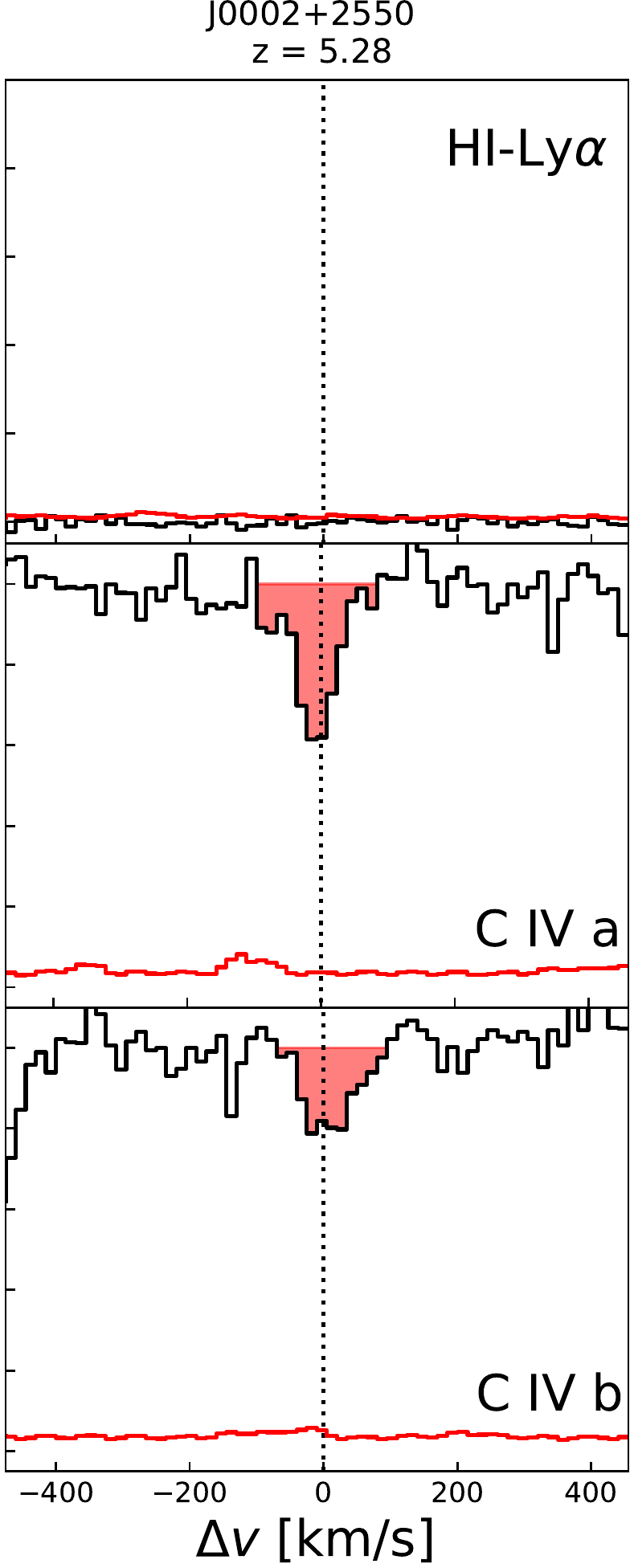}
\includegraphics[height=0.24 \textheight]{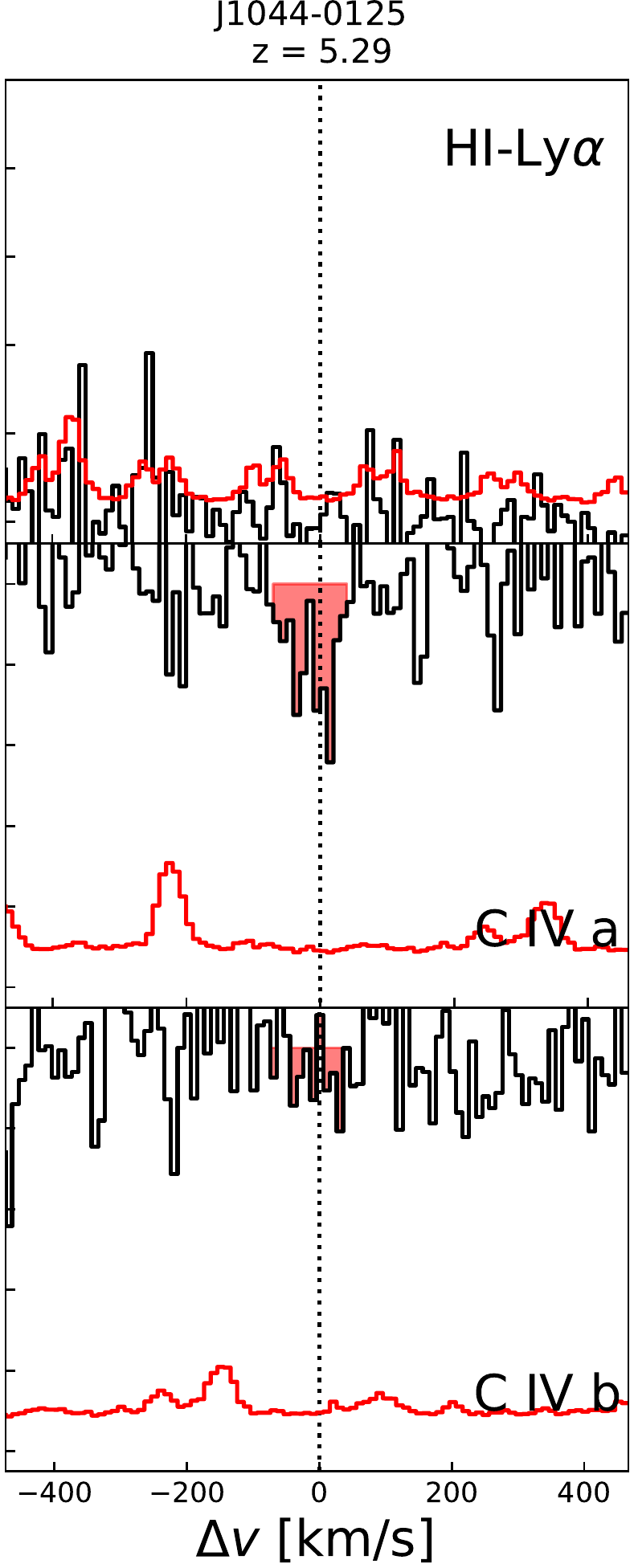}
\includegraphics[height=0.24 \textheight]{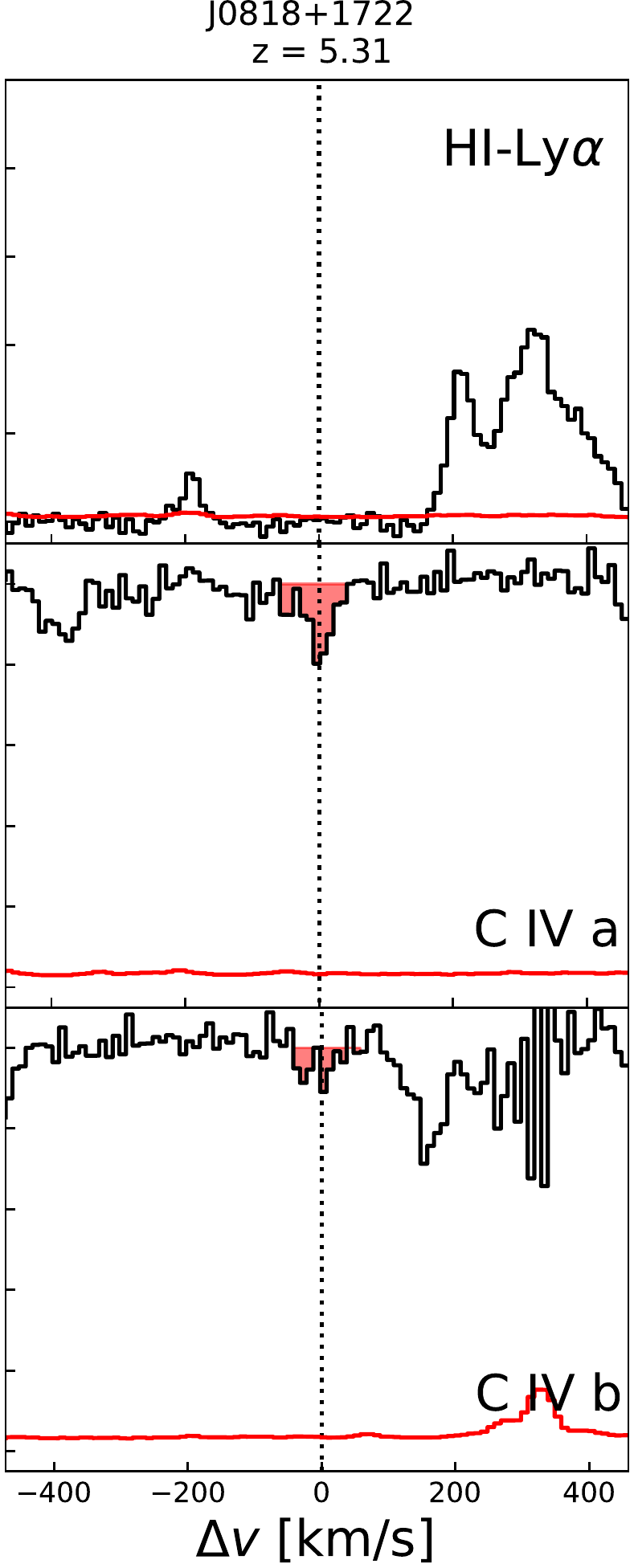}
\includegraphics[height=0.24 \textheight]{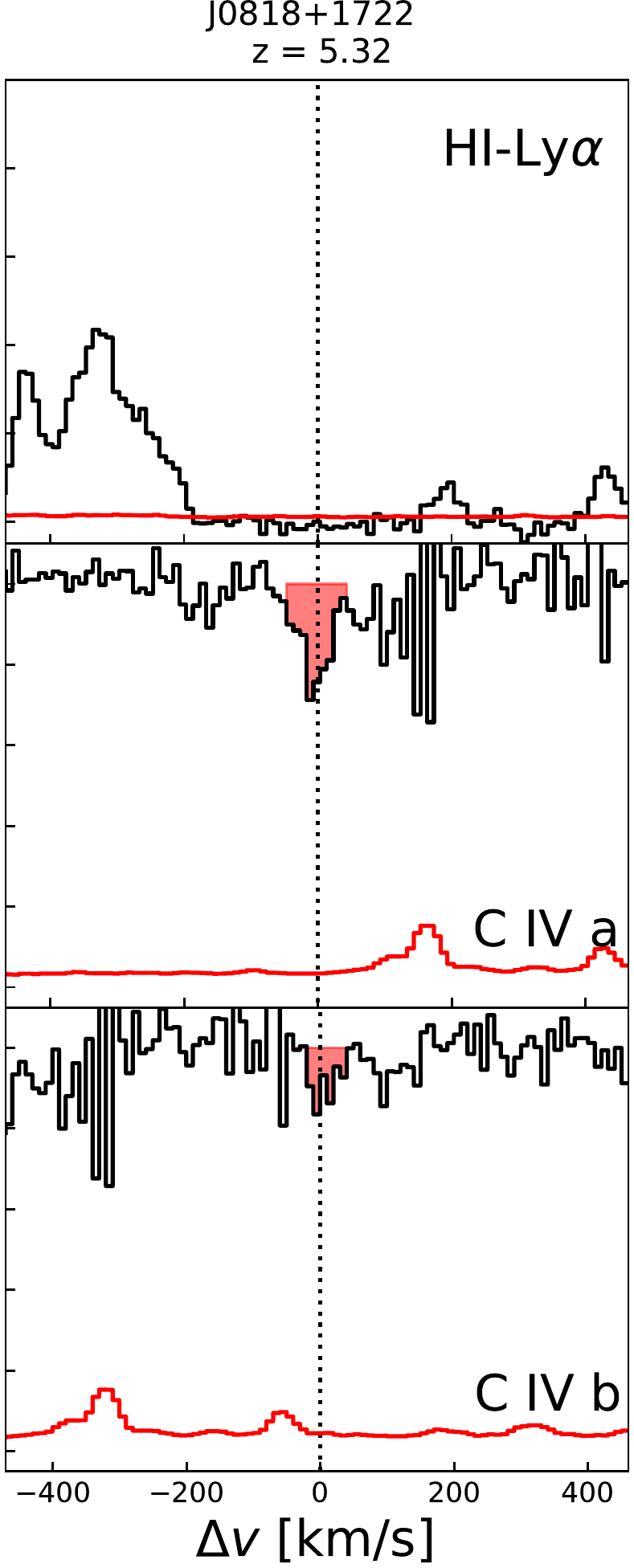}
\includegraphics[height=0.24 \textheight]{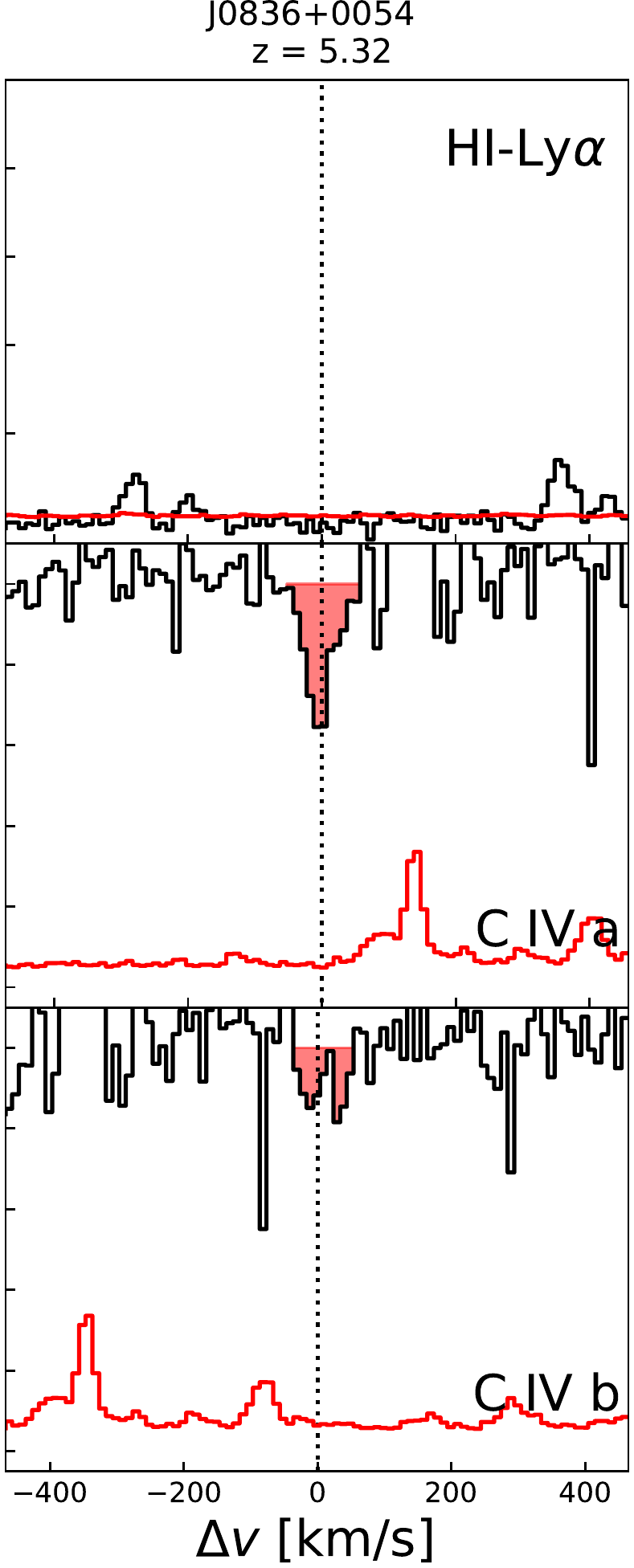}
\includegraphics[height=0.24 \textheight]{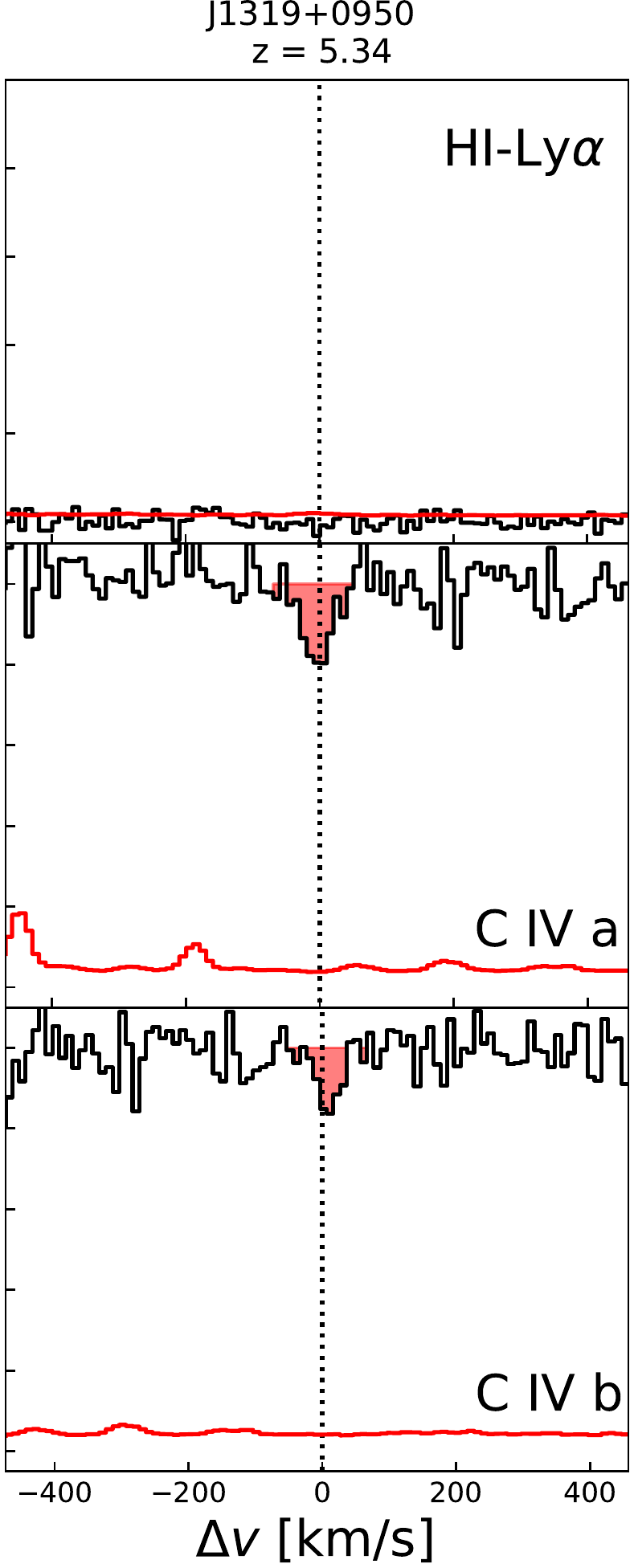} \\
\includegraphics[height=0.24 \textheight]{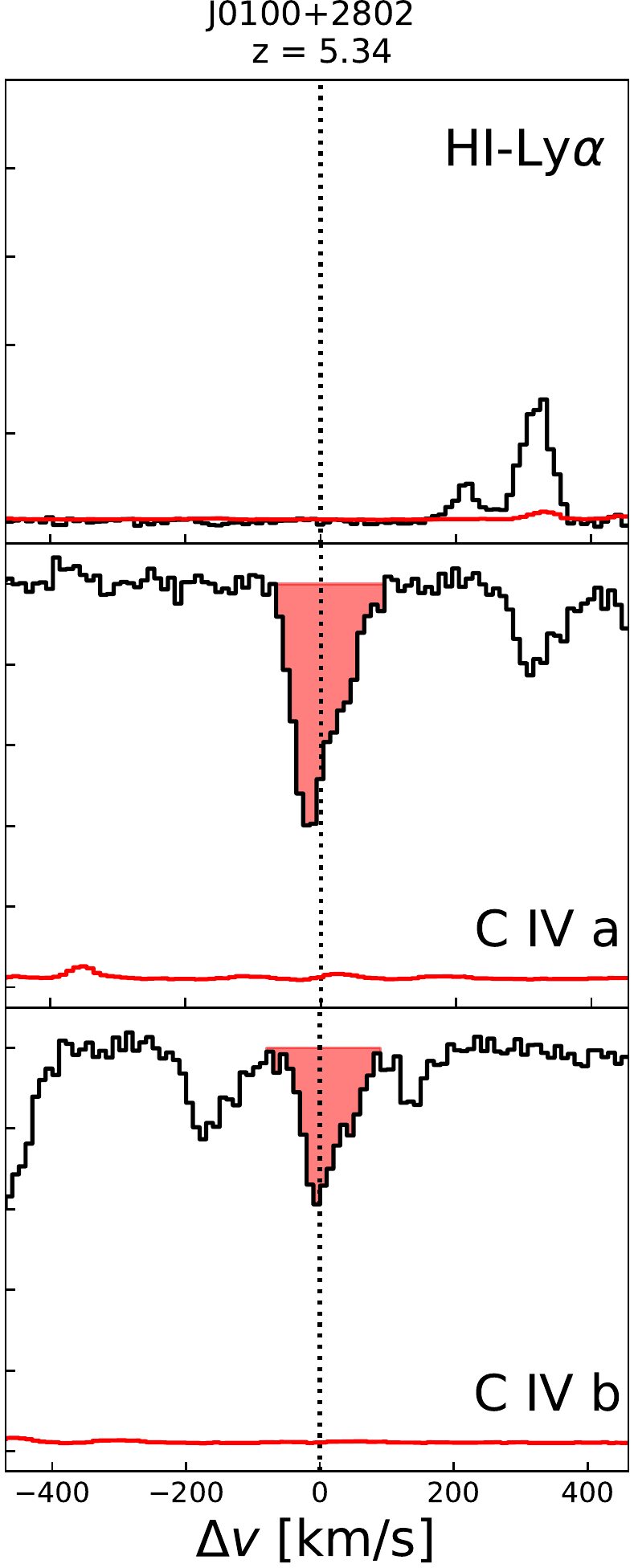} 
\includegraphics[height=0.24 \textheight]{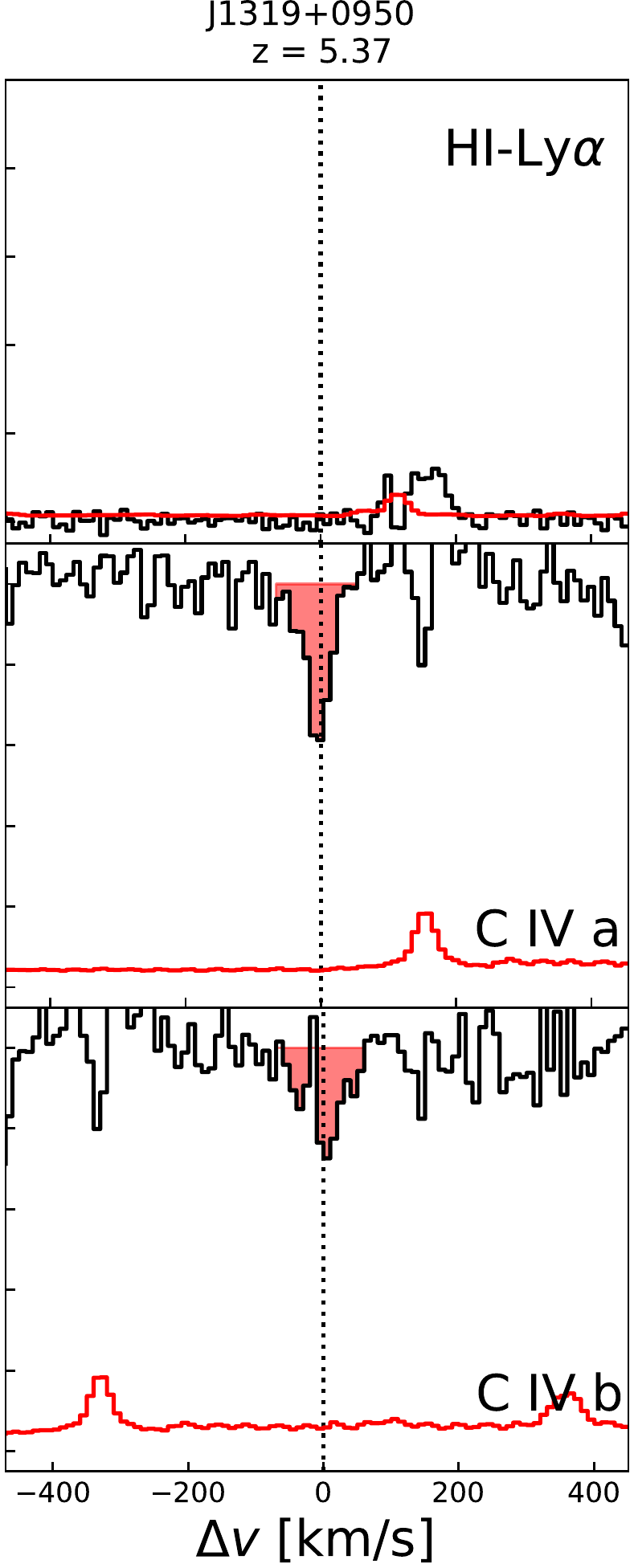}
\includegraphics[height=0.24 \textheight]{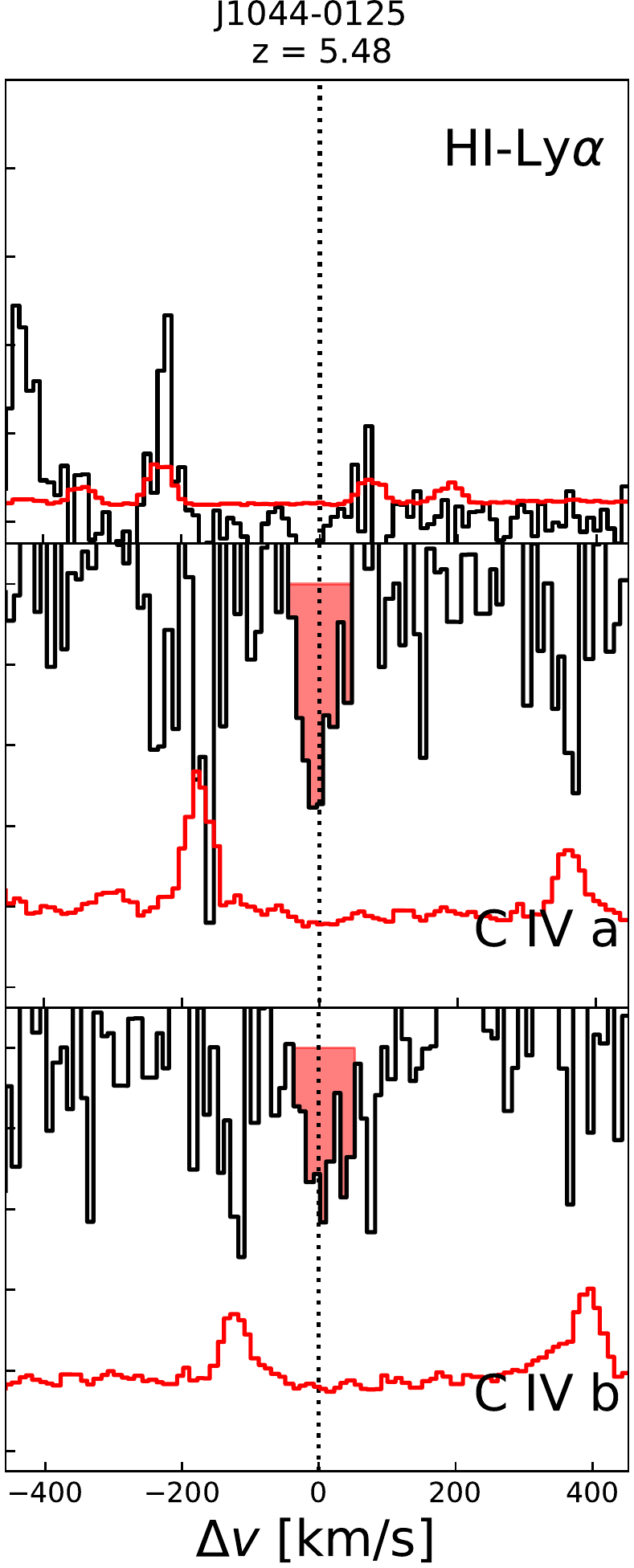}
\includegraphics[height=0.24 \textheight]{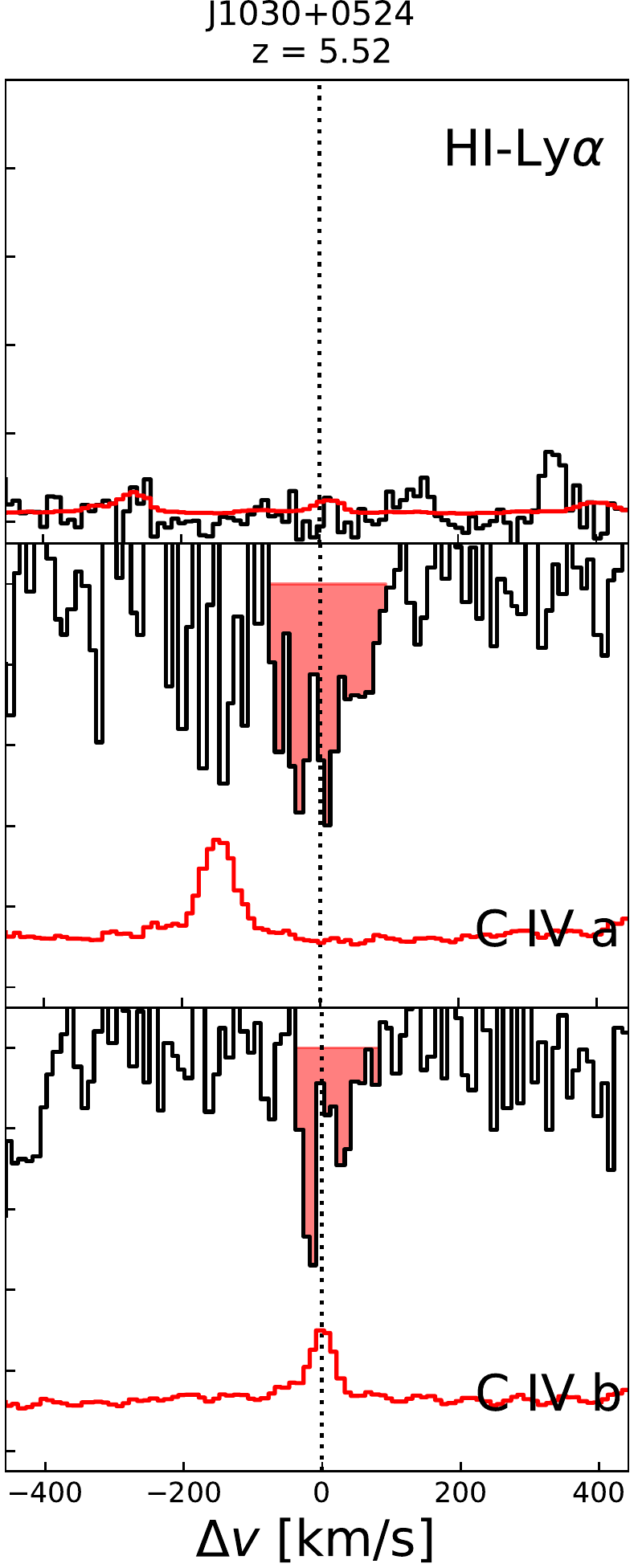}
\includegraphics[height=0.24 \textheight]{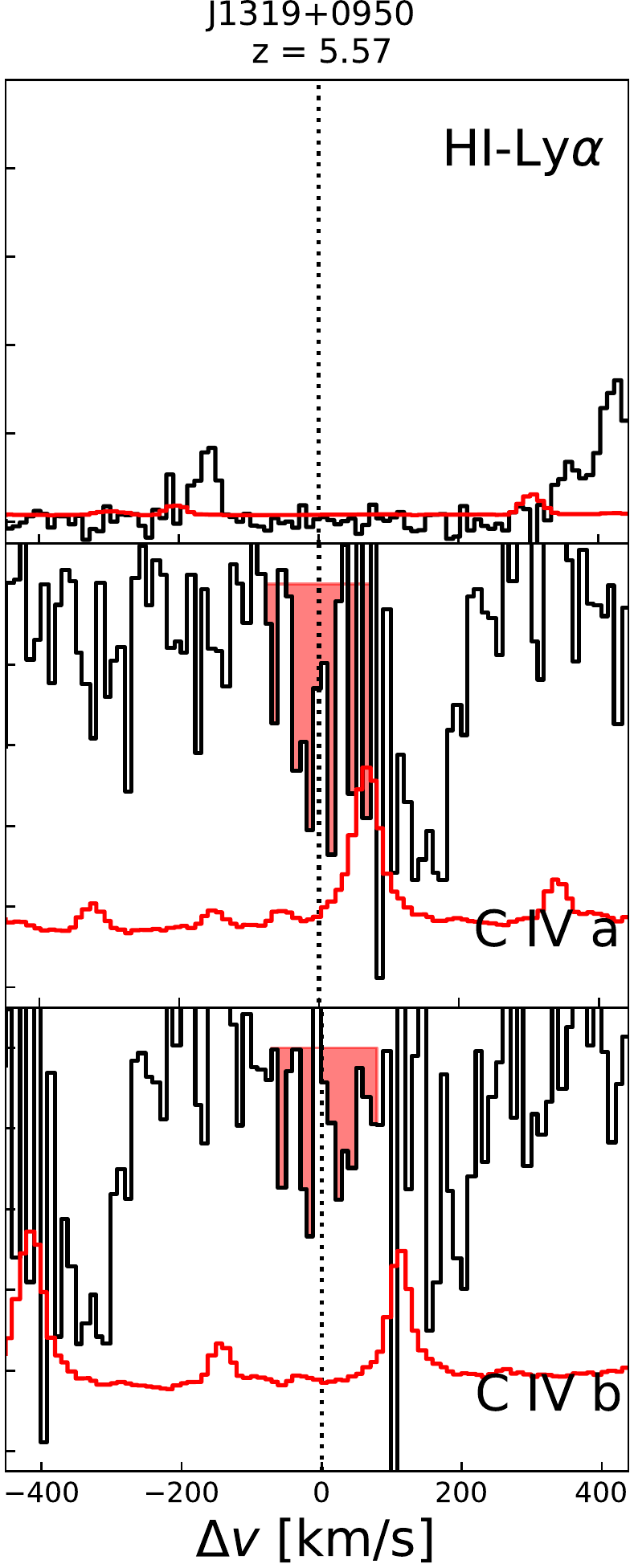}
\includegraphics[height=0.24 \textheight]{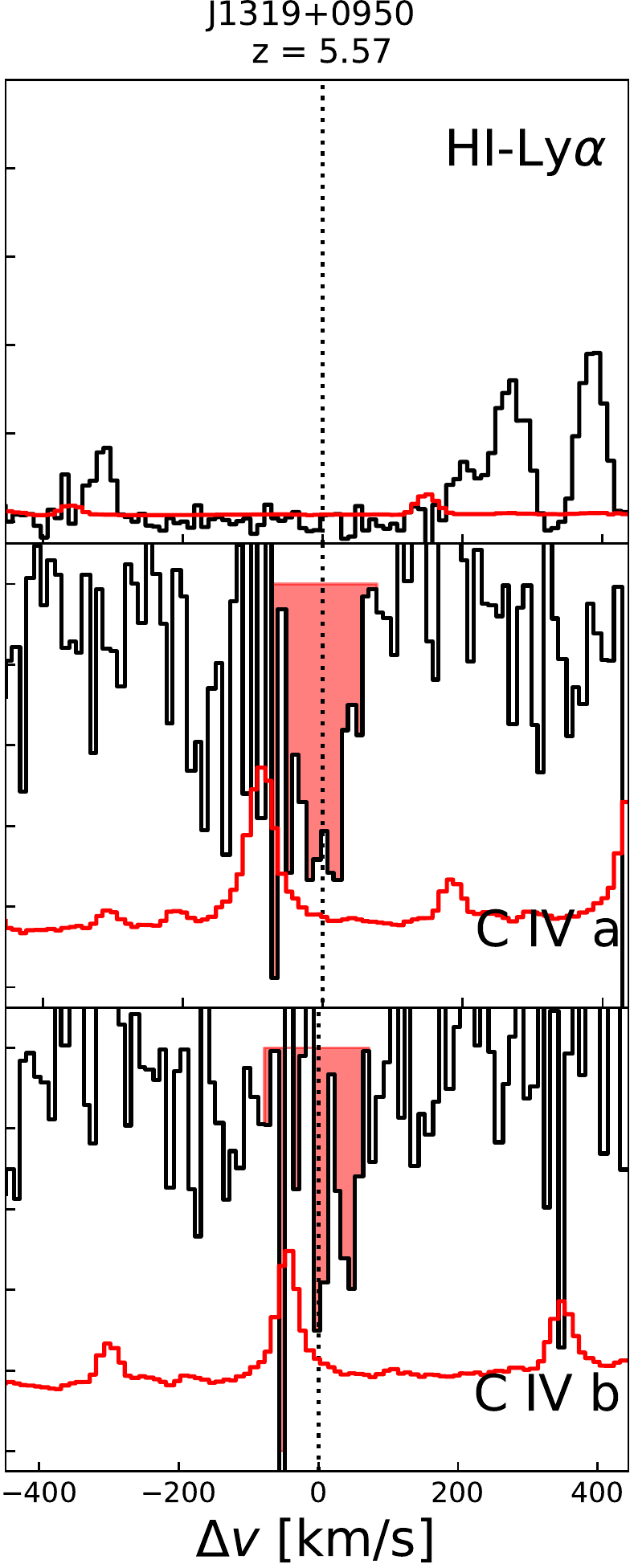} \\
\includegraphics[height=0.24 \textheight]{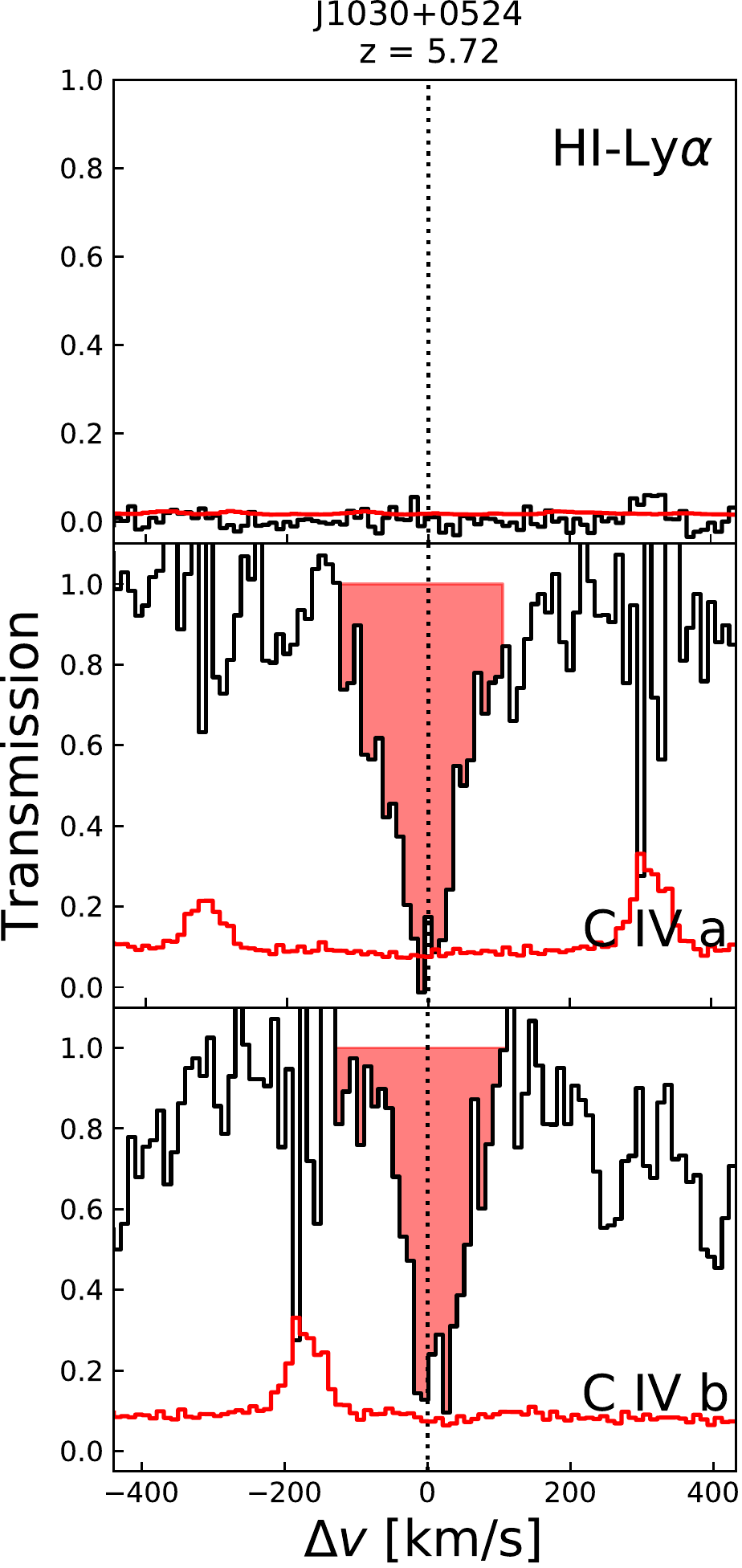} \\
\contcaption{\label{fig:mosaic_transmission_2}}
\end{figure*}

\section{Relaxing the bias parameters of the Lyman-$\alpha$ forest}\label{appendix:params}
\begin{figure*}
\hspace{-0.5cm}
\includegraphics[width=0.9\textwidth]{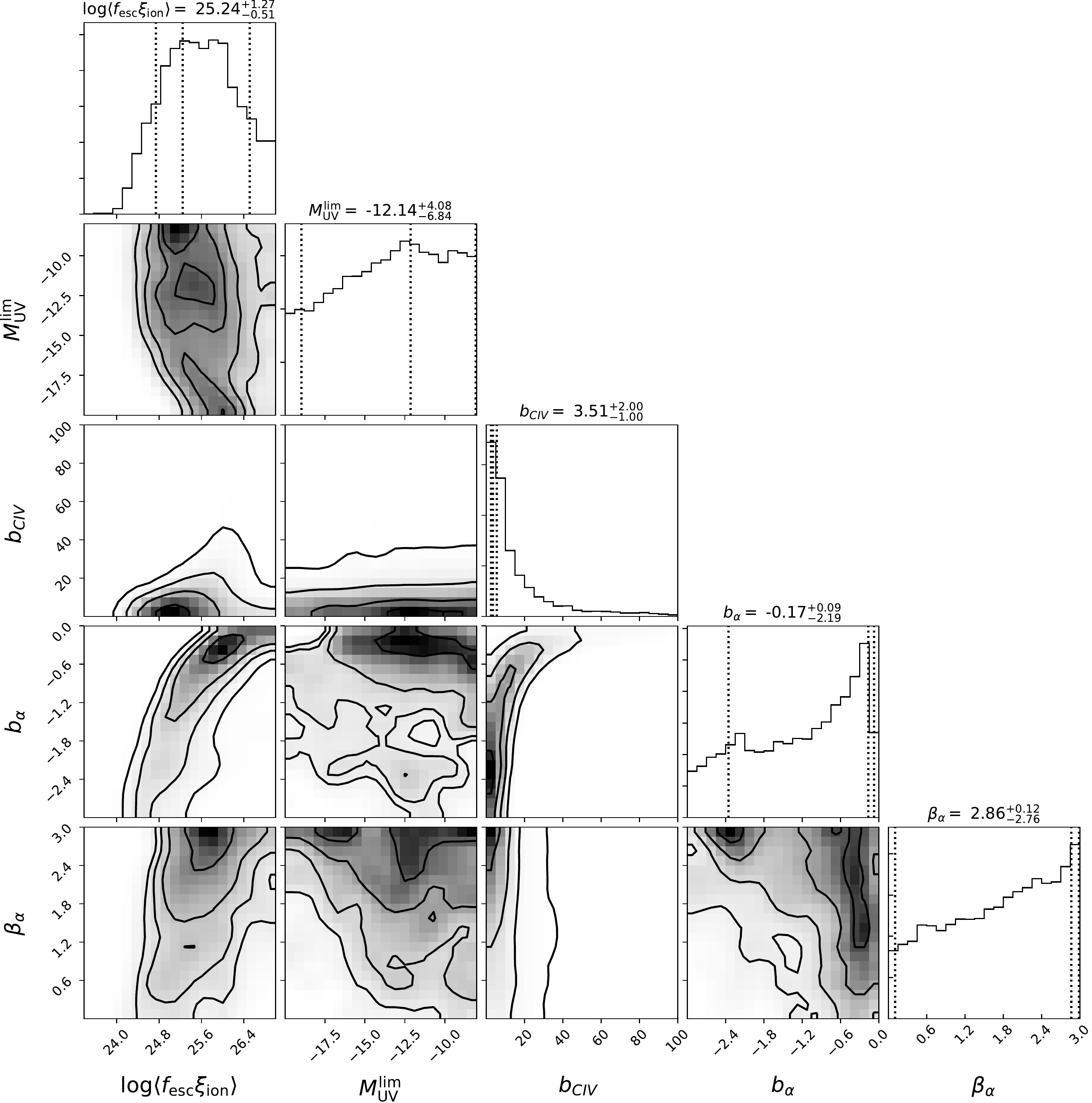}
\caption{Posterior distribution for the $3$ original parameters of the linear model to which have been added the Lyman-$\alpha$ bias parameters $(b_\alpha,\beta_\alpha)$ with appropriate priors. The quoted numbers give the best-fit (maximum likelihood) and the 1-$\sigma$ credibility intervals on the marginalized distributions. The constraints on $\langle f_{\text{esc}} \xi_{\text{ion}}\rangle$ and $b_{\cfourmath}$ are in good agreement with the more restricted fit presented in the main text (Fig.\ref{fig:posterior_linear}). There is a clear degeneracy between these two parameters and $b_\alpha$.  A wide variety of  combinations of the $4$ parameters (excluding $M_{\text{UV}}^{\text{lim}}$) can produce a similar fit the data. Hence a choice of $(b_\alpha, \beta_\alpha)$ must be made in order to infer the remaining parameters.\label{fig:posterior_bias}}
\end{figure*}

We present the posterior probability distribution of the parameters of our linear model including the bias of the Lyman-$\alpha$ and the associated RSD parameter as free parameters with flat priors in $-3 <b_\alpha<0 $ and $-3<\beta_f < 0$, respectively. The result is presented in Fig. \ref{fig:posterior_bias}. The bias is in a degenerate state with all other parameters, parameters can be tuned to compensate the bias parameter changes and still produce the same fit to the data.
We note that higher values of the bias than the one extrapolated from low-redshift measurements would yield in turn lower values of the host halo mass, and lower values of the escape fractions and LyC photon production product (see Table \ref{table:bias_beta_free}). 

\begin{table}
\caption{Inferred model parameters for different choices of Lyman-$\alpha$ bias parameters $(b_\alpha,\beta_\alpha)$. Notably, the choice of $b_\alpha =-1.3$ corresponds to our fiducial choice on the Lyman-$\alpha$ power spectrum measurement whereas $b_\alpha=-0.75$ corresponds the extrapolation of the evolution of the bias from low-redshift values \citep{McDonald06, duMasdesBourboux17}. \label{table:bias_beta_free}
}
\centering
\begin{tabular}{llrrcc}
\hline\hline
$b_\alpha$& $\beta_\alpha$ & $\log \langle f_{\text{esc}} \xi_{\text{ion}}\rangle$ & $M_{\text{UV}}^{\text{lim}}$ & $b_{\cfourmath}$ \\[0.1 cm]
\hline
-0.5 & 1.0  & $25.55^{+0.14}_{-0.11}$ & $-11.66^{+3.47}_{-8.16}$  & $22.15_{-8.29}^{+7.26}$  \\[0.1 cm]
-0.5 & 1.5 & $25.66^{+0.21}_{-0.19}$ & $-9.98^{+1.92}_{-9.96}$  & $18.48_{-5.75}^{+7.67}$   \\[0.1 cm]
-0.5 & 2.0  & $25.66^{+0.32}_{-0.14}$ & $-8.90^{+0.84}_{-10.92}$  & $16.65_{-5.92}^{+6.38}$   \\[0.1 cm]
-0.75 &1.0 & $25.25^{+0.19}_{-0.09}$ & $-19.94^{+11.87}_{-0.00}$  & $11.89_{-4.96}^{+7.80}$   \\[0.1 cm]
-0.75 & 1.5   &$25.35^{+0.23}_{-0.11}$ & $-19.58^{+11.40}_{-0.36}$  & $15.96_{-8.04}^{+2.25}$   \\[0.1 cm]
-0.75 & 2.0  & $25.54^{+0.22}_{-0.22}$ & $-8.78^{+0.72}_{-11.16}$  & $8.97_{-2.84}^{+7.95}$   \\[0.1 cm]
-1.3 & 1.0 & $24.92^{+0.23}_{-0.15}$ & $-10.22^{+2.16}_{-9.72}$  & $8.01_{-3.72}^{+2.84}$   \\[0.1 cm]
-1.3 & 1.5 & $25.01^{+0.30}_{-0.19}$ & $-10.82^{+2.76}_{-9.12}$  & $7.09_{-2.86}^{+3.29}$   \\[0.1 cm]
-1.3 & 2.0 & $25.17^{+0.28}_{-0.22}$ & $-8.06^{+0.00}_{-11.88}$  & $6.40_{-2.15}^{+2.69}$   \\[0.1 cm]
-2.0 & 1.0 & $24.63^{+0.15}_{-0.15}$ & $-8.06^{+0.00}_{-11.87}$  & $4.46_{-2.07}^{+2.85}$   \\[0.1 cm]
-2.0 & 1.5 &  $24.80^{+0.22}_{-0.26}$ & $-8.78^{+0.72}_{-11.16}$  & $4.33_{-2.20}^{+1.95}$   \\[0.1 cm]
-2.0 & 2.0 &  $24.74^{+0.55}_{-0.10}$ & $-11.90^{+3.84}_{-8.04}$  & $4.02_{-1.58}^{+1.69}$   \\[0.1 cm]
\hline
\end{tabular}

\end{table}

%%%%%%%%%%%%%%%%%%%%%%%%%%%%%%%%%%%%%%%%%%%%%%%%%%

% Don't change these lines
\bsp	% typesetting comment
\label{lastpage}
\end{document}